\documentclass[aps,prl,reprint,superscriptaddress,amsmath,amssymb,longbibliography,nofootinbib]{revtex4-2}
\usepackage[T1]{fontenc}
\usepackage{lmodern}
\usepackage{graphicx,xcolor,microtype}
\usepackage[colorlinks=true,linkcolor=blue,citecolor=blue,urlcolor=blue]{hyperref}
\hypersetup{pdftitle={Information Causality Defeats Every Finite NPA Level},pdfauthor={Anubhav Chaturvedi and Marcin Paw{\l}owski}}
\newcommand{\nsection}[1]{\section*{#1}}
\newcommand{\msection}[1]{\par\smallskip\noindent\textbf{#1}\enspace\ignorespaces}
\newcommand{\Q}{\mathcal Q}
\newcommand{\IC}{\mathcal I}
\newcommand{\id}{\mathbb I}
\newcommand{\ip}[2]{\langle #1,#2\rangle}
\newcommand{\norm}[1]{\lVert#1\rVert}

\newif\ifSupplementContents
\begin{document}
\title{Information Causality Defeats Every Finite NPA Level}
\author{Anubhav Chaturvedi}
\thanks{Both authors contributed equally to this work.}
\email{anubhav.chaturvedi@ug.edu.pl}
\affiliation{Institute of Theoretical Physics and Astrophysics, Faculty of Mathematics, Physics and Informatics, University of Gda\'nsk, 80-308 Gda\'nsk, Poland}
\author{Marcin Paw{\l}owski}
\affiliation{International Centre for Theory of Quantum Technologies, University of Gda\'nsk, 80-309 Gda\'nsk, Poland}
\date{16 November 2026}
\begin{abstract}
For every finite level of the Navascu\'es--Pironio--Ac\'in (NPA) hierarchy, we construct correlations that satisfy all its constraints but violate information causality. At each fixed level, one use of a binary-outcome correlation and one classical bit suffice for a violation that grows without bound as the number of measurement settings increases. Information causality bounds the information gained about independent database bits, summed over possible requests, by the length of the message. Quantum correlations obey the one-bit bound; the optimal information gain at any fixed NPA level grows logarithmically with the number of settings. The violation remains unbounded even when all outcome probabilities are positive and, for any fixed number of measurement settings on each side, every choice of those settings admits a local model. One communicated bit therefore exposes an unbounded gap between information causality and every finite NPA level.
\end{abstract}
\maketitle
\noindent With independent copies of a Popescu--Rohrlich correlation~\cite{PopescuRohrlich1994}, Bob can recover whichever bit he requests from an arbitrarily long database of independent bits using one bit from Alice. Alice never learns which bit Bob wants, and the correlations respect the prohibition on faster-than-light signalling~\cite{Pawlowski2009,ChaturvediPawlowskiHorodecki2017}. Information causality (IC) requires that the information Bob gains about the requested bit, summed over all possible requests, be at most the number of bits Alice sends~\cite{Pawlowski2009}. Quantum theory obeys this bound, whose derivation uses nonsignalling and entropy rules that can be stated independently of quantum theory (Methods).

The Navascu\'es--Pironio--Ac\'in (NPA) hierarchy bounds the correlations between distant measurements by imposing successively stronger consistency conditions on the measured probabilities~\cite{NPA2007,NPA2008}. For a fixed Bell scenario, its levels converge to the commuting-operator correlation set, in which the parties' measurements act on a common Hilbert space and commute across parties. The first level, $\Q_1$, coincides with macroscopic locality, which requires the fluctuations in outcome counts from many independent pairs to admit a local description~\cite{NavascuesWunderlich2010}. Both macroscopic locality and IC give the quantum bound on the Clauser--Horne--Shimony--Holt (CHSH) expression.

Cavalcanti, Salles and Scarani found IC violations in $\Q_1$ using larger outcome alphabets and concatenation~\cite{Cavalcanti2010}. The almost-quantum set $\Q_{1+AB}$ lies between $\Q_1$ and $\Q_2$ and satisfied the numerical IC tests reported when it was introduced~\cite{Navascues2015}. Das Bhowmik and Banik obtained unbounded IC violations with a single use of an almost-quantum correlation and one communicated bit~\cite{DasBhowmikBanik2026}, using the correlations of Gazi \emph{et al.}~\cite{Gazi2026}. The question is whether stronger consistency conditions can enforce IC at a finite level that works independently of the number of measurement settings.

For each fixed finite NPA level, we construct a family of binary-outcome correlations that satisfy every constraint at that level and violate IC by an amount that grows without bound as the number of settings increases. Each protocol uses the correlation once and sends one bit from Alice to Bob. The optimal information gain grows logarithmically with the number of settings at each fixed level. To enforce the quantum one-bit bound across these Bell scenarios, the hierarchy level must grow with the number of settings.

\nsection{Reading a database with one bit}

We construct binary-outcome correlations in which Alice and Bob can both obtain zero only when their measurement settings agree. At each fixed NPA level $L\geq2$, this event has the same probability $s_L>0$ at every pair of equal settings, independent of how many settings each party has. Let $x,y\in\{1,\ldots,N\}$ be Alice's and Bob's settings, and let $a,b\in\{0,1\}$ be their outcomes. Either party obtains zero with probability $3s_L$ at every setting, as in the almost-quantum correlations of Gazi \emph{et al.}~\cite{Gazi2026}. Write $\Q_L(N,N,2,2)$ for the level-$L$ set in this scenario and $\delta_{xy}$ for the indicator of $x=y$. For each integer $L\geq2$, we choose a rational $s_L>1/(185L^3)$ so that, for every $N\geq2$, the following correlation belongs to $\Q_L(N,N,2,2)$:
\begin{equation}
 p_{L,N}(a,b\mid x,y)=
 \begin{cases}
 s_L\delta_{xy},& a=b=0,\\
 3s_L-s_L\delta_{xy},& a\ne b,\\
 1-6s_L+s_L\delta_{xy},& a=b=1 .
 \end{cases}
 \label{eq:nat-family}
\end{equation}

A nonsignalling correlation cannot by itself transmit information from Alice to Bob. Information causality constrains how much information Bob can gain when Alice also sends him a classical message. To test IC, we therefore need a communication task and a protocol that uses the correlation.

In the IC task, Alice holds independent database bits $R_i$ and Bob is asked for one of them, without revealing the requested position $i$ to Alice. Their shared nonsignalling resource is independent of the database and request. She sends him a classical message $M$ of at most $m$ bits. Let $G_i$ be Bob's output for request $i$, obtained using the message and their shared correlation. Information causality requires that the Shannon mutual information, measured in bits and summed over possible requests, obey~\cite{Pawlowski2009}
\begin{equation}
 \IC:=\sum_i I(R_i:G_i)\leq m.
 \label{eq:nat-ic}
\end{equation}
This bound must hold for every communication protocol. To exhibit a violation, we therefore need to specify how Alice encodes the database, how the parties use the correlation, and how Bob produces his output. We do so with one use of Eq.~\eqref{eq:nat-family} and one communicated bit, for which the IC bound is $\IC\leq1$.

\begin{figure*}[t]
\centering
\includegraphics[width=\linewidth]{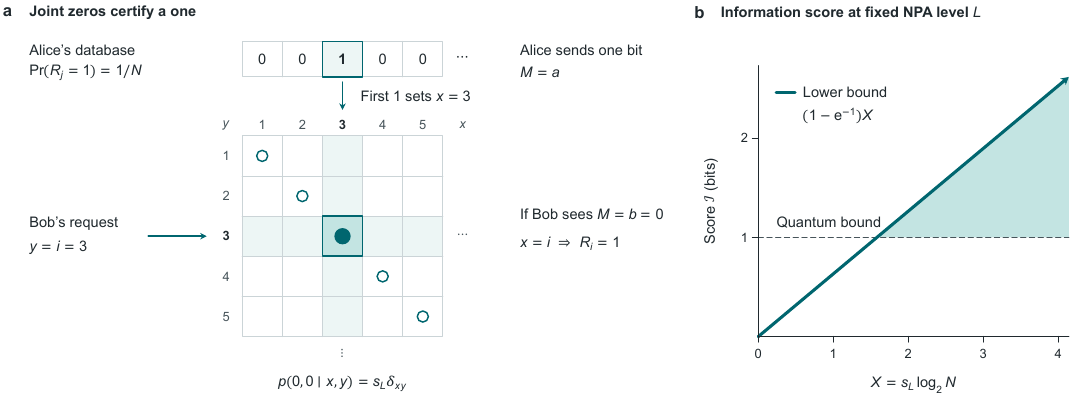}
\caption{\textbf{One-bit communication and matching settings.}
\textbf{a}, Alice's independent database bits are one with probability $1/N$. She uses the position $x$ of the first one as her setting and sends her outcome $M=a$. She sends $M=1$ on an all-zero database. Bob uses the requested position $i$ as his setting. The circles mark equal settings, where both outcomes are zero with probability $s_L$. The filled circle marks the example $x=y=i=3$. Bob records $G_i=1$ if $M=b=0$, which implies $R_i=1$, and $G_i=0$ otherwise.
\textbf{b}, The solid line gives the lower bound $(1-\mathrm e^{-1})X$ on the information score for $X=s_L\log_2 N$. The dashed line marks the one-bit quantum bound. The shading marks the amount by which the lower bound exceeds it. At fixed NPA level $L$, the rate $s_L>0$ is independent of $N$, so increasing $N$ makes the lower bound arbitrarily large.}
\label{fig:mechanism}
\end{figure*}

Take $N$ database bits, each equal to one with probability $r$, where $0<r<1$. Alice uses the position of the first one as her setting $x$ and sends her outcome $M=a$, adapting the block-selection protocol of Ref.~\cite{DasBhowmikBanik2026} to biased bits. If there is no one, she uses any fixed setting and sends $M=1$. Bob uses the requested position $i$ as his setting. He sets $G_i=1$ when $M=b=0$ and $G_i=0$ otherwise (Fig.~\ref{fig:mechanism}a). Joint zeros are impossible for $x\ne i$, so $G_i=1$ certifies that Alice's first-one position is $i$ and hence that $R_i=1$.

Upon seeing $G_i=1$, Bob learns that $R_i=1$, which had prior probability $r$. The relative entropy of this posterior from the prior is $\log_2(1/r)$ bits. For request $i$, this outcome has probability $s_Lr(1-r)^{i-1}$, since $i$ must be the first position with a one. For a given database, only a request for that position can give $G_i=1$. Summing over requests therefore gives $s_L$ times the chance that there is a one in the database. The output $G_i=0$ also contributes a nonnegative term, so the total information $\IC$ obeys
\begin{equation}
 \IC\ \geq\ s_L\bigl[1-(1-r)^N\bigr]\log_2(1/r).
 \label{eq:nat-score}
\end{equation}
At $r=1/N$, a one is rare at any given address and the probability of at least one in the database is at least $1-\mathrm e^{-1}$. The lower bound is $(1-\mathrm e^{-1})s_L\log_2 N$ (Fig.~\ref{fig:mechanism}b).

The same encoding works with fair bits by grouping them into blocks. Alice chooses the first all-ones block from $N$ blocks of $k$ independent fair bits and sends her outcome; if no such block exists, she sends one. Bob chooses the block containing his requested bit. This is the original block protocol~\cite{DasBhowmikBanik2026}. Joint zeros imply that every bit of the selected block is one. For each of the $k$ requests within that block, the probability of a one changes from the prior $1/2$ to certainty, giving one bit of relative entropy. An all-ones block has probability $2^{-k}$, so the score is at least $k s_L[1-(1-2^{-k})^N]$. Taking $k=\lfloor\log_2N\rfloor$ also gives logarithmic growth.

If joint zeros never occur at unequal settings, quantum correlations have average joint-zero probability at most $1/N$ across the $N$ equal settings. The same bound holds in the commuting-operator set $\Q_c$. In this model, Alice's and Bob's projectors act on a common Hilbert space and commute across parties; measurements on separate tensor factors are included. Let $s_i=p(0,0\mid i,i)$, with no restriction on the marginals. Let Alice measure her projectors in order, stopping at the first zero. Conditional on Bob obtaining zero at setting $i$, Alice cannot obtain zero at any earlier setting because joint zeros are absent at unequal settings. Each earlier measurement therefore gives outcome one with certainty and leaves her conditional state unchanged. The joint probability of Bob's zero and Alice's first zero at $i$ is therefore $s_i$. Alice's distribution for the position of her first zero is independent of Bob's setting by nonsignalling. The unconditional probability of her first zero at $i$ is at least $s_i$. These events are disjoint, so $\sum_i s_i\leq1$~\cite{KlauckDeWolf2013,Gazi2026}. A shared uniform label attains the bound if each party outputs zero exactly when its setting equals that label.

To determine how fast the information can grow, let $F_L(N)$ be the supremum of $\IC$ over all correlations in $\Q_L(N,N,2,2)$ and all protocols with at most one use of the correlation and at most one classical bit from Alice to Bob. The database may be any finite collection of independent bits with arbitrary biases. Shared randomness is independent of the database, the request and the correlation. The message alphabet has at most two values, including on runs where the correlation is unused. Bob may choose his setting using the message, and his output $G_i$ may be his full transcript. In quantum theory this optimum is one, attained by sending a fair database bit.

For a nonsignalling correlation, transmitting Alice's setting as well as her bit is sufficient to reproduce Bob's transcript classically. Before either party receives an input, draw a table with one sample of Alice's outcome at each of her settings. For each sampled outcome, draw a response for every Bob setting, conditioned on that Alice setting and outcome. In the simulation Alice generates her message using the table outcome at her actual setting. Bob receives the setting and the message, applies his original setting rule and reads the corresponding response. The table and all random controls are independent of the database. Data processing and independence of the bits bound the score by the conditional entropy of the setting and bit given this shared information, at most $\log_2 N+1$. At powers of two this bound is attained by a nonsignalling correlation (Methods). Combining this upper bound with Eq.~\eqref{eq:nat-score} and $s_L>1/(185L^3)$ gives, for $N,L\geq2$,
\begin{equation}
 \frac{(1-\mathrm e^{-1})\log_2N}{185 L^3}
 \ \leq\ F_L(N)\ \leq\ \log_2 N+1 .
 \label{eq:nat-rate}
\end{equation}
Hence $F_L(N)=\Theta_L(\log N)$ at each fixed $L\geq2$. By containment of level two in level one, this order also holds for $L=1$.

\nsection{The optimal information gain}

For the correlation in Eq.~\eqref{eq:nat-family}, the probability $s_L$ of two zeros on equal settings fixes the leading coefficient of the optimal information gain. Let $J_L(N)$ be the supremum over the same one-copy, one-bit protocols with this correlation fixed. Then
\begin{equation}
 J_L(N)=s_L\log_2N+O(1).
 \label{eq:nat-box-upper}
\end{equation}
The additive bounds are uniform in $L$. To make Alice use every setting equally often, give her $N-1$ independent bits with $\Pr(R_i=1)=1/(N-i+1)$. She uses the first-one position, or setting $N$ on an all-zero database, and sends her outcome in either case. The chosen setting is uniform over all $N$ labels. For each request $i<N$, the outcome $G_i=1$ occurs with probability $s_L/N$ and implies $R_i=1$, an event of prior probability $1/(N-i+1)$. The information from these outcomes sums to $(s_L/N)\log_2(N!)\geq s_L\log_2N-s_L\log_2\mathrm e$.

The upper bound uses a table that reproduces Bob's possible responses. An outcome impossible at unequal settings identifies Alice's setting in this table. For Eq.~\eqref{eq:nat-family}, that outcome is two zeros, with probability $s_L$ on equal settings. In the remaining cases the table contains only a bounded amount of information about Alice's setting (Methods and Supplementary Information).

\nsection{False coincidences and the optimal score}

If two zero outcomes can occur at unequal settings, Bob can see $M=b=0$ even when the addresses differ. Mix the correlation in Eq.~\eqref{eq:nat-family} with a fixed fraction $0<\epsilon<1$ of a local product correlation in which each party independently outputs zero with probability $0<u_N\leq1/2$. Neither party knows which correlation was used on a given run. Write $c=(1-\epsilon)s_L$ and $t_N=\epsilon u_N^2$, so joint zeros have probability $c+t_N$ on equal settings and $t_N$ on unequal settings. The optimal one-copy, one-bit score for this mixture is
\begin{equation}
 J_N=c\log_2\!\left(\frac{N}{1+Nt_N/c}\right)+O(1),
 \label{eq:nat-noise-crossover}
\end{equation}
with the remainder bounded uniformly in $N$ and $u_N$ at fixed $s_L,\epsilon$ (Supplementary Information). The database with a uniform first-one index attains the lower bound up to an additive constant.

For this database the event $M=b=0$ has probability $t_N+c/N$ at each request. The signal contributes $c/N$ because Alice uses each address with probability $1/N$. If $Nt_N\ll c$, the optimum grows as $c\log_2N$. If $Nt_N\gg c$, the scale is $c\log_2(c/t_N)$ up to an additive constant. In particular $u_N=\tfrac12N^{-\alpha/2}$ gives a coefficient $c\min\{\alpha,1\}$ of $\log_2N$ in the optimal score. Every $\alpha>0$ allows unbounded growth with positive probabilities for all outcomes. Uniform binary product noise ($u_N=1/2$, $\alpha=0$) bounds the optimum independently of $N$ for each fixed $\epsilon$.

To preserve a score of at least two, the noise may vary across settings. At each fixed level there is a positive tolerance in total variation, independent of $N$, such that any nonsignalling correlation within that tolerance of Eq.~\eqref{eq:nat-family} at every setting pair achieves this score for all sufficiently large $N$ (Supplementary Information).

\nsection{Locality on prescribed sets of settings}

How many settings must a Bell test use to detect the nonlocality of a correlation that violates the one-bit budget? Fix a level $L$, a number $K$ of settings per party and a target score $V$. For sufficiently large $N$, a level-$L$ correlation gives a one-copy, one-bit score above $V$, with Bell-local models on every restriction to at most $K$ settings per party and all outcome probabilities at least $1/(2N)$ (Supplementary Information).

Mixing Eq.~\eqref{eq:nat-family} with the deterministic outcome $(1,1)$ lowers the probability of two zeros on equal settings until its restriction to any choice of at most $K$ settings on each side admits a local model. Mixing the diluted correlation in equal proportions with local product noise of zero-outcome probability $1/\sqrt N$ preserves those local models and gives every joint outcome probability at least $1/(2N)$. The first-one protocol with $r=1/N$ on the resulting correlation achieves at least
\[
 \frac{(1-\mathrm e^{-1})\log_2N}{2\max\{185\max(L,2)^3,K+5\}}-1.
\]

\nsection{Constructing correlations at a fixed NPA level}

We must construct a positive moment matrix for Eq.~\eqref{eq:nat-family} while keeping $s_L$ independent of $N$. The proof has three steps: choose reference vectors with the required unnormalized probabilities, replace the vector for the identity while preserving every tested inner product, and bound the replacement norm independently of $N$. Fix $L\geq2$, and let $A_x$ and $B_y$ be the projectors for outcome zero, with complements $\id-A_x$ and $\id-B_y$. An ordered product of the outcome-zero projectors is a word $w$, of length $|w|$. At level $L$, the measured probabilities must admit a positive semidefinite moment matrix indexed by words of length at most $L$, with entries consistent with the projector identities and commutation across parties. The outcome-one projectors are recovered as complements. For a reference vector $R_N$, the vectors $wR_N$ with $|w|\leq L$ define an unnormalized Gram matrix. The empty word is the identity, so its vector is $R_N$. Normalizing the corresponding diagonal entry to one divides every moment by $\|R_N\|^2$.

\msection{1. Fix the unnormalized probabilities.}
Choose projectors and $R_N$ with the desired unnormalized moments. In a tensor-product block, $R_N$ pairs orthogonal directions of Alice and Bob bearing the same setting label. Each outcome-zero projector has a component along its own labelled direction. The two projectors select a matched component exactly when $x=y$, giving $\langle R_N,A_xB_yR_N\rangle=\delta_{xy}$. Two auxiliary blocks, one for each party, raise the unnormalized single-party zero moments from two to three and leave the joint-zero moments unchanged (Methods). The tensor-product block contributes $4N$ to $\|R_N\|^2$, so ordinary normalization would send the probability on equal settings to zero.

\msection{2. Replace the identity vector.}
At level $L$, only the vectors with $|w|\leq L$ enter the Gram matrix. Replace the identity vector $R_N$ by $S_L$, leaving the nonempty-word vectors and all their mutual inner products unchanged. Matching the inner products of $S_L$ with those vectors to the original ones then changes only the identity diagonal. For words in the outcome-zero projectors, we require
\begin{equation}
 \begin{gathered}
 \ip{S_L}{wR_N}=\ip{R_N}{wR_N}\\
 (1\leq |w|\leq L),\qquad C_L=\norm{S_L}^2.
 \end{gathered}
 \label{eq:nat-pairing}
\end{equation}
These conditions fix the projection of $S_L$ onto the span of the nonempty-word vectors, leaving its orthogonal component free. The condition at $w=A_xB_x$ is one, so $C_L>0$. Because the unnormalized joint-zero moment is one on equal settings, normalization by $C_L$ gives $s_L=1/C_L$. It remains to make $C_L$ independent of $N$.

The deterministic assignment in which both parties always output one gives an evaluation map $\varepsilon$ that obeys the projector identities. This multiplicative map sends the identity to one and every nonempty word in the outcome-zero projectors to zero. We use it to change the identity moment from $\|R_N\|^2$ to $C_L$. For a polynomial $p$ in the projectors, define
\begin{equation}
 C_L\ell(p)=\ip{R_N}{pR_N}
           +(C_L-\norm{R_N}^2)\,\varepsilon(p).
 \label{eq:nat-functional}
\end{equation}
The two terms obey the projector identities, including those for the complementary outcomes. The correction gives $\ell(\id)=1$ and vanishes on each nonempty word in the outcome-zero projectors.

The projector identities impose linear constraints on the moments; positivity still has to hold for every tested polynomial. Write a polynomial $f$ of degree at most $L$ as $f=c\id+g$, with $c=\varepsilon(f)$ and $\varepsilon(g)=0$. Multiplicativity gives $\varepsilon(g^\dagger g)=0$, so the quadratic term remains $\|gR_N\|^2$, involving reference moments through degree $2L$. Equation~\eqref{eq:nat-pairing} replaces $\langle R_N,gR_N\rangle$ by $\langle S_L,gR_N\rangle$ in the cross terms, which involve words only through degree $L$. The scalar term is $|c|^2C_L$. Thus
\begin{equation}
 C_L\ell(f^\dagger f)=\norm{c S_L+gR_N}^2 .
 \label{eq:nat-squared}
\end{equation}
The identity gives $\ell(f^\dagger f)\geq0$ for every polynomial of degree at most $L$, making the moment matrix $\Gamma_{u,v}=\ell(u^\dagger v)$ positive semidefinite. Expanding each outcome-one projector as the identity minus its outcome-zero projector preserves the degree bound. Each joint-outcome projector $E$ has degree at most two, so at $L\geq2$ its probability is $\ell(E)=\ell(E^\dagger E)\geq0$. Summing over either party's outcomes gives the other party's marginal, and the probabilities sum to $\ell(\id)=1$. The joint-zero moments and zero-outcome marginals are the reference values divided by $C_L$, giving Eq.~\eqref{eq:nat-family} with $s_L=1/C_L$. Thus a vector satisfying Eq.~\eqref{eq:nat-pairing} gives a valid level-$L$ moment matrix. It remains to construct such a vector with norm bounded independently of $N$.

\msection{3. Bound the normalization independently of the number of settings.}

To choose $S_L$ with bounded norm, give each local outcome-zero projector a direction with a component unique to its setting and components in modes common to all settings. The common components carry independent random signs, identical for Alice and Bob at each setting and mode. A product of rank-one projectors contains overlaps between consecutive directions along a sequence of settings. In the tensor-product block, a word involving both parties annihilates the reference vector unless the terminal setting in each party's sequence is the same. When they agree, join the sequences there, reversing Bob's. The reference pairing then requires the other endpoints to agree as well. The replacement vector pairs shared modes and inserts the corresponding signs at these endpoints. The auxiliary blocks supply the same endpoint condition for words from only one party (Methods).

For a closed sequence, or walk, the two endpoint signs in a mode multiply to one. For an open walk, expanding the overlaps assigns a mode to each step. Any overlap term with no step in the marked mode is independent of the two inserted endpoint signs, whose product has zero mean. The surviving average is a polynomial in the marked mode weight with zero constant term and degree at most the number of overlaps (Methods). Thus preserving closed-walk pairings and cancelling open-walk pairings reduces all the required inner products to finitely many conditions on the mode weights.

The longest walk has at most $L-1$ overlaps. It suffices to choose signed coefficients whose weighted sum evaluates every polynomial of degree at most $L-1$ at zero: the sum preserves the closed-walk averages and cancels the open-walk averages. Lagrange interpolation supplies such coefficients at $L$ positive nodes proportional to $(j-\tfrac12)^2$, for $j=1,\ldots,L$. Splitting a coefficient equally among orthogonal modes of the same weight preserves the evaluating sum and reduces its contribution to the squared tensor-product norm in proportion to the number of modes. A mode of coefficient zero supplies the remaining weight so that the common component has squared norm $1/2$. The multiplicities in Methods give $C_L<185L^3$ for every $N$. Since $s_L=1/C_L$, 
\begin{equation}
 s_L>\frac{1}{185L^3}.
 \label{eq:nat-quantitative}
\end{equation}
The matching conditions and this norm bound give a normalized level-$L$ correlation for every $N$, with $s_L$ independent of $N$. All weights and the normalization can be chosen rational. Allowing the ratio of the zero-outcome marginals to the diagonal probability to depend on $L$ gives a certificate with rate $\widetilde s_L>1/(165L^3)$ (Supplementary Information). With the rate in Eq.~\eqref{eq:nat-quantitative}, Eq.~\eqref{eq:nat-rate} guarantees a violation whenever $185L^3<(1-\mathrm e^{-1})\log_2 N$. The score also diverges along sequences with $L=o((\log_2 N)^{1/3})$.

\nsection{An upper bound on exact equality certificates}

For any level-$L$ correlation in which joint zeros are impossible at unequal settings, write $s_i=p(0,0\mid i,i)$. Such an outcome is an exact equality certificate: observing it proves that the settings agree. A polynomial in $J=\sum_{j=1}^{N}A_j$ bounds the average of the $s_i$, without assuming equal diagonal probabilities or marginals $3s_L$ (Methods).

For each positive odd integer $K\leq2L-3$ satisfying $K^2<6N+1$, the average diagonal rate satisfies
\begin{equation}
 \frac1N\sum_{i=1}^N s_i
 \leq
 \frac{1}{K^2}\left(1-\frac{K^2-1}{6N}\right)^{-2}.
 \label{eq:nat-pack}
\end{equation}
A rate $s_L$ feasible uniformly in $N$ then obeys $s_L\leq(2L-3)^{-2}$, by taking $K=2L-3$ and letting $N$ grow. At finite $N$, use the largest odd $K\leq\sqrt{2N}$. For $L\geq(K+3)/2$, this bounds the average rate by $(9/8+O(N^{-1}))/N$. At level $N$, the sequential projector argument recovers the exact quantum bound $1/N$ on the average joint-zero probability on equal settings for an exact certificate. For $N\geq L\geq2$, the same argument gives $s_L\leq1/(L+2)$ for the family with marginals $3s_L$ (Supplementary Information).

\nsection{Computing equality without errors}

The same joint outcome can be used to decide whether two labels agree. Alice receives $x$ and Bob receives $y$ in $\{1,\ldots,N\}$. Bob may declare them equal or unequal, or abstain. Every conclusive answer must be correct. For each input pair, the probability of giving one must be at least $1/2$. Using the labels as measurement settings, two zero outcomes certify equality. To certify inequality, the parties share a random binary hash with independently sampled fair values at the labels, independent of the copies. Equal labels have the same hash bit; each unequal pair has different hash bits with probability $1/2$.

Independent copies form a correlation in the same NPA level, with tuples of copy outcomes as the local outcomes (Supplementary Information). The parties use their labels as settings in every copy. For equal labels, at least one joint zero occurs with probability $1-(1-s_L)^m$, which is at least $1/2$ for $m=\lceil\ln 2/[-\ln(1-s_L)]\rceil$. Alice sends her $m$ outcomes and her hash bit in one message. Bob declares equality if any copy gives a joint zero, declares inequality if the hash bits differ, and otherwise abstains. The declarations are always correct, and the probability of a conclusive result is at least $1/2$ on each input pair. Equation~\eqref{eq:nat-quantitative} bounds the message length $m+1$ by $O(L^3)$ bits, independently of $N$. For $N=2^n$, the minimum one-way cost with quantum or commuting-operator correlations is $n$ bits at this worst-case conclusive probability~\cite{Gazi2026}. For a related equality task with one-sided error, extended IC yields a linear one-way lower bound~\cite{MiklinJainGachechiladze2026}.

Within the three-block rank-one construction specified in Supplementary Information, with zero-outcome marginals fixed at $3s_2$, the maximum certificate probability at level two is $s_2^{\mathrm{ref}}=1/(35+18\sqrt{2})$ (Supplementary Information). A rational choice gives $s_2^{\mathrm{rat}}=7/424$. This takes 42 copies, for a total of 43 bits including the hash, irrespective of $N$.

Hardy and Cabello tests also use forbidden events to draw conclusions from observed probabilities. They have been developed into device-independent dimension witnesses~\cite{Mukherjee2015} and self-tests of quantum states~\cite{Rai2021,Rai2022}. Hardy correlations can also assist reverse zero-error channel coding: a noiseless channel with smaller input and output alphabets can simulate a larger noisy channel with their help~\cite{Alimuddin2023}. In our equality task, the zeros at unequal settings ensure that a joint-zero event cannot give a false equality declaration.

\nsection{Information causality and finite-level moment tests}

With the Bell scenario held fixed, the NPA hierarchy converges to $\Q_c$~\cite{NPA2008}. Even there, the upper bound on a fixed Bell functional can remain above the commuting-operator value at every finite level~\cite{Fanizza2025}. Self-testing analyses of doubly-tilted CHSH inequalities have already revealed gaps between quantum maxima and high finite NPA levels~\cite{Gigena2025}. In the binary-input, binary-output scenario, the symmetrically doubly-tilted CHSH family separates each finite level from the quantum set near its critical tilt~\cite{Chaturvedi2026}. With the level fixed and the number of settings growing, our one-bit protocol gives an arbitrarily large IC violation. For sufficiently large $N$, the functional in Eq.~\eqref{eq:nat-functional} assigns a negative value to a square of higher degree (Supplementary Information).

Some correlations excluded by macroscopic locality obey the necessary inequalities used in equal-bias tests~\cite{Gazi2013}. Earlier comparisons also used Hardy's paradox: local orthogonality tightened the available IC and macroscopic-locality bounds on its success probability~\cite{Das2013}. Allowing a noisy communication channel yields additional IC constraints~\cite{MiklinPawlowski2021,Gachechiladze2022}. With correlated database entries, IC gives polynomial inequalities across bipartite scenarios~\cite{JainGachechiladzeMiklin2024}. A noisy-channel formulation with correlated entries also implies the Tsirelson--Landau--Masanes condition for the four normalized covariances in the binary-input, binary-output scenario~\cite{GachechiladzeMiklin2026}. The full behavior also depends on the marginals. Variational and matrix inequalities give single-copy tests in fixed scenarios~\cite{PompiliBudroni2026}. Nonsignalling also imposes complementarity relations between genuine multipartite nonlocality and the nonlocality of subsets~\cite{Sami2017}. In multipartite scenarios IC constrains the sharing of Bell violations and device-independent key rates against individual attacks~\cite{PollycenoChavesRabelo2023,Pollyceno2025}. Communication constraints also exclude correlations through tasks beyond random access codes, independently of a particular encoding and decoding strategy~\cite{Pollyceno2026}.

The communication task determines which property of a correlation is tested. Bell inequalities can be tailored to bounded-message tasks so that their violation expresses an advantage from entanglement-assisted classical communication~\cite{Rout2025}. Bounding how well the message reveals the sender's input gives a different measure of communication: quantum advantage can then grow without bound in random-access and equality tasks~\cite{Manna2024}. With prior entanglement, the comparison also depends on whether message dimension or input distinguishability is constrained~\cite{Manna2026}. Allowing the communicated system itself to vary, IC-based random-access tests also constrain how locally quantum systems may be composed~\cite{Patra2023}.

Random-access tasks admit further operational restrictions. The receiver may request functions of the input string~\cite{Vaisakh2021}, or the message may be required to conceal parities of that string~\cite{Ambainis2019}. Preparation contextuality can be witnessed through quantum advantages in oblivious communication tasks~\cite{SahaChaturvedi2019}. Semidefinite relaxations also bound quantum behaviours in prepare-and-measure contextuality scenarios~\cite{ChaturvediFarkasWright2021}. Random-access performance also supports semi-device-independent key distribution under a dimension bound~\cite{Chaturvedi2018Security} or operational constraints on the source~\cite{Chaturvedi2026OperationalQKD}. In the present task, the fixed classical message length gives the IC bound, and a single protocol exceeding it excludes the shared correlation.

Pompili and Weilenmann give correlations outside the almost-quantum set that obey IC for every one-way protocol using finitely many independent copies, including adaptive protocols~\cite{PompiliWeilenmann2026}. They lie outside every standard NPA level $L\geq2$, since these levels are contained in the almost-quantum set. Duplicating their settings puts the example in the same scenarios as our construction: a protocol using a duplicate can use the original setting, and restricting to the original settings still gives a correlation outside $\Q_L$. For each fixed $L\geq2$ and sufficiently large $N$, our correlation belongs to $\Q_L$ and violates IC in one use, whereas their extended correlation obeys IC for every one-way protocol using finitely many independent copies, including adaptive protocols, and lies outside $\Q_L$. The two sets are therefore incomparable in these scenarios.
\nsection{Methods}

\msection{The information budget from nonsignalling and entropy rules.}
Let $R=(R_j)_j$ be a database of independent classical entries, $B$ Bob's system before receipt of the message, and $M$ Alice's classical message. Assume a symmetric, nonnegative mutual information on these systems and their composites. It agrees with Shannon mutual information on classical registers and obeys data processing under local transformations. We also assume the chain rule,
\[
 I(A:BC)-I(A:C)=I(B:AC)-I(B:C).
\]
This difference defines $I(A:B\mid C)$~\cite{Pawlowski2009,AlSafiShort2011,WakakuwaMurao2012}. For request $j$, Bob obtains $G_j$ by local processing of $(M,B)$. Write $R_{<j}$ for the database entries before $j$. Data processing and the chain rule give
\begin{align*}
 \sum_j I(R_j:G_j)&\leq\sum_j I(R_j:MB)\\
 &\leq\sum_j I(R_j:R_{<j}MB)\\
 &=I(R:MB).
\end{align*}
Independence gives $I(R_j:R_{<j})=0$, so each term on the second line equals $I(R_j:MB\mid R_{<j})$. Summing over $j$ gives $I(R:MB)$ by the chain rule. Before communication, nonsignalling keeps Bob's system independent of the database. Data processing from a trivial register then gives $I(R:B)=0$. By the chain rule and $I(B:M)\geq0$, we have $I(R:MB)=I(R:M\mid B)\leq I(RB:M)$.

Because $M$ is classical, we may introduce a copy $M'$ to bound its information. Data processing under local copying and discarding gives $I(RB:MM')=I(RB:M')$, so $I(RB:M\mid M')=0$. Therefore
\begin{align*}
 I(RB:M)&\leq I(RBM':M)\\
 &=I(M':M)+I(RB:M\mid M')=H(M).
\end{align*}
For an $m$-bit message, $H(M)\leq m$, proving the information-causality bound.

\msection{Simulating one copy with a setting label and one bit.}
Condition on the classical randomness used by the original protocol, which is independent of the supplied correlation. Represent a run that skips the query by a fixed setting whose outcome is ignored. Write $X,A,M$ for Alice's setting, outcome and one-bit message. Before the database and request arrive, use additional shared randomness to draw one outcome $a_x\sim p_A(\cdot\mid x)$ at each Alice setting $x$. Nonsignalling makes $p_A$ independent of Bob's setting. Conditional on each $a_x$, draw responses $g_x(y)$ independently from $p(\cdot\mid a_x,x,y)$ for all of Bob's settings $y$.

Alice uses $a_X$ to generate $M$ by her original rule and sends $(X,M)$. Bob chooses $y$ by his original rule, which may depend on $M$, and uses $g_X(y)$ as his outcome. For each request this reproduces the original joint distribution of the database and Bob's transcript. Let $T$ collect the original classical randomness and the response table. At a fixed request, Bob's transcript is a function of $(X,M,T)$. Since $T$ is independent of the database,
\begin{align*}
 \sum_j I(R_j:G_j)
 &\leq\sum_j I(R_j:X,M\mid T)\\
 &\leq I(R:X,M\mid T)\\
 &\leq H(X,M\mid T)\leq\log_2 N+1 .
\end{align*}
The database entries are independent conditional on $T$, giving the second inequality.

For $N=2^k$ with $k\geq1$, a nonsignalling correlation attains this bound. Give Alice $k+1$ independent fair bits $R_0,\ldots,R_k$. She chooses $x=(R_1\oplus R_0,\ldots,R_k\oplus R_0)$ and sets $x_0=0$. Alice's outcome $a$ is uniform, and Bob's outcome is $b=a\oplus x_y$ for his setting $y\in\{0,\ldots,k\}$. Since $k+1\leq2^k$, any remaining Bob settings can duplicate setting zero. Both outcome marginals are uniform, so the correlation is nonsignalling. Alice sends $M=a\oplus R_0$ and Bob obtains $M\oplus b=R_y$ for every request, for a score of $k+1$.

For the correlation in Eq.~\eqref{eq:nat-family}, fair database bits also attain the leading coefficient $s_L$. The all-ones block protocol gives a score of at least $k s_L[1-(1-2^{-k})^N]$. For large $N$, take $k=\lfloor\log_2N-\log_2\log_2N\rfloor$. Then $N2^{-k}\geq\log_2N$, and the score is at least $s_L\log_2 N-O_L(\log\log N)$.

\msection{The leading coefficient of the information gain.}
Fix two joint-outcome distributions with identical single-party marginals, using one on equal settings and the other on unequal settings. Let $\rho$ be the equal-setting probability of outcome pairs impossible at unequal settings. For a given Alice setting and outcome, draw a table of independent responses for Bob's possible settings. The entry at Alice's setting uses the equal-setting conditional distribution; all other entries use the unequal-setting distribution. For each request and fixed classical randomness in the original protocol, Bob chooses his setting by the original rule, which may depend on the message, and reads that entry.

An entry impossible under the unequal-setting conditional distribution identifies Alice's setting. Averaged over Alice's outcome, this event has probability $\rho$. On the remaining runs, fix Alice's outcome $a$. Let $V_a$ be Bob's unequal-setting outcome law and $U_a$ his equal-setting law conditioned on outcomes possible under $V_a$. For any fixed Alice setting, the table differs from the reference law $V_a^{\otimes N}$ in just one entry. Its relative entropy from that law is $D(U_a\Vert V_a)$, independent of $N$. The relative-entropy chain rule therefore bounds the table's information about Alice's setting by this same quantity. Averaging over Alice's outcomes and the two kinds of runs gives an upper bound $\rho\log_2N+O(1)$ for any fixed message alphabet (Supplementary Information, Sec.~\ref{supp:optimal-equality-score}). Choosing the independent bit biases so that Alice's setting is uniform attains the coefficient whenever the message can encode her outcome. In Eq.~\eqref{eq:nat-family}, two zeros are the only outcome pair impossible at unequal settings, so $\rho=s_L$.

\msection{Projectors and reference vectors.}
We construct the reference vector $R_N$ and replacement vector $S_L$ required by Eq.~\eqref{eq:nat-pairing}. Choose finitely many modes $\alpha$ with positive weights $\lambda_\alpha$ summing to $1/2$, and independent fair signs $\sigma_{i,\alpha}\in\{-1,1\}$ for every setting $i$ and mode $\alpha$. The mode weights and coefficients below depend on $L$ alone. Alice and Bob use separate local spaces with the same sign assignment.

In Alice's space, let $e_i$ be orthonormal setting-specific directions and $g_{A,\alpha}$ orthonormal shared directions, orthogonal to all the $e_i$. Bob has corresponding directions $f_i$ and $g_{B,\alpha}$. Put $h_{A,i}=\sum_\alpha\sqrt{\lambda_\alpha}\,\sigma_{i,\alpha}g_{A,\alpha}$ and define $h_{B,i}$ with the same signs. In the tensor-product block, $A_i$ and $B_i$ project onto the unit vectors $u_i=h_{A,i}+e_i/\sqrt2$ and $v_i=h_{B,i}+f_i/\sqrt2$ in their respective factors. For $i\ne j$, only the shared components contribute to the overlap,
\begin{equation}
 X_{ij}=\sum_\alpha\lambda_\alpha\sigma_{i,\alpha}\sigma_{j,\alpha}.
 \label{eq:nat-overlap}
\end{equation}
Use $T_N=2\sum_i e_i\otimes f_i$ as the reference vector in this block. It has squared norm $4N$ and satisfies $A_iB_jT_N=\delta_{ij}u_i\otimes v_i$. Its unnormalized joint-zero moments on equal settings are one, and its single-party zero moments are two.

The replacement vector pairs shared modes. Let $\nu_\alpha$ be real coefficients, not all zero, to be chosen so that the word pairings agree. An endpoint in mode $\alpha$ contributes $\sqrt{\lambda_\alpha}$; dividing by $\lambda_\alpha$ cancels the two endpoint factors and leaves coefficient $\nu_\alpha$. Thus take
\[
 S_0=\sum_\alpha (\nu_\alpha/\lambda_\alpha)\,g_{A,\alpha}\otimes g_{B,\alpha}.
\]
For a word from only one party, the other party's factor remains in its setting-specific subspace, so its pairing with $S_0$ vanishes. Add one auxiliary block with Alice's projectors as before and all $B_i=0$, and a second block with the parties interchanged. In each block the inactive party's outcome-one projector is the identity. Choose positive $q_\alpha$ on the modes with $\nu_\alpha\ne0$, with $\sum_\alpha q_\alpha^2=1$, and use
\begin{align*}
 R_{\mathrm{aux}}&=\sum_{\alpha:\nu_\alpha\ne0}\frac{q_\alpha}{\sqrt{\lambda_\alpha}}g_\alpha,\\
 S_{\mathrm{aux}}&=\sum_{\alpha:\nu_\alpha\ne0}
             \frac{q_\alpha+2\nu_\alpha/q_\alpha}{\sqrt{\lambda_\alpha}}g_\alpha .
\end{align*}
In a fixed mode, these coefficients change the product of endpoint amplitudes from $q_\alpha^2$ in the reference pairing to $q_\alpha^2+2\nu_\alpha$ in the replacement pairing. The extra term will supply the missing tensor-product contribution for one-party words.

Take $R_N=T_N\oplus R_{\mathrm{aux}}\oplus R_{\mathrm{aux}}$ and $S_L=S_0\oplus S_{\mathrm{aux}}\oplus S_{\mathrm{aux}}$, using the modes in their respective spaces. The squared reference norm is $4N+2\|R_{\mathrm{aux}}\|^2$. Average over signs by taking a finite direct sum over all sign assignments, weighting each vector by the square root of the assignment's uniform probability. Each auxiliary block adds one to the active party's unnormalized marginal and leaves all mixed moments unchanged. The complete reference therefore has joint-zero moments $\delta_{ij}$ and single-party zero moments three, as required.

\msection{Reducing word pairings to endpoint averages.}
Commute projectors of different parties and remove consecutive repeated settings by idempotence. Each party's product gives one overlap from Eq.~\eqref{eq:nat-overlap} for every step along its sequence of settings. We call this sequence a walk. A one-party word has at most $L-1$ steps. A mixed word gives two walks with at most $L-2$ steps in total and acts nontrivially on $T_N$ only when their terminal labels agree. Join Alice's walk to the reverse of Bob's at that label. For either kind of word, write $W=(i_0,\ldots,i_e)$ for the resulting walk and $M_W$ for its product of overlaps.

For a mixed word with matching terminal labels, pairing with $T_N$ also requires $i_0=i_e$, giving $\delta_{i_0i_e}\mathbb E M_W$. Pairing with $S_0$ inserts the two endpoint signs in the same mode, giving
\[
 \sum_\alpha\nu_\alpha\,\mathbb E[\sigma_{i_0,\alpha}\sigma_{i_e,\alpha} M_W].
\]
A one-party word gives $2\delta_{i_0i_e}\mathbb E M_W$ in the tensor-product reference and zero in its replacement. In its auxiliary block, a term with distinct endpoint modes averages to zero: flipping all signs in one mode reverses its sign and leaves every overlap unchanged. For equal endpoint modes, the difference between the auxiliary replacement and reference pairings is twice the sum above. It therefore suffices to make this sum equal to $\delta_{i_0i_e}\mathbb E M_W$ for every walk of at most $L-1$ steps.

\msection{Choosing the coefficients to match every word.}
When the walk closes, the endpoint signs multiply to one, so $\sum_\alpha\nu_\alpha=1$ preserves its reference average. For a single open step, $\mathbb E[\sigma_{i,\alpha}\sigma_{j,\alpha}X_{ij}]=\lambda_\alpha$, and cancellation requires $\sum_\alpha\nu_\alpha\lambda_\alpha=0$. Longer open walks give polynomials in the marked mode weight; their degree is bounded by the number of steps.

Fix an open walk $W=(i_0,\ldots,i_e)$ with $i_0\ne i_e$ and write
$F_\alpha=\mathbb E[\sigma_{i_0,\alpha}\sigma_{i_e,\alpha}M_W]$.
Expanding the overlap product assigns a mode to each edge. A term
with no edge in mode $\alpha$ averages to zero because the product
is independent of the two inserted signs. Group the other terms by
the number $r$ of edges in mode $\alpha$ to obtain
\[
 F_\alpha=\sum_{r=1}^{e}\lambda_\alpha^r
 Q_{W,r}\bigl((\lambda_\beta)_{\beta\ne\alpha}\bigr).
\]
Permuting the unmarked modes leaves the sign average unchanged.
Thus $Q_{W,r}$ is the same symmetric polynomial for each marked
mode $\alpha$, homogeneous of degree $e-r$ in the other weights.

Write $p_k=\sum_\beta\lambda_\beta^k$ for the full power sums.
We express $Q_{W,r}$ using the power sums over the list that omits
$\alpha$. For a list with power sums $P_k$, let $D(d_1,\ldots,d_m)$ be the sum of
$\lambda_{b_1}^{d_1}\cdots\lambda_{b_m}^{d_m}$ over distinct
indices. Multiplying
$D(d_2,\ldots,d_m)$ by $P_{d_1}$ allows the new index to equal one
of the others. Subtract the terms with these coincidences to obtain
\[
 \begin{aligned}
 D(d_1,\ldots,d_m)
 &=P_{d_1}D(d_2,\ldots,d_m)\\
 &\quad-\sum_{j=2}^{m}D(d_2,\ldots,d_j+d_1,\ldots,d_m).
 \end{aligned}
\]
Starting from $D(d_1)=P_{d_1}$ gives a power-sum expression for each
such symmetric sum that preserves its total degree. Grouping the
monomials by their permutations writes $Q_{W,r}$ as a linear
combination of these sums.

For the list excluding $\alpha$, use $P_k=p_k-\lambda_\alpha^k$.
With the full sums $p_k$ held fixed, this gives
\[
 F_\alpha=P_W(\lambda_\alpha),\qquad
 \deg P_W\leq e,\qquad P_W(0)=0.
\]
The coefficients of $P_W$ are independent of $\alpha$. For the
three-edge walk $(i,j,i,j)$ the expression is
\[
 \mathbb E[\sigma_{i,\alpha}\sigma_{j,\alpha}X_{ij}^{3}]
 =3p_2\lambda_\alpha-2\lambda_\alpha^3.
\]
To preserve closed-walk averages and cancel every open-walk polynomial of degree at most $L-1$, choose
\begin{equation}
 \sum_\alpha \nu_\alpha=1,\qquad
 \sum_\alpha \nu_\alpha\lambda_\alpha^k=0\quad (1\leq k\leq L-1).
 \label{eq:nat-power}
\end{equation}
These conditions make the signed sum evaluate every such polynomial at zero. Since $P_W(0)=0$ for an open walk, its pairing vanishes; for a closed walk, the average is unchanged. The reference and replacement vectors therefore satisfy Eq.~\eqref{eq:nat-pairing} for every nonempty word of length at most $L$.

\msection{Bounding the replacement norm.}
We choose the interpolation weights so that $\|S_L\|^2$ grows at most
cubically with $L$ and is independent of $N$.
For $1\leq j\leq L$, let $t_j=j-\tfrac12$,
$T=4\sum_jt_j^2=(4L^3-L)/3$ and $\lambda_j=3t_j^2/(2T)$.
The Lagrange coefficients
$\mu_j=\prod_{k\ne j}\lambda_k/(\lambda_k-\lambda_j)$ evaluate
polynomials of degree at most $L-1$ at zero. They obey
$|\mu_j|\leq2/(\pi t_j)$ and $\sum_j\mu_j^2\leq2$
(Supplementary Information).

Splitting a coefficient among orthogonal modes of equal weight leaves
the signed moments unchanged and reduces its contribution to the norm.
At each node, take $n_j=\lceil T/(60t_j^4)\rceil$ independent modes
of weight $\lambda_j$, each with signed coefficient $\mu_j/n_j$.
Their total weight is at most
$3/8+\sum_j t_j^{-2}/40\leq 3/8+\pi^2/80<1/2$.
Add one mode of the remaining weight with signed coefficient zero.
The signed moments in Eq.~\eqref{eq:nat-power} are unchanged by
splitting the coefficients, and the squared norm of $S_0$ satisfies
\[
 \|S_0\|^2=\sum_j\frac{\mu_j^2}{n_j\lambda_j^2}
 \leq\frac{160}{3}T.
\]
For the two auxiliary blocks, set $H_*=\sum_\alpha|\nu_\alpha|/\sqrt{\lambda_\alpha}$
and $q_\alpha^2=|\nu_\alpha|/(H_*\sqrt{\lambda_\alpha})$ on the
active modes. Then $\sum_\alpha q_\alpha^2=1$ and
$H_*^2\leq2\pi^2T/3$. To evaluate the signed sum that enters the
auxiliary norm, use the polynomial $[1-\prod_j(1-z/\lambda_j)]/z$.
It takes the value $1/\lambda_j$ at node $j$ and $\sum_j1/\lambda_j$
at zero. Interpolation gives $\sum_\alpha\nu_\alpha/\lambda_\alpha
=\sum_j1/\lambda_j\leq\pi^2T/3$. The remaining sum
$\sum_\alpha q_\alpha^2/\lambda_\alpha$ is at most
$1/\lambda_1=8T/3$. The expanded auxiliary norm then obeys
\begin{align*}
 \|S_{\mathrm{aux}}\|^2
 &=\sum_\alpha\frac{q_\alpha^2}{\lambda_\alpha}
    +4\sum_\alpha\frac{\nu_\alpha}{\lambda_\alpha}+4H_*^2\\
 &\leq\left(\frac83+4\pi^2\right)T ,
\end{align*}
so adding the three blocks gives
\[
 C_L\leq \left(\frac{176}{3}+8\pi^2\right)T
 <\frac{1664}{9}L^3<185L^3.
\]
For a mode of weight $\lambda_j$, the factor
$H_*\sqrt{\lambda_j}=t_j\sum_k|\mu_k|/t_k$ is rational.
All weights, signed coefficients and $q_\alpha^2$ are rational,
so the norm formula gives rational $C_L$ and $s_L=1/C_L$.

\msection{Polynomial bound for exact certificates.}
Suppose joint zeros are impossible on unequal settings. At level $L\geq2$, the moment vector for $A_jB_i$ then vanishes whenever $j\ne i$, since its squared norm is $p(0,0\mid j,i)$. With $J=\sum_{j=1}^N A_j$, the vectors for $JB_i$ and $A_iB_i$ coincide. This fixes the first two moments of $J$ at $B_i$ to $s_i$. We use a polynomial $q(J)$ to make the sum of squared moment norms of $A_iq(J)$ small. The same polynomial keeps the inner products of $A_iq(J)B_i$ with the diagonal vectors $A_iB_i$ large. Summing these overlaps gives a lower bound in terms of $\sum_i s_i$. The polynomial's degree determines the level needed for both bounds.

Let $\ell$ be the normalized level-$L$ moment functional of an exact equality certificate and write $\|f\|_\ell^2=\ell(f^\dagger f)$ for the moment seminorm of a polynomial with $\deg f\leq L$. Choose $K=2d+1$ with $0\leq d\leq L-2$ and $K^2<6N+1$. For $x=N\sin^2\theta$, $0\leq\theta\leq\pi/2$, define the degree-$d$ polynomials
\[
 q(x)=\frac{\sin(K\theta)}{K\sin\theta},\qquad
 t(x)=\frac{\cos(K\theta)}{\cos\theta}.
\]
Both quotients have continuous extensions at the endpoints. The sum of squared moment norms of $A_iq(J)$ is $\ell(Jq(J)^2)$. To bound it, use
\[
 \frac N{K^2}-xq(x)^2=\frac{N-x}{K^2}t(x)^2
\]
At $J$, the right-hand side is a sum of squares of $K^{-1}(1-A_j)t(J)$, giving $\ell(Jq(J)^2)\leq N/K^2$.

To control the diagonal overlaps, set $\Delta=1-(K^2-1)/(6N)>0$ and $h(x)=x[q(x)-1+(K^2-1)x/(6N)]$. The first two moments of $J$ at $B_i$ both equal $s_i$, so we need $\ell(B_i h(J)B_i)\geq0$. The cosine expansion of $q$ writes $h$ as a positive weighted sum of $x^2\sin^2(r\theta)$ for integers $1\leq r\leq d-1$. These terms are nonnegative multiples of $x^3P(x)^2$ for odd $r$ and $x^3(N-x)P(x)^2$ for even $r$, for polynomials $P$ of suitable degree. With $J=\sum_j A_j^2$ and $J(N-J)=\sum_{j<k}(A_j-A_k)^2$, the expression $B_i h(J)B_i$ is a sum of squares of polynomials of degree at most $d+2$ (Supplementary Information). Since $d+2\leq L$, level-$L$ positivity applies to these squares.

Using $\ell(B_i JB_i)=\ell(B_i J^2B_i)=s_i$ now yields $\ell(B_i Jq(J) B_i)\geq\Delta s_i$. Since the vectors for $JB_i$ and $A_iB_i$ coincide, idempotence of $A_i$ makes this last moment the inner product of $A_iB_i$ with $A_iq(J)B_i$. Both vectors have degree at most $d+2\leq L$. Cauchy--Schwarz gives $\|A_i q(J)B_i\|_\ell^2\geq\Delta^2 s_i$. By commutation across parties, the squared seminorm $\|A_i q(J)\|_\ell^2$ splits into contributions from $B_i$ and its complementary outcome, so it bounds the $B_i$ contribution from above. Summing over $i$ gives $\Delta^2\sum_i s_i\leq\ell(Jq(J)^2)\leq N/K^2$.
\begin{acknowledgments}
M.P. acknowledges support from the International Research Agendas programme of the Foundation for Polish Science (FNP), through the ICTQT 2.0 project (FENG.02.01-IP.05-0006/23), financed by the European Funds for Smart Economy 2021--2027, Priority FENG.02, Measure FENG.02.01.
\end{acknowledgments}
\makeatletter
\immediate\write\@auxout{\string\citation{apsrev42Control}}
\makeatother
\bibliographystyle{apsrev4-2}
\bibliography{references}

\begin{thebibliography}{43}%
\makeatletter
\providecommand \@ifxundefined [1]{%
 \@ifx{#1\undefined}
}%
\providecommand \@ifnum [1]{%
 \ifnum #1\expandafter \@firstoftwo
 \else \expandafter \@secondoftwo
 \fi
}%
\providecommand \@ifx [1]{%
 \ifx #1\expandafter \@firstoftwo
 \else \expandafter \@secondoftwo
 \fi
}%
\providecommand \natexlab [1]{#1}%
\providecommand \enquote  [1]{``#1''}%
\providecommand \bibnamefont  [1]{#1}%
\providecommand \bibfnamefont [1]{#1}%
\providecommand \citenamefont [1]{#1}%
\providecommand \href@noop [0]{\@secondoftwo}%
\providecommand \href [0]{\begingroup \@sanitize@url \@href}%
\providecommand \@href[1]{\@@startlink{#1}\@@href}%
\providecommand \@@href[1]{\endgroup#1\@@endlink}%
\providecommand \@sanitize@url [0]{\catcode `\\12\catcode `\$12\catcode
  `\&12\catcode `\#12\catcode `\^12\catcode `\_12\catcode `\%12\relax}%
\providecommand \@@startlink[1]{}%
\providecommand \@@endlink[0]{}%
\providecommand \url  [0]{\begingroup\@sanitize@url \@url }%
\providecommand \@url [1]{\endgroup\@href {#1}{\urlprefix }}%
\providecommand \urlprefix  [0]{URL }%
\providecommand \Eprint [0]{\href }%
\providecommand \doibase [0]{https://doi.org/}%
\providecommand \selectlanguage [0]{\@gobble}%
\providecommand \bibinfo  [0]{\@secondoftwo}%
\providecommand \bibfield  [0]{\@secondoftwo}%
\providecommand \translation [1]{[#1]}%
\providecommand \BibitemOpen [0]{}%
\providecommand \bibitemStop [0]{}%
\providecommand \bibitemNoStop [0]{.\EOS\space}%
\providecommand \EOS [0]{\spacefactor3000\relax}%
\providecommand \BibitemShut  [1]{\csname bibitem#1\endcsname}%
\let\auto@bib@innerbib\@empty
\bibitem [{\citenamefont {Popescu}\ and\ \citenamefont
  {Rohrlich}(1994)}]{PopescuRohrlich1994}%
  \BibitemOpen
  \bibfield  {author} {\bibinfo {author} {\bibfnamefont {S.}~\bibnamefont
  {Popescu}}\ and\ \bibinfo {author} {\bibfnamefont {D.}~\bibnamefont
  {Rohrlich}},\ }\bibfield  {title} {\bibinfo {title} {Quantum nonlocality as
  an axiom},\ }\href {https://doi.org/10.1007/BF02058098} {\bibfield  {journal}
  {\bibinfo  {journal} {Found. Phys.}\ }\textbf {\bibinfo {volume} {24}},\
  \bibinfo {pages} {379} (\bibinfo {year} {1994})}\BibitemShut {NoStop}%
\bibitem [{\citenamefont {Paw{\l}owski}\ \emph {et~al.}(2009)\citenamefont
  {Paw{\l}owski}, \citenamefont {Paterek}, \citenamefont {Kaszlikowski},
  \citenamefont {Scarani}, \citenamefont {Winter},\ and\ \citenamefont
  {{\.Z}ukowski}}]{Pawlowski2009}%
  \BibitemOpen
  \bibfield  {author} {\bibinfo {author} {\bibfnamefont {M.}~\bibnamefont
  {Paw{\l}owski}}, \bibinfo {author} {\bibfnamefont {T.}~\bibnamefont
  {Paterek}}, \bibinfo {author} {\bibfnamefont {D.}~\bibnamefont
  {Kaszlikowski}}, \bibinfo {author} {\bibfnamefont {V.}~\bibnamefont
  {Scarani}}, \bibinfo {author} {\bibfnamefont {A.}~\bibnamefont {Winter}},\
  and\ \bibinfo {author} {\bibfnamefont {M.}~\bibnamefont {{\.Z}ukowski}},\
  }\bibfield  {title} {\bibinfo {title} {Information causality as a physical
  principle},\ }\href {https://doi.org/10.1038/nature08400} {\bibfield
  {journal} {\bibinfo  {journal} {Nature}\ }\textbf {\bibinfo {volume} {461}},\
  \bibinfo {pages} {1101} (\bibinfo {year} {2009})},\ \Eprint
  {https://arxiv.org/abs/0905.2292} {arXiv:0905.2292 [quant-ph]} \BibitemShut
  {NoStop}%
\bibitem [{\citenamefont {Chaturvedi}\ \emph {et~al.}(2017)\citenamefont
  {Chaturvedi}, \citenamefont {Pawlowski},\ and\ \citenamefont
  {Horodecki}}]{ChaturvediPawlowskiHorodecki2017}%
  \BibitemOpen
  \bibfield  {author} {\bibinfo {author} {\bibfnamefont {A.}~\bibnamefont
  {Chaturvedi}}, \bibinfo {author} {\bibfnamefont {M.}~\bibnamefont
  {Pawlowski}},\ and\ \bibinfo {author} {\bibfnamefont {K.}~\bibnamefont
  {Horodecki}},\ }\bibfield  {title} {\bibinfo {title} {Random access codes and
  nonlocal resources},\ }\href {https://doi.org/10.1103/PhysRevA.96.022125}
  {\bibfield  {journal} {\bibinfo  {journal} {Phys. Rev. A}\ }\textbf {\bibinfo
  {volume} {96}},\ \bibinfo {pages} {022125} (\bibinfo {year} {2017})},\
  \Eprint {https://arxiv.org/abs/1610.01268} {arXiv:1610.01268 [quant-ph]}
  \BibitemShut {NoStop}%
\bibitem [{\citenamefont {Navascu{\'e}s}\ \emph {et~al.}(2007)\citenamefont
  {Navascu{\'e}s}, \citenamefont {Pironio},\ and\ \citenamefont
  {Ac{\'i}n}}]{NPA2007}%
  \BibitemOpen
  \bibfield  {author} {\bibinfo {author} {\bibfnamefont {M.}~\bibnamefont
  {Navascu{\'e}s}}, \bibinfo {author} {\bibfnamefont {S.}~\bibnamefont
  {Pironio}},\ and\ \bibinfo {author} {\bibfnamefont {A.}~\bibnamefont
  {Ac{\'i}n}},\ }\bibfield  {title} {\bibinfo {title} {Bounding the set of
  quantum correlations},\ }\href
  {https://doi.org/10.1103/PhysRevLett.98.010401} {\bibfield  {journal}
  {\bibinfo  {journal} {Phys. Rev. Lett.}\ }\textbf {\bibinfo {volume} {98}},\
  \bibinfo {pages} {010401} (\bibinfo {year} {2007})},\ \Eprint
  {https://arxiv.org/abs/quant-ph/0607119} {arXiv:quant-ph/0607119 [quant-ph]}
  \BibitemShut {NoStop}%
\bibitem [{\citenamefont {Navascu{\'e}s}\ \emph {et~al.}(2008)\citenamefont
  {Navascu{\'e}s}, \citenamefont {Pironio},\ and\ \citenamefont
  {Ac{\'i}n}}]{NPA2008}%
  \BibitemOpen
  \bibfield  {author} {\bibinfo {author} {\bibfnamefont {M.}~\bibnamefont
  {Navascu{\'e}s}}, \bibinfo {author} {\bibfnamefont {S.}~\bibnamefont
  {Pironio}},\ and\ \bibinfo {author} {\bibfnamefont {A.}~\bibnamefont
  {Ac{\'i}n}},\ }\bibfield  {title} {\bibinfo {title} {A convergent hierarchy
  of semidefinite programs characterizing the set of quantum correlations},\
  }\href {https://doi.org/10.1088/1367-2630/10/7/073013} {\bibfield  {journal}
  {\bibinfo  {journal} {New J. Phys.}\ }\textbf {\bibinfo {volume} {10}},\
  \bibinfo {pages} {073013} (\bibinfo {year} {2008})},\ \Eprint
  {https://arxiv.org/abs/0803.4290} {arXiv:0803.4290 [quant-ph]} \BibitemShut
  {NoStop}%
\bibitem [{\citenamefont {Navascu{\'e}s}\ and\ \citenamefont
  {Wunderlich}(2010)}]{NavascuesWunderlich2010}%
  \BibitemOpen
  \bibfield  {author} {\bibinfo {author} {\bibfnamefont {M.}~\bibnamefont
  {Navascu{\'e}s}}\ and\ \bibinfo {author} {\bibfnamefont {H.}~\bibnamefont
  {Wunderlich}},\ }\bibfield  {title} {\bibinfo {title} {A glance beyond the
  quantum model},\ }\href {https://doi.org/10.1098/rspa.2009.0453} {\bibfield
  {journal} {\bibinfo  {journal} {Proc. R. Soc. A}\ }\textbf {\bibinfo {volume}
  {466}},\ \bibinfo {pages} {881} (\bibinfo {year} {2010})},\ \Eprint
  {https://arxiv.org/abs/0907.0372} {arXiv:0907.0372 [quant-ph]} \BibitemShut
  {NoStop}%
\bibitem [{\citenamefont {Cavalcanti}\ \emph {et~al.}(2010)\citenamefont
  {Cavalcanti}, \citenamefont {Salles},\ and\ \citenamefont
  {Scarani}}]{Cavalcanti2010}%
  \BibitemOpen
  \bibfield  {author} {\bibinfo {author} {\bibfnamefont {D.}~\bibnamefont
  {Cavalcanti}}, \bibinfo {author} {\bibfnamefont {A.}~\bibnamefont {Salles}},\
  and\ \bibinfo {author} {\bibfnamefont {V.}~\bibnamefont {Scarani}},\
  }\bibfield  {title} {\bibinfo {title} {Macroscopically local correlations can
  violate information causality},\ }\href {https://doi.org/10.1038/ncomms1138}
  {\bibfield  {journal} {\bibinfo  {journal} {Nat. Commun.}\ }\textbf {\bibinfo
  {volume} {1}},\ \bibinfo {pages} {136} (\bibinfo {year} {2010})},\ \Eprint
  {https://arxiv.org/abs/1008.2624} {arXiv:1008.2624 [quant-ph]} \BibitemShut
  {NoStop}%
\bibitem [{\citenamefont {Navascu{\'e}s}\ \emph {et~al.}(2015)\citenamefont
  {Navascu{\'e}s}, \citenamefont {Guryanova}, \citenamefont {Hoban},\ and\
  \citenamefont {Ac{\'i}n}}]{Navascues2015}%
  \BibitemOpen
  \bibfield  {author} {\bibinfo {author} {\bibfnamefont {M.}~\bibnamefont
  {Navascu{\'e}s}}, \bibinfo {author} {\bibfnamefont {Y.}~\bibnamefont
  {Guryanova}}, \bibinfo {author} {\bibfnamefont {M.~J.}\ \bibnamefont
  {Hoban}},\ and\ \bibinfo {author} {\bibfnamefont {A.}~\bibnamefont
  {Ac{\'i}n}},\ }\bibfield  {title} {\bibinfo {title} {Almost quantum
  correlations},\ }\href {https://doi.org/10.1038/ncomms7288} {\bibfield
  {journal} {\bibinfo  {journal} {Nat. Commun.}\ }\textbf {\bibinfo {volume}
  {6}},\ \bibinfo {pages} {6288} (\bibinfo {year} {2015})},\ \Eprint
  {https://arxiv.org/abs/1403.4621} {arXiv:1403.4621 [quant-ph]} \BibitemShut
  {NoStop}%
\bibitem [{\citenamefont {Das~Bhowmik}\ and\ \citenamefont
  {Banik}(2026)}]{DasBhowmikBanik2026}%
  \BibitemOpen
  \bibfield  {author} {\bibinfo {author} {\bibfnamefont {A.}~\bibnamefont
  {Das~Bhowmik}}\ and\ \bibinfo {author} {\bibfnamefont {M.}~\bibnamefont
  {Banik}},\ }\href@noop {} {\bibinfo {title} {Almost-quantum correlations
  violate information causality}} (\bibinfo {year} {2026}),\ \Eprint
  {https://arxiv.org/abs/2610.41444} {arXiv:2610.41444 [quant-ph]} \BibitemShut
  {NoStop}%
\bibitem [{\citenamefont {Gazi}\ \emph {et~al.}(2026)\citenamefont {Gazi},
  \citenamefont {Chatterjee}, \citenamefont {Karmakar}, \citenamefont {Patra},
  \citenamefont {Banik},\ and\ \citenamefont {Das~Bhowmik}}]{Gazi2026}%
  \BibitemOpen
  \bibfield  {author} {\bibinfo {author} {\bibfnamefont {R.}~\bibnamefont
  {Gazi}}, \bibinfo {author} {\bibfnamefont {B.}~\bibnamefont {Chatterjee}},
  \bibinfo {author} {\bibfnamefont {J.}~\bibnamefont {Karmakar}}, \bibinfo
  {author} {\bibfnamefont {R.~K.}\ \bibnamefont {Patra}}, \bibinfo {author}
  {\bibfnamefont {M.}~\bibnamefont {Banik}},\ and\ \bibinfo {author}
  {\bibfnamefont {A.}~\bibnamefont {Das~Bhowmik}},\ }\href@noop {} {\bibinfo
  {title} {An unbounded separation of almost-quantum and commuting-operator
  correlations in zero-error communication}} (\bibinfo {year} {2026}),\ \Eprint
  {https://arxiv.org/abs/2610.41448} {arXiv:2610.41448 [quant-ph]} \BibitemShut
  {NoStop}%
\bibitem [{\citenamefont {Klauck}\ and\ \citenamefont
  {de~Wolf}(2013)}]{KlauckDeWolf2013}%
  \BibitemOpen
  \bibfield  {author} {\bibinfo {author} {\bibfnamefont {H.}~\bibnamefont
  {Klauck}}\ and\ \bibinfo {author} {\bibfnamefont {R.}~\bibnamefont
  {de~Wolf}},\ }\bibfield  {title} {\bibinfo {title} {{Fooling One-Sided
  Quantum Protocols}},\ }in\ \href
  {https://doi.org/10.4230/LIPIcs.STACS.2013.424} {\emph {\bibinfo {booktitle}
  {30th International Symposium on Theoretical Aspects of Computer Science
  (STACS 2013)}}},\ \bibinfo {series} {Leibniz International Proceedings in
  Informatics (LIPIcs)}, Vol.~\bibinfo {volume} {20},\ \bibinfo {editor}
  {edited by\ \bibinfo {editor} {\bibfnamefont {N.}~\bibnamefont {Portier}}\
  and\ \bibinfo {editor} {\bibfnamefont {T.}~\bibnamefont {Wilke}}}\ (\bibinfo
  {publisher} {Schloss Dagstuhl -- Leibniz-Zentrum f{\"u}r Informatik},\
  \bibinfo {year} {2013})\ pp.\ \bibinfo {pages} {424--433},\ \Eprint
  {https://arxiv.org/abs/1204.4619} {arXiv:1204.4619 [quant-ph]} \BibitemShut
  {NoStop}%
\bibitem [{\citenamefont {Miklin}\ \emph {et~al.}(2026)\citenamefont {Miklin},
  \citenamefont {Jain},\ and\ \citenamefont
  {Gachechiladze}}]{MiklinJainGachechiladze2026}%
  \BibitemOpen
  \bibfield  {author} {\bibinfo {author} {\bibfnamefont {N.}~\bibnamefont
  {Miklin}}, \bibinfo {author} {\bibfnamefont {P.}~\bibnamefont {Jain}},\ and\
  \bibinfo {author} {\bibfnamefont {M.}~\bibnamefont {Gachechiladze}},\
  }\href@noop {} {\bibinfo {title} {Communication complexity bounds from
  information causality}} (\bibinfo {year} {2026}),\ \Eprint
  {https://arxiv.org/abs/2602.10206} {arXiv:2602.10206 [quant-ph]} \BibitemShut
  {NoStop}%
\bibitem [{\citenamefont {Mukherjee}\ \emph {et~al.}(2015)\citenamefont
  {Mukherjee}, \citenamefont {Roy}, \citenamefont {Bhattacharya}, \citenamefont
  {Das}, \citenamefont {Gazi},\ and\ \citenamefont {Banik}}]{Mukherjee2015}%
  \BibitemOpen
  \bibfield  {author} {\bibinfo {author} {\bibfnamefont {A.}~\bibnamefont
  {Mukherjee}}, \bibinfo {author} {\bibfnamefont {A.}~\bibnamefont {Roy}},
  \bibinfo {author} {\bibfnamefont {S.~S.}\ \bibnamefont {Bhattacharya}},
  \bibinfo {author} {\bibfnamefont {S.}~\bibnamefont {Das}}, \bibinfo {author}
  {\bibfnamefont {M.~R.}\ \bibnamefont {Gazi}},\ and\ \bibinfo {author}
  {\bibfnamefont {M.}~\bibnamefont {Banik}},\ }\bibfield  {title} {\bibinfo
  {title} {Hardy's test as a device-independent dimension witness},\ }\href
  {https://doi.org/10.1103/PhysRevA.92.022302} {\bibfield  {journal} {\bibinfo
  {journal} {Phys. Rev. A}\ }\textbf {\bibinfo {volume} {92}},\ \bibinfo
  {pages} {022302} (\bibinfo {year} {2015})}\BibitemShut {NoStop}%
\bibitem [{\citenamefont {Rai}\ \emph {et~al.}(2021)\citenamefont {Rai},
  \citenamefont {Pivoluska}, \citenamefont {Plesch}, \citenamefont {Sasmal},
  \citenamefont {Banik},\ and\ \citenamefont {Ghosh}}]{Rai2021}%
  \BibitemOpen
  \bibfield  {author} {\bibinfo {author} {\bibfnamefont {A.}~\bibnamefont
  {Rai}}, \bibinfo {author} {\bibfnamefont {M.}~\bibnamefont {Pivoluska}},
  \bibinfo {author} {\bibfnamefont {M.}~\bibnamefont {Plesch}}, \bibinfo
  {author} {\bibfnamefont {S.}~\bibnamefont {Sasmal}}, \bibinfo {author}
  {\bibfnamefont {M.}~\bibnamefont {Banik}},\ and\ \bibinfo {author}
  {\bibfnamefont {S.}~\bibnamefont {Ghosh}},\ }\bibfield  {title} {\bibinfo
  {title} {Device-independent bounds from {Cabello}'s nonlocality argument},\
  }\href {https://doi.org/10.1103/PhysRevA.103.062219} {\bibfield  {journal}
  {\bibinfo  {journal} {Phys. Rev. A}\ }\textbf {\bibinfo {volume} {103}},\
  \bibinfo {pages} {062219} (\bibinfo {year} {2021})}\BibitemShut {NoStop}%
\bibitem [{\citenamefont {Rai}\ \emph {et~al.}(2022)\citenamefont {Rai},
  \citenamefont {Pivoluska}, \citenamefont {Sasmal}, \citenamefont {Banik},
  \citenamefont {Ghosh},\ and\ \citenamefont {Plesch}}]{Rai2022}%
  \BibitemOpen
  \bibfield  {author} {\bibinfo {author} {\bibfnamefont {A.}~\bibnamefont
  {Rai}}, \bibinfo {author} {\bibfnamefont {M.}~\bibnamefont {Pivoluska}},
  \bibinfo {author} {\bibfnamefont {S.}~\bibnamefont {Sasmal}}, \bibinfo
  {author} {\bibfnamefont {M.}~\bibnamefont {Banik}}, \bibinfo {author}
  {\bibfnamefont {S.}~\bibnamefont {Ghosh}},\ and\ \bibinfo {author}
  {\bibfnamefont {M.}~\bibnamefont {Plesch}},\ }\bibfield  {title} {\bibinfo
  {title} {Self-testing quantum states via nonmaximal violation in {Hardy}'s
  test of nonlocality},\ }\href {https://doi.org/10.1103/PhysRevA.105.052227}
  {\bibfield  {journal} {\bibinfo  {journal} {Phys. Rev. A}\ }\textbf {\bibinfo
  {volume} {105}},\ \bibinfo {pages} {052227} (\bibinfo {year}
  {2022})}\BibitemShut {NoStop}%
\bibitem [{\citenamefont {Alimuddin}\ \emph {et~al.}(2023)\citenamefont
  {Alimuddin}, \citenamefont {Chakraborty}, \citenamefont {Sidhardh},
  \citenamefont {Patra}, \citenamefont {Sen}, \citenamefont {Chowdhury},
  \citenamefont {Naik},\ and\ \citenamefont {Banik}}]{Alimuddin2023}%
  \BibitemOpen
  \bibfield  {author} {\bibinfo {author} {\bibfnamefont {M.}~\bibnamefont
  {Alimuddin}}, \bibinfo {author} {\bibfnamefont {A.}~\bibnamefont
  {Chakraborty}}, \bibinfo {author} {\bibfnamefont {G.~L.}\ \bibnamefont
  {Sidhardh}}, \bibinfo {author} {\bibfnamefont {R.~K.}\ \bibnamefont {Patra}},
  \bibinfo {author} {\bibfnamefont {S.}~\bibnamefont {Sen}}, \bibinfo {author}
  {\bibfnamefont {S.~R.}\ \bibnamefont {Chowdhury}}, \bibinfo {author}
  {\bibfnamefont {S.~G.}\ \bibnamefont {Naik}},\ and\ \bibinfo {author}
  {\bibfnamefont {M.}~\bibnamefont {Banik}},\ }\bibfield  {title} {\bibinfo
  {title} {Advantage of {Hardy}'s nonlocal correlation in reverse zero-error
  channel coding},\ }\href {https://doi.org/10.1103/PhysRevA.108.052430}
  {\bibfield  {journal} {\bibinfo  {journal} {Phys. Rev. A}\ }\textbf {\bibinfo
  {volume} {108}},\ \bibinfo {pages} {052430} (\bibinfo {year}
  {2023})}\BibitemShut {NoStop}%
\bibitem [{\citenamefont {Fanizza}\ \emph {et~al.}(2025)\citenamefont
  {Fanizza}, \citenamefont {Kroell}, \citenamefont {Mehta}, \citenamefont
  {Paddock}, \citenamefont {Rochette}, \citenamefont {Slofstra},\ and\
  \citenamefont {Zhao}}]{Fanizza2025}%
  \BibitemOpen
  \bibfield  {author} {\bibinfo {author} {\bibfnamefont {M.}~\bibnamefont
  {Fanizza}}, \bibinfo {author} {\bibfnamefont {L.}~\bibnamefont {Kroell}},
  \bibinfo {author} {\bibfnamefont {A.}~\bibnamefont {Mehta}}, \bibinfo
  {author} {\bibfnamefont {C.}~\bibnamefont {Paddock}}, \bibinfo {author}
  {\bibfnamefont {D.}~\bibnamefont {Rochette}}, \bibinfo {author}
  {\bibfnamefont {W.}~\bibnamefont {Slofstra}},\ and\ \bibinfo {author}
  {\bibfnamefont {Y.}~\bibnamefont {Zhao}},\ }\href@noop {} {\bibinfo {title}
  {The {NPA} hierarchy does not always attain the commuting operator value}}
  (\bibinfo {year} {2025}),\ \Eprint {https://arxiv.org/abs/2510.04943}
  {arXiv:2510.04943 [quant-ph]} \BibitemShut {NoStop}%
\bibitem [{\citenamefont {Gigena}\ \emph {et~al.}(2025)\citenamefont {Gigena},
  \citenamefont {Panwar}, \citenamefont {Scala}, \citenamefont {Ara{\'{u}}jo},
  \citenamefont {Farkas},\ and\ \citenamefont {Chaturvedi}}]{Gigena2025}%
  \BibitemOpen
  \bibfield  {author} {\bibinfo {author} {\bibfnamefont {N.}~\bibnamefont
  {Gigena}}, \bibinfo {author} {\bibfnamefont {E.}~\bibnamefont {Panwar}},
  \bibinfo {author} {\bibfnamefont {G.}~\bibnamefont {Scala}}, \bibinfo
  {author} {\bibfnamefont {M.}~\bibnamefont {Ara{\'{u}}jo}}, \bibinfo {author}
  {\bibfnamefont {M.}~\bibnamefont {Farkas}},\ and\ \bibinfo {author}
  {\bibfnamefont {A.}~\bibnamefont {Chaturvedi}},\ }\bibfield  {title}
  {\bibinfo {title} {Self-testing tilted strategies for maximal loophole-free
  nonlocality},\ }\href {https://doi.org/10.1038/s41534-025-01029-6} {\bibfield
   {journal} {\bibinfo  {journal} {npj Quantum Inf.}\ }\textbf {\bibinfo
  {volume} {11}},\ \bibinfo {pages} {82} (\bibinfo {year} {2025})},\ \Eprint
  {https://arxiv.org/abs/2405.08743} {arXiv:2405.08743 [quant-ph]} \BibitemShut
  {NoStop}%
\bibitem [{\citenamefont {Chaturvedi}(2026)}]{Chaturvedi2026}%
  \BibitemOpen
  \bibfield  {author} {\bibinfo {author} {\bibfnamefont {A.}~\bibnamefont
  {Chaturvedi}},\ }\bibfield  {title} {\bibinfo {title} {No finite {NPA} level
  characterizes the complete quantum set in the simplest {Bell} scenario},\
  }\bibfield  {journal} {\bibinfo  {journal} {Phys. Rev. Lett.}\ }\href
  {https://doi.org/10.1103/9298-2bth} {10.1103/9298-2bth} (\bibinfo {year}
  {2026}),\ \bibinfo {note} {accepted 6 November 2026},\ \Eprint
  {https://arxiv.org/abs/2607.14569v2} {arXiv:2607.14569v2 [quant-ph]}
  \BibitemShut {NoStop}%
\bibitem [{\citenamefont {Gazi}\ \emph {et~al.}(2013)\citenamefont {Gazi},
  \citenamefont {Banik}, \citenamefont {Das}, \citenamefont {Rai},\ and\
  \citenamefont {Kunkri}}]{Gazi2013}%
  \BibitemOpen
  \bibfield  {author} {\bibinfo {author} {\bibfnamefont {M.~R.}\ \bibnamefont
  {Gazi}}, \bibinfo {author} {\bibfnamefont {M.}~\bibnamefont {Banik}},
  \bibinfo {author} {\bibfnamefont {S.}~\bibnamefont {Das}}, \bibinfo {author}
  {\bibfnamefont {A.}~\bibnamefont {Rai}},\ and\ \bibinfo {author}
  {\bibfnamefont {S.}~\bibnamefont {Kunkri}},\ }\bibfield  {title} {\bibinfo
  {title} {Macroscopic locality with equal bias reproduces with high fidelity a
  quantum distribution achieving the {Tsirelson}'s bound},\ }\href
  {https://doi.org/10.1103/PhysRevA.88.052115} {\bibfield  {journal} {\bibinfo
  {journal} {Phys. Rev. A}\ }\textbf {\bibinfo {volume} {88}},\ \bibinfo
  {pages} {052115} (\bibinfo {year} {2013})},\ \Eprint
  {https://arxiv.org/abs/1308.4003} {arXiv:1308.4003 [quant-ph]} \BibitemShut
  {NoStop}%
\bibitem [{\citenamefont {Das}\ \emph {et~al.}(2013)\citenamefont {Das},
  \citenamefont {Banik}, \citenamefont {Gazi}, \citenamefont {Rai},\ and\
  \citenamefont {Kunkri}}]{Das2013}%
  \BibitemOpen
  \bibfield  {author} {\bibinfo {author} {\bibfnamefont {S.}~\bibnamefont
  {Das}}, \bibinfo {author} {\bibfnamefont {M.}~\bibnamefont {Banik}}, \bibinfo
  {author} {\bibfnamefont {M.~R.}\ \bibnamefont {Gazi}}, \bibinfo {author}
  {\bibfnamefont {A.}~\bibnamefont {Rai}},\ and\ \bibinfo {author}
  {\bibfnamefont {S.}~\bibnamefont {Kunkri}},\ }\bibfield  {title} {\bibinfo
  {title} {Local orthogonality provides a tight upper bound for {Hardy}'s
  nonlocality},\ }\href {https://doi.org/10.1103/PhysRevA.88.062101} {\bibfield
   {journal} {\bibinfo  {journal} {Phys. Rev. A}\ }\textbf {\bibinfo {volume}
  {88}},\ \bibinfo {pages} {062101} (\bibinfo {year} {2013})}\BibitemShut
  {NoStop}%
\bibitem [{\citenamefont {Miklin}\ and\ \citenamefont
  {Paw{\l}owski}(2021)}]{MiklinPawlowski2021}%
  \BibitemOpen
  \bibfield  {author} {\bibinfo {author} {\bibfnamefont {N.}~\bibnamefont
  {Miklin}}\ and\ \bibinfo {author} {\bibfnamefont {M.}~\bibnamefont
  {Paw{\l}owski}},\ }\bibfield  {title} {\bibinfo {title} {Information
  causality without concatenation},\ }\href
  {https://doi.org/10.1103/PhysRevLett.126.220403} {\bibfield  {journal}
  {\bibinfo  {journal} {Phys. Rev. Lett.}\ }\textbf {\bibinfo {volume} {126}},\
  \bibinfo {pages} {220403} (\bibinfo {year} {2021})},\ \Eprint
  {https://arxiv.org/abs/2101.12710} {arXiv:2101.12710 [quant-ph]} \BibitemShut
  {NoStop}%
\bibitem [{\citenamefont {Gachechiladze}\ \emph {et~al.}(2022)\citenamefont
  {Gachechiladze}, \citenamefont {B{\k{a}}k}, \citenamefont {Paw{\l}owski},\
  and\ \citenamefont {Miklin}}]{Gachechiladze2022}%
  \BibitemOpen
  \bibfield  {author} {\bibinfo {author} {\bibfnamefont {M.}~\bibnamefont
  {Gachechiladze}}, \bibinfo {author} {\bibfnamefont {B.}~\bibnamefont
  {B{\k{a}}k}}, \bibinfo {author} {\bibfnamefont {M.}~\bibnamefont
  {Paw{\l}owski}},\ and\ \bibinfo {author} {\bibfnamefont {N.}~\bibnamefont
  {Miklin}},\ }\bibfield  {title} {\bibinfo {title} {Quantum {Bell}
  inequalities from {Information Causality} -- tight for {Macroscopic
  Locality}},\ }\href {https://doi.org/10.22331/q-2022-05-24-717} {\bibfield
  {journal} {\bibinfo  {journal} {Quantum}\ }\textbf {\bibinfo {volume} {6}},\
  \bibinfo {pages} {717} (\bibinfo {year} {2022})},\ \Eprint
  {https://arxiv.org/abs/2103.05029} {arXiv:2103.05029 [quant-ph]} \BibitemShut
  {NoStop}%
\bibitem [{\citenamefont {Jain}\ \emph {et~al.}(2024)\citenamefont {Jain},
  \citenamefont {Gachechiladze},\ and\ \citenamefont
  {Miklin}}]{JainGachechiladzeMiklin2024}%
  \BibitemOpen
  \bibfield  {author} {\bibinfo {author} {\bibfnamefont {P.}~\bibnamefont
  {Jain}}, \bibinfo {author} {\bibfnamefont {M.}~\bibnamefont
  {Gachechiladze}},\ and\ \bibinfo {author} {\bibfnamefont {N.}~\bibnamefont
  {Miklin}},\ }\bibfield  {title} {\bibinfo {title} {Information causality as a
  tool for bounding the set of quantum correlations},\ }\href
  {https://doi.org/10.1103/PhysRevLett.133.160201} {\bibfield  {journal}
  {\bibinfo  {journal} {Phys. Rev. Lett.}\ }\textbf {\bibinfo {volume} {133}},\
  \bibinfo {pages} {160201} (\bibinfo {year} {2024})},\ \Eprint
  {https://arxiv.org/abs/2308.02478} {arXiv:2308.02478 [quant-ph]} \BibitemShut
  {NoStop}%
\bibitem [{\citenamefont {Gachechiladze}\ and\ \citenamefont
  {Miklin}(2026)}]{GachechiladzeMiklin2026}%
  \BibitemOpen
  \bibfield  {author} {\bibinfo {author} {\bibfnamefont {M.}~\bibnamefont
  {Gachechiladze}}\ and\ \bibinfo {author} {\bibfnamefont {N.}~\bibnamefont
  {Miklin}},\ }\href@noop {} {\bibinfo {title} {Information causality
  characterizes the set of quantum correlations in the simplest {Bell}
  scenario}} (\bibinfo {year} {2026}),\ \Eprint
  {https://arxiv.org/abs/2609.10508} {arXiv:2609.10508 [quant-ph]} \BibitemShut
  {NoStop}%
\bibitem [{\citenamefont {Pompili}\ and\ \citenamefont
  {Budroni}(2026)}]{PompiliBudroni2026}%
  \BibitemOpen
  \bibfield  {author} {\bibinfo {author} {\bibfnamefont {M.}~\bibnamefont
  {Pompili}}\ and\ \bibinfo {author} {\bibfnamefont {C.}~\bibnamefont
  {Budroni}},\ }\href@noop {} {\bibinfo {title} {Bounding single-copy
  information causality violations}} (\bibinfo {year} {2026}),\ \Eprint
  {https://arxiv.org/abs/2610.32783} {arXiv:2610.32783 [quant-ph]} \BibitemShut
  {NoStop}%
\bibitem [{\citenamefont {Sami}\ \emph {et~al.}(2017)\citenamefont {Sami},
  \citenamefont {Chakrabarty},\ and\ \citenamefont {Chaturvedi}}]{Sami2017}%
  \BibitemOpen
  \bibfield  {author} {\bibinfo {author} {\bibfnamefont {S.}~\bibnamefont
  {Sami}}, \bibinfo {author} {\bibfnamefont {I.}~\bibnamefont {Chakrabarty}},\
  and\ \bibinfo {author} {\bibfnamefont {A.}~\bibnamefont {Chaturvedi}},\
  }\bibfield  {title} {\bibinfo {title} {Complementarity of genuine
  multipartite {Bell} nonlocality},\ }\href
  {https://doi.org/10.1103/physreva.96.022121} {\bibfield  {journal} {\bibinfo
  {journal} {Phys. Rev. A}\ }\textbf {\bibinfo {volume} {96}},\ \bibinfo
  {pages} {022121} (\bibinfo {year} {2017})},\ \Eprint
  {https://arxiv.org/abs/1603.09120} {arXiv:1603.09120 [quant-ph]} \BibitemShut
  {NoStop}%
\bibitem [{\citenamefont {Pollyceno}\ \emph {et~al.}(2023)\citenamefont
  {Pollyceno}, \citenamefont {Chaves},\ and\ \citenamefont
  {Rabelo}}]{PollycenoChavesRabelo2023}%
  \BibitemOpen
  \bibfield  {author} {\bibinfo {author} {\bibfnamefont {L.}~\bibnamefont
  {Pollyceno}}, \bibinfo {author} {\bibfnamefont {R.}~\bibnamefont {Chaves}},\
  and\ \bibinfo {author} {\bibfnamefont {R.}~\bibnamefont {Rabelo}},\
  }\bibfield  {title} {\bibinfo {title} {Information causality in multipartite
  scenarios},\ }\href {https://doi.org/10.1103/PhysRevA.107.042203} {\bibfield
  {journal} {\bibinfo  {journal} {Phys. Rev. A}\ }\textbf {\bibinfo {volume}
  {107}},\ \bibinfo {pages} {042203} (\bibinfo {year} {2023})},\ \Eprint
  {https://arxiv.org/abs/2212.05601} {arXiv:2212.05601 [quant-ph]} \BibitemShut
  {NoStop}%
\bibitem [{\citenamefont {Pollyceno}\ \emph {et~al.}(2025)\citenamefont
  {Pollyceno}, \citenamefont {Chaturvedi}, \citenamefont {Raj}, \citenamefont
  {Dieguez},\ and\ \citenamefont {Paw{\l}owski}}]{Pollyceno2025}%
  \BibitemOpen
  \bibfield  {author} {\bibinfo {author} {\bibfnamefont {L.}~\bibnamefont
  {Pollyceno}}, \bibinfo {author} {\bibfnamefont {A.}~\bibnamefont
  {Chaturvedi}}, \bibinfo {author} {\bibfnamefont {C.}~\bibnamefont {Raj}},
  \bibinfo {author} {\bibfnamefont {P.~R.}\ \bibnamefont {Dieguez}},\ and\
  \bibinfo {author} {\bibfnamefont {M.}~\bibnamefont {Paw{\l}owski}},\
  }\bibfield  {title} {\bibinfo {title} {Security of device-independent quantum
  key distribution via monogamy relations from multipartite information
  causality},\ }\href {https://doi.org/10.1103/jnng-m87v} {\bibfield  {journal}
  {\bibinfo  {journal} {Phys. Rev. A}\ }\textbf {\bibinfo {volume} {112}},\
  \bibinfo {pages} {042201} (\bibinfo {year} {2025})},\ \Eprint
  {https://arxiv.org/abs/2405.20115} {arXiv:2405.20115 [quant-ph]} \BibitemShut
  {NoStop}%
\bibitem [{\citenamefont {Pollyceno}\ \emph {et~al.}(2026)\citenamefont
  {Pollyceno}, \citenamefont {Freudenheim}, \citenamefont {Nogueira},
  \citenamefont {Chaturvedi}, \citenamefont {Rabelo},\ and\ \citenamefont
  {Paw{\l}owski}}]{Pollyceno2026}%
  \BibitemOpen
  \bibfield  {author} {\bibinfo {author} {\bibfnamefont {L.}~\bibnamefont
  {Pollyceno}}, \bibinfo {author} {\bibfnamefont {D.}~\bibnamefont
  {Freudenheim}}, \bibinfo {author} {\bibfnamefont {J.}~\bibnamefont
  {Nogueira}}, \bibinfo {author} {\bibfnamefont {A.}~\bibnamefont
  {Chaturvedi}}, \bibinfo {author} {\bibfnamefont {R.}~\bibnamefont {Rabelo}},\
  and\ \bibinfo {author} {\bibfnamefont {M.}~\bibnamefont {Paw{\l}owski}},\
  }\href@noop {} {\bibinfo {title} {Communication-constrained nonlocal
  correlations}} (\bibinfo {year} {2026}),\ \Eprint
  {https://arxiv.org/abs/2603.08610} {arXiv:2603.08610 [quant-ph]} \BibitemShut
  {NoStop}%
\bibitem [{\citenamefont {Rout}\ \emph {et~al.}(2025)\citenamefont {Rout},
  \citenamefont {Chaturvedi}, \citenamefont {Bhattacharya},\ and\ \citenamefont
  {Horodecki}}]{Rout2025}%
  \BibitemOpen
  \bibfield  {author} {\bibinfo {author} {\bibfnamefont {S.}~\bibnamefont
  {Rout}}, \bibinfo {author} {\bibfnamefont {A.}~\bibnamefont {Chaturvedi}},
  \bibinfo {author} {\bibfnamefont {S.~S.}\ \bibnamefont {Bhattacharya}},\ and\
  \bibinfo {author} {\bibfnamefont {P.}~\bibnamefont {Horodecki}},\ }\href
  {https://arxiv.org/abs/2507.10830} {\bibinfo {title} {Facets of non-locality
  and advantage in entanglement-assisted classical communication tasks}}
  (\bibinfo {year} {2025}),\ \Eprint {https://arxiv.org/abs/2507.10830}
  {arXiv:2507.10830 [quant-ph]} \BibitemShut {NoStop}%
\bibitem [{\citenamefont {Manna}\ \emph {et~al.}(2024)\citenamefont {Manna},
  \citenamefont {Chaturvedi},\ and\ \citenamefont {Saha}}]{Manna2024}%
  \BibitemOpen
  \bibfield  {author} {\bibinfo {author} {\bibfnamefont {S.}~\bibnamefont
  {Manna}}, \bibinfo {author} {\bibfnamefont {A.}~\bibnamefont {Chaturvedi}},\
  and\ \bibinfo {author} {\bibfnamefont {D.}~\bibnamefont {Saha}},\ }\bibfield
  {title} {\bibinfo {title} {Unbounded quantum advantage in communication
  complexity measured by distinguishability},\ }\href
  {https://doi.org/10.1103/physrevresearch.6.043269} {\bibfield  {journal}
  {\bibinfo  {journal} {Phys. Rev. Research}\ }\textbf {\bibinfo {volume}
  {6}},\ \bibinfo {pages} {043269} (\bibinfo {year} {2024})},\ \Eprint
  {https://arxiv.org/abs/2401.12903} {arXiv:2401.12903 [quant-ph]} \BibitemShut
  {NoStop}%
\bibitem [{\citenamefont {Manna}\ \emph {et~al.}(2026)\citenamefont {Manna},
  \citenamefont {Pandit}, \citenamefont {Chaturvedi},\ and\ \citenamefont
  {Saha}}]{Manna2026}%
  \BibitemOpen
  \bibfield  {author} {\bibinfo {author} {\bibfnamefont {S.}~\bibnamefont
  {Manna}}, \bibinfo {author} {\bibfnamefont {A.}~\bibnamefont {Pandit}},
  \bibinfo {author} {\bibfnamefont {A.}~\bibnamefont {Chaturvedi}},\ and\
  \bibinfo {author} {\bibfnamefont {D.}~\bibnamefont {Saha}},\ }\href
  {https://arxiv.org/abs/2603.19105} {\bibinfo {title} {Entanglement assisted
  communication complexity measured by distinguishability}} (\bibinfo {year}
  {2026}),\ \Eprint {https://arxiv.org/abs/2603.19105} {arXiv:2603.19105
  [quant-ph]} \BibitemShut {NoStop}%
\bibitem [{\citenamefont {Patra}\ \emph {et~al.}(2023)\citenamefont {Patra},
  \citenamefont {Naik}, \citenamefont {Lobo}, \citenamefont {Sen},
  \citenamefont {Sidhardh}, \citenamefont {Alimuddin},\ and\ \citenamefont
  {Banik}}]{Patra2023}%
  \BibitemOpen
  \bibfield  {author} {\bibinfo {author} {\bibfnamefont {R.~K.}\ \bibnamefont
  {Patra}}, \bibinfo {author} {\bibfnamefont {S.~G.}\ \bibnamefont {Naik}},
  \bibinfo {author} {\bibfnamefont {E.~P.}\ \bibnamefont {Lobo}}, \bibinfo
  {author} {\bibfnamefont {S.}~\bibnamefont {Sen}}, \bibinfo {author}
  {\bibfnamefont {G.~L.}\ \bibnamefont {Sidhardh}}, \bibinfo {author}
  {\bibfnamefont {M.}~\bibnamefont {Alimuddin}},\ and\ \bibinfo {author}
  {\bibfnamefont {M.}~\bibnamefont {Banik}},\ }\bibfield  {title} {\bibinfo
  {title} {Principle of information causality rationalizes quantum
  composition},\ }\href {https://doi.org/10.1103/PhysRevLett.130.110202}
  {\bibfield  {journal} {\bibinfo  {journal} {Phys. Rev. Lett.}\ }\textbf
  {\bibinfo {volume} {130}},\ \bibinfo {pages} {110202} (\bibinfo {year}
  {2023})},\ \Eprint {https://arxiv.org/abs/2208.13996} {arXiv:2208.13996
  [quant-ph]} \BibitemShut {NoStop}%
\bibitem [{\citenamefont {M}\ \emph {et~al.}(2021)\citenamefont {M},
  \citenamefont {Patra}, \citenamefont {Janpandit}, \citenamefont {Sen},
  \citenamefont {Banik},\ and\ \citenamefont {Chaturvedi}}]{Vaisakh2021}%
  \BibitemOpen
  \bibfield  {author} {\bibinfo {author} {\bibfnamefont {V.}~\bibnamefont {M}},
  \bibinfo {author} {\bibfnamefont {R.~k.}\ \bibnamefont {Patra}}, \bibinfo
  {author} {\bibfnamefont {M.}~\bibnamefont {Janpandit}}, \bibinfo {author}
  {\bibfnamefont {S.}~\bibnamefont {Sen}}, \bibinfo {author} {\bibfnamefont
  {M.}~\bibnamefont {Banik}},\ and\ \bibinfo {author} {\bibfnamefont
  {A.}~\bibnamefont {Chaturvedi}},\ }\bibfield  {title} {\bibinfo {title}
  {Mutually unbiased balanced functions and generalized random access codes},\
  }\href {https://doi.org/10.1103/physreva.104.012420} {\bibfield  {journal}
  {\bibinfo  {journal} {Phys. Rev. A}\ }\textbf {\bibinfo {volume} {104}},\
  \bibinfo {pages} {012420} (\bibinfo {year} {2021})},\ \Eprint
  {https://arxiv.org/abs/2105.03932} {arXiv:2105.03932 [quant-ph]} \BibitemShut
  {NoStop}%
\bibitem [{\citenamefont {Ambainis}\ \emph {et~al.}(2019)\citenamefont
  {Ambainis}, \citenamefont {Banik}, \citenamefont {Chaturvedi}, \citenamefont
  {Kravchenko},\ and\ \citenamefont {Rai}}]{Ambainis2019}%
  \BibitemOpen
  \bibfield  {author} {\bibinfo {author} {\bibfnamefont {A.}~\bibnamefont
  {Ambainis}}, \bibinfo {author} {\bibfnamefont {M.}~\bibnamefont {Banik}},
  \bibinfo {author} {\bibfnamefont {A.}~\bibnamefont {Chaturvedi}}, \bibinfo
  {author} {\bibfnamefont {D.}~\bibnamefont {Kravchenko}},\ and\ \bibinfo
  {author} {\bibfnamefont {A.}~\bibnamefont {Rai}},\ }\bibfield  {title}
  {\bibinfo {title} {Parity oblivious d-level random access codes and class of
  noncontextuality inequalities},\ }\href
  {https://doi.org/10.1007/s11128-019-2228-3} {\bibfield  {journal} {\bibinfo
  {journal} {Quantum Inf. Process.}\ }\textbf {\bibinfo {volume} {18}},\
  \bibinfo {pages} {111} (\bibinfo {year} {2019})},\ \Eprint
  {https://arxiv.org/abs/1607.05490} {arXiv:1607.05490 [quant-ph]} \BibitemShut
  {NoStop}%
\bibitem [{\citenamefont {Saha}\ and\ \citenamefont
  {Chaturvedi}(2019)}]{SahaChaturvedi2019}%
  \BibitemOpen
  \bibfield  {author} {\bibinfo {author} {\bibfnamefont {D.}~\bibnamefont
  {Saha}}\ and\ \bibinfo {author} {\bibfnamefont {A.}~\bibnamefont
  {Chaturvedi}},\ }\bibfield  {title} {\bibinfo {title} {Preparation
  contextuality as an essential feature underlying quantum communication
  advantage},\ }\href {https://doi.org/10.1103/physreva.100.022108} {\bibfield
  {journal} {\bibinfo  {journal} {Phys. Rev. A}\ }\textbf {\bibinfo {volume}
  {100}},\ \bibinfo {pages} {022108} (\bibinfo {year} {2019})},\ \Eprint
  {https://arxiv.org/abs/1802.07215} {arXiv:1802.07215 [quant-ph]} \BibitemShut
  {NoStop}%
\bibitem [{\citenamefont {Chaturvedi}\ \emph {et~al.}(2021)\citenamefont
  {Chaturvedi}, \citenamefont {Farkas},\ and\ \citenamefont
  {Wright}}]{ChaturvediFarkasWright2021}%
  \BibitemOpen
  \bibfield  {author} {\bibinfo {author} {\bibfnamefont {A.}~\bibnamefont
  {Chaturvedi}}, \bibinfo {author} {\bibfnamefont {M.}~\bibnamefont {Farkas}},\
  and\ \bibinfo {author} {\bibfnamefont {V.~J.}\ \bibnamefont {Wright}},\
  }\bibfield  {title} {\bibinfo {title} {Characterising and bounding the set of
  quantum behaviours in contextuality scenarios},\ }\href
  {https://doi.org/10.22331/q-2021-06-29-484} {\bibfield  {journal} {\bibinfo
  {journal} {Quantum}\ }\textbf {\bibinfo {volume} {5}},\ \bibinfo {pages}
  {484} (\bibinfo {year} {2021})},\ \Eprint {https://arxiv.org/abs/2010.05853}
  {arXiv:2010.05853 [quant-ph]} \BibitemShut {NoStop}%
\bibitem [{\citenamefont {Chaturvedi}\ \emph {et~al.}(2018)\citenamefont
  {Chaturvedi}, \citenamefont {Ray}, \citenamefont {Veynar},\ and\
  \citenamefont {Paw{\l}owski}}]{Chaturvedi2018Security}%
  \BibitemOpen
  \bibfield  {author} {\bibinfo {author} {\bibfnamefont {A.}~\bibnamefont
  {Chaturvedi}}, \bibinfo {author} {\bibfnamefont {M.}~\bibnamefont {Ray}},
  \bibinfo {author} {\bibfnamefont {R.}~\bibnamefont {Veynar}},\ and\ \bibinfo
  {author} {\bibfnamefont {M.}~\bibnamefont {Paw{\l}owski}},\ }\bibfield
  {title} {\bibinfo {title} {On the security of semi-device-independent {QKD}
  protocols},\ }\href {https://doi.org/10.1007/s11128-018-1892-z} {\bibfield
  {journal} {\bibinfo  {journal} {Quantum Inf. Process.}\ }\textbf {\bibinfo
  {volume} {17}},\ \bibinfo {pages} {131} (\bibinfo {year} {2018})},\ \Eprint
  {https://arxiv.org/abs/1504.00939} {arXiv:1504.00939 [quant-ph]} \BibitemShut
  {NoStop}%
\bibitem [{\citenamefont {Chaturvedi}\ \emph {et~al.}(2026)\citenamefont
  {Chaturvedi}, \citenamefont {Viola}, \citenamefont {Panwar}, \citenamefont
  {Prasad},\ and\ \citenamefont {Saha}}]{Chaturvedi2026OperationalQKD}%
  \BibitemOpen
  \bibfield  {author} {\bibinfo {author} {\bibfnamefont {A.}~\bibnamefont
  {Chaturvedi}}, \bibinfo {author} {\bibfnamefont {G.}~\bibnamefont {Viola}},
  \bibinfo {author} {\bibfnamefont {E.}~\bibnamefont {Panwar}}, \bibinfo
  {author} {\bibfnamefont {T.}~\bibnamefont {Prasad}},\ and\ \bibinfo {author}
  {\bibfnamefont {D.}~\bibnamefont {Saha}},\ }\href
  {https://arxiv.org/abs/2607.06682} {\bibinfo {title} {Semi-device-independent
  quantum key distribution from operational assumptions}} (\bibinfo {year}
  {2026}),\ \Eprint {https://arxiv.org/abs/2607.06682} {arXiv:2607.06682
  [quant-ph]} \BibitemShut {NoStop}%
\bibitem [{\citenamefont {Pompili}\ and\ \citenamefont
  {Weilenmann}(2026)}]{PompiliWeilenmann2026}%
  \BibitemOpen
  \bibfield  {author} {\bibinfo {author} {\bibfnamefont {M.}~\bibnamefont
  {Pompili}}\ and\ \bibinfo {author} {\bibfnamefont {M.}~\bibnamefont
  {Weilenmann}},\ }\href@noop {} {\bibinfo {title} {Information causality does
  not single out the set of quantum correlations}} (\bibinfo {year} {2026}),\
  \Eprint {https://arxiv.org/abs/2610.06604} {arXiv:2610.06604 [quant-ph]}
  \BibitemShut {NoStop}%
\bibitem [{\citenamefont {Al-Safi}\ and\ \citenamefont
  {Short}(2011)}]{AlSafiShort2011}%
  \BibitemOpen
  \bibfield  {author} {\bibinfo {author} {\bibfnamefont {S.~W.}\ \bibnamefont
  {Al-Safi}}\ and\ \bibinfo {author} {\bibfnamefont {A.~J.}\ \bibnamefont
  {Short}},\ }\bibfield  {title} {\bibinfo {title} {Information causality from
  an entropic and a probabilistic perspective},\ }\href
  {https://doi.org/10.1103/PhysRevA.84.042323} {\bibfield  {journal} {\bibinfo
  {journal} {Phys. Rev. A}\ }\textbf {\bibinfo {volume} {84}},\ \bibinfo
  {pages} {042323} (\bibinfo {year} {2011})},\ \Eprint
  {https://arxiv.org/abs/1107.4031} {arXiv:1107.4031 [quant-ph]} \BibitemShut
  {NoStop}%
\bibitem [{\citenamefont {Wakakuwa}\ and\ \citenamefont
  {Murao}(2012)}]{WakakuwaMurao2012}%
  \BibitemOpen
  \bibfield  {author} {\bibinfo {author} {\bibfnamefont {E.}~\bibnamefont
  {Wakakuwa}}\ and\ \bibinfo {author} {\bibfnamefont {M.}~\bibnamefont
  {Murao}},\ }\bibfield  {title} {\bibinfo {title} {The chain rule implies
  {Tsirelson}'s bound: an approach from generalized mutual information},\
  }\href {https://doi.org/10.1088/1367-2630/14/11/113037} {\bibfield  {journal}
  {\bibinfo  {journal} {New J. Phys.}\ }\textbf {\bibinfo {volume} {14}},\
  \bibinfo {pages} {113037} (\bibinfo {year} {2012})},\ \Eprint
  {https://arxiv.org/abs/1207.2286} {arXiv:1207.2286 [quant-ph]} \BibitemShut
  {NoStop}%
\end{thebibliography}%

\clearpage
\onecolumngrid
\fontsize{11}{13.2}\selectfont
\section*{Supplementary Information}
\setcounter{secnumdepth}{3}
\setcounter{section}{0}
\setcounter{subsection}{0}
\setcounter{equation}{0}
\renewcommand{\thesection}{S\arabic{section}}
\renewcommand{\thesubsection}{\thesection.\arabic{subsection}}
\renewcommand{\thesubsubsection}{\thesubsection.\arabic{subsubsection}}
\makeatletter
\renewcommand{\p@subsection}{}
\renewcommand{\p@subsubsection}{}
\makeatother
\renewcommand{\theequation}{S\arabic{equation}}
\providecommand{\theHsection}{}
\providecommand{\theHsubsection}{}
\providecommand{\theHsubsubsection}{}
\providecommand{\theHequation}{}
\renewcommand{\theHsection}{supp.\arabic{section}}
\renewcommand{\theHsubsection}{supp.\arabic{section}.\arabic{subsection}}
\renewcommand{\theHsubsubsection}{supp.\arabic{section}.\arabic{subsection}.\arabic{subsubsection}}
\renewcommand{\theHequation}{supp.\arabic{equation}}
\noindent
Alice sends Bob her outcome as a single bit. When that bit and his own outcome are both zero, Bob knows that their settings agree. At each fixed NPA level, we construct correlations for which the probability of two zero outcomes on equal settings stays constant as the number of settings grows. In the database protocol, a joint zero outcome certifies that the requested bit is one and gives more information when the prior probability of a one is smaller. With independent copies and a shared random hash, Bob can decide equality of long strings without error on conclusive runs.

\begingroup
\small
\SupplementContentsfalse
\let\FullContentsLine\contentsline
\renewcommand{\contentsline}[4]{%
  \ifSupplementContents
    \FullContentsLine{#1}{#2}{#3}{#4}%
  \fi}
\makeatletter
\let\l@subsubsection\@gobbletwo
\let\l@paragraph\@gobbletwo
\makeatother
\tableofcontents
\endgroup
\addtocontents{toc}{\protect\SupplementContentstrue}
\clearpage
\section{The finite-level equality certificate}\label{supp:certificate}

Fix an integer hierarchy level \(L\ge2\). We allow the number of settings \(N\) to be any positive integer and seek one certificate probability that works for all of them. Alice's and Bob's settings are \(i,j\in\{1,\ldots,N\}\), and their respective outcomes are \(a,b\in\{0,1\}\). Write \(p(a,b\mid i,j)\) for the conditional probabilities and \(p_A,p_B\) for their marginals. The joint outcome \((0,0)\) must be impossible for unequal settings. On equal settings it will have the same probability \(s_L\) for every label. We construct a rational \(s_L>(185L^3)^{-1}\) and a correlation at level \(L\) of the Navascu\'es--Pironio--Ac\'in (NPA) hierarchy with the following probabilities, where \(\delta_{ij}=1\) if \(i=j\) and \(\delta_{ij}=0\) otherwise,
\[
 p(00\mid i,j)=s_L\delta_{ij},\qquad
 p_A(0\mid i)=p_B(0\mid i)=3s_L.
\]

Write \(Q_L(N,N,2,2)\) for the set of correlations that satisfy the level-\(L\) NPA moment constraints.

\subsection{Positivity at a finite level}\label{supp:npa}

In a projector model, apply an ordered product of measurement projectors
to the state. The resulting vectors give a Gram matrix, with an entry
for each inner product between two such vectors. The level-\(L\) test asks
for a positive semidefinite matrix of this form through words of length
\(L\), with entries obeying all equalities implied by the projector
relations.

Let \([N]=\{1,\ldots,N\}\). We use the unital complex \(*\)-algebra generated
by \(A_i\) and \(B_j\), the projectors for outcome \(0\) at Alice's setting
\(i\) and Bob's setting \(j\). The local generators are self-adjoint and
idempotent, and projectors belonging to different parties commute,
\begin{equation}
 A_i^*=A_i^2=A_i,\qquad B_j^*=B_j^2=B_j,\qquad A_iB_j=B_jA_i .
 \label{supp:relations}
\end{equation}
The outcome-\(1\) projectors are \(1-A_i\) and \(1-B_j\). Words in the
generators multiply by concatenation, and the adjoint reverses the order of
the letters. The empty word is \(1\), and the length of a word is its number
of letters. A degree-at-most-\(L\) polynomial is a linear combination of
words of length at most \(L\). Using a complementary outcome in a word
preserves this bound after expansion.

In a commuting-operator model with unit state vector \(\psi\), the moment
of a polynomial \(f\) is \(\langle\psi,f\psi\rangle\), with the inner
product linear in its second argument. At level \(L\) of the NPA
hierarchy~\cite{NPA2008}, candidate moments are the values of a linear
functional \(\ell\) on the projector algebra. The functional satisfies
\(\ell(1)=1\) and \(\ell(f^*f)\ge 0\) for every \(f\) of degree at most
\(L\). Thus \(\Gamma_{u,v}=\ell(u^*v)\) is a positive semidefinite matrix
on the words of length at most \(L\). For \(f=\sum_v \alpha_v v\), the
quadratic form is
\(\ell(f^*f)=\sum_{u,v}\overline{\alpha_u}\Gamma_{u,v}\alpha_v\).
The products \(u^*v\) can have length up to \(2L\). If two products agree
in the algebra, evaluating \(\ell\) gives the same entry for both. We
write \(Q_L\) for the set of correlations with such a matrix. For
example, the entry for \(u=A_i\) and \(v=B_j\) is the joint zero-outcome
probability. The entries for \(u=1\) and \(v=A_i\) or \(B_j\) give the
single-party zero-outcome probabilities.

We start with projectors and an unnormalized reference vector \(R_N\)
whose joint moments are \(\langle R_N,A_iB_jR_N\rangle=\delta_{ij}\).
The construction assigns a private direction to each setting, so the
squared norm of the reference vector grows with \(N\). Normalizing the
vectors \(wR_N\) by \(\|R_N\|\) would make the certificate probability
\(\|R_N\|^{-2}\) tend to zero. The normalization comes entirely from
the entry for the empty word paired with itself. We look for a
replacement \(S_L\) of bounded norm for that word, preserving its
pairings with every nonempty word \(w\) of length \(|w|\le L\),
\begin{equation}
 \langle S_L,wR_N\rangle=\langle R_N,wR_N\rangle
 \qquad (0<|w|\le L).
 \label{supp:identity-match}
\end{equation}
Equation~\eqref{supp:identity-match} leaves all entries involving
nonempty words at their reference values before normalization. Dividing
the Gram vectors by \(\|S_L\|\) now gives a joint diagonal probability
of \(\|S_L\|^{-2}\). The replacement must match a separate pairing for
every retained word. We use independent signs to enforce their
common dependence on the first and last labels of the word.

\subsection{Averaging the random signs}

In a product of rank-one projectors, each pair of successive projectors
supplies the overlap of their unit vectors. The settings in a word can
therefore be read as a walk, with an overlap for each step. The private
directions in \(R_N\) select matching endpoint labels. To reproduce that
selection using common directions, we assign each common direction
its own independent signs and form a weighted average with the signs
at both endpoints inserted.

Index these common directions, or modes, by \(\alpha\). Choose finitely
many weights \(\lambda_{\alpha}>0\) with total \(1/2\). The real, signed
coefficients \(\nu_{\alpha}\) will reproduce the reference endpoint condition.
A word of length at most \(L\) gives at most \(L-1\) successive overlaps.
The sign average below is a polynomial in the marked mode weight of at
most this degree. We therefore require the signed coefficients to
preserve its constant term and cancel powers one through \(L-1\),
\begin{equation}
\sum_{\alpha}\nu_{\alpha}=1,\qquad \sum_{\alpha}\nu_{\alpha}\lambda_{\alpha}^k=0\quad(1\le k\le L-1).
\label{eq:phase-power-moments}
\end{equation}

For each pair \((i,\alpha)\), draw an independent uniform sign
\(\sigma_{i,\alpha}\in\{-1,1\}\). In the orthonormal common basis \(g_{\alpha}\),
the vector \(h_i=\sum_{\alpha}\sqrt{\lambda_{\alpha}}\sigma_{i,\alpha}g_{\alpha}\) has squared
norm \(1/2\) and overlaps \(X_{ij}=\langle h_i,h_j\rangle
=\sum_{\alpha}\lambda_{\alpha}\sigma_{i,\alpha}\sigma_{j,\alpha}\). Write \(\mathbb E\) for
expectation over the signs. For a one-step walk between distinct
settings, inserting the signs from mode \(\alpha\) at both ends leaves only
the term \(\lambda_{\alpha}\) after averaging. Its signed sum is zero by
Eq.~\eqref{eq:phase-power-moments}. For any walk
\(W=(i_0,\ldots,i_e)\) of \(0\le e\le L-1\) steps, let
\(M_W=\prod_{t=1}^e X_{i_{t-1},i_t}\), with the empty product equal
to one. Write \(\mathbf 1_E\) for the indicator of a condition \(E\).
The same signed average gives the ordinary overlap average when the
walk closes and zero when its endpoints differ,
\begin{equation}
\sum_{\alpha}\nu_{\alpha}\mathbb E[\sigma_{i_0,\alpha}\sigma_{i_e,\alpha}M_W]
=\mathbf{1}_{i_0=i_e}\mathbb E M_W.
\label{eq:phase-walk-pairing}
\end{equation}

Give each step \(t\) its own formal variable \(z_t\), including when
the walk revisits a setting. The exponential
\(\exp(\sum_t z_t X_{i_{t-1},i_t})\) factors into one exponential
for each step. Taking the coefficient of \(z_1\cdots z_e\) selects
the linear term from every factor and therefore gives \(M_W\), even
when two steps use the same pair of settings. For independent
uniform signs \(\epsilon_i\) in a single mode, write \(G\) for the
series averaging the exponential and \(H\) for the series averaging it
with the two endpoint signs inserted,
\[
G(z)=\mathbb E e^{\sum_t z_t\epsilon_{i_{t-1}}\epsilon_{i_t}},\qquad
H(z)=\mathbb E\left[\epsilon_{i_0}\epsilon_{i_e}
e^{\sum_t z_t\epsilon_{i_{t-1}}\epsilon_{i_t}}\right].
\]
Independence across modes factors the expectation. An insertion in
mode \(\alpha\) replaces its \(G\) factor by \(H\), so the two averages
have the same coefficient formula with a single extra quotient.
Write \([z_1\cdots z_e]\) for the coefficient of that monomial,
\[
\begin{split}
\mathbb E M_W&=[z_1\cdots z_e]\prod_{\beta} G(\lambda_{\beta} z),\\
\mathbb E[\sigma_{i_0,\alpha}\sigma_{i_e,\alpha}M_W]
&=[z_1\cdots z_e]\left(\prod_{\beta} G(\lambda_{\beta} z)\right)
\frac{H(\lambda_{\alpha} z)}{G(\lambda_{\alpha} z)}.
\end{split}
\]

Since \(G(0)=1\), the quotient is a formal power series. For each monomial
of degree \(k\), its coefficient in the quotient for mode \(\alpha\) is its
coefficient in \(H(z)/G(z)\) times \(\lambda_{\alpha}^k\). Equation
\eqref{eq:phase-power-moments} cancels degrees \(1\) through \(L-1\) in
the signed sum of quotients. The remaining constant term is
\(H(0)=\mathbb E[\epsilon_{i_0}\epsilon_{i_e}]=\mathbf{1}_{i_0=i_e}\).
Multiplying the sum by \(\prod_{\beta} G(\lambda_{\beta} z)\) and extracting the
coefficient of \(z_1\cdots z_e\) gives Eq.~\eqref{eq:phase-walk-pairing}.
The product can only use quotient terms of degree at most \(e\), so
the higher powers of the mode weights are immaterial at level \(L\).

\subsubsection{Polynomial in the marked weight}

For an open walk, the endpoint average is a polynomial in the marked
mode's weight, with coefficients common to all marked modes.
Let $W=(i_0,\ldots,i_e)$ have distinct endpoints, and mark a mode
$\alpha$. Expand the product $M_W$ by assigning a mode to each of its
$e$ edge positions. A term that never uses $\alpha$ vanishes after
averaging the inserted endpoint signs. Every remaining term contains
at least one factor $\lambda_\alpha$.

Fix the nonempty set of positions that use $\alpha$, and partition the
remaining positions according to equality of their mode labels. Order
these blocks by their first position. If their sizes are
$d_1,\ldots,d_r$, the sum of weights over the possible distinct
unmarked labels is
\[
 D_\alpha(d_1,\ldots,d_r)=
 \sum_{\substack{\beta_1,\ldots,\beta_r\ne\alpha\\
                  \text{pairwise distinct}}}
 \lambda_{\beta_1}^{d_1}\cdots\lambda_{\beta_r}^{d_r}.
\]
The sign expectation for this assignment pattern is independent of
the names of these labels. Modes with equal numerical weights can have distinct labels in this
sum.

Put $q_k=\sum_{\beta\ne\alpha}\lambda_\beta^k$ and
$D_\alpha(\varnothing)=1$. Multiplying $q_{d_1}$ by the distinct-index
sum for the other blocks includes all allowed assignments. It also
includes assignments in which the first index coincides with exactly
one of the other, mutually distinct indices. Subtracting them gives
\[
\begin{split}
 D_\alpha(d_1,\ldots,d_r)
 ={}&q_{d_1}D_\alpha(d_2,\ldots,d_r)\\
 &-\sum_{j=2}^{r}
 D_\alpha(d_2,\ldots,d_j+d_1,\ldots,d_r).
\end{split}
\]
Induction on the number of blocks expresses every such sum in the
$q_k$. Giving $q_k$ degree $k$ preserves the total degree
$d_1+\cdots+d_r$.

Now define the full sums $p_k=\sum_\beta\lambda_\beta^k$ and replace
$q_k$ by $p_k-t^k$. If the marked mode occupied $k$ edge positions,
its pattern contributes $t^kD_\alpha(d_1,\ldots,d_r)$ after this
substitution, with degree at most $e$ and a factor $t$.
Summing the patterns gives one polynomial $P_W(t)$ for which
\[
 \mathbb E[\sigma_{i_0,\alpha}\sigma_{i_e,\alpha}M_W]
 =P_W(\lambda_\alpha),\qquad
 \deg P_W\leq e,\qquad P_W(0)=0.
\]
All coefficients depend on the walk and on the full sums, which are
fixed for every evaluation. The conditions in
Eq.~\eqref{eq:phase-power-moments} therefore cancel the polynomial
when $e\leq L-1$. Closed walks need only
$\sum_\alpha\nu_\alpha=1$, since their endpoint signs multiply to one.

\subsection{The projector construction}

The outcome-zero projectors need a common part to carry the overlap
products and a private part to select matching labels. In Alice's
space let \(g_{A,\alpha}\) be the orthonormal common modes and \(e_i\) the
orthonormal private directions, perpendicular to all the common modes.
Use \(g_{B,\alpha}\) and \(f_i\) in Bob's space. The parties use the same
signs to construct their common vectors \(h_{A,i}\) and \(h_{B,i}\)
by the formula for \(h_i\). In the tensor-product block, \(A_i,B_i\)
project onto \(u_i=h_{A,i}+e_i/\sqrt2\) and \(v_i=h_{B,i}+f_i/\sqrt2\)
in their respective factors and act as the identity on the other
factor. Both vectors are unit. Their within-party overlaps are
\(X_{ij}+\delta_{ij}/2\), so only \(X_{ij}\) enters when successive
labels are distinct. The reference vector pairs
matching private directions. The replacement pairs common
directions, with \(1/\lambda_{\alpha}\) cancelling the two endpoint
factors of \(\sqrt{\lambda_{\alpha}}\),
\[
R_0=2\sum_i e_i\otimes f_i,\qquad S_0=\sum_{\alpha}(\nu_{\alpha}/\lambda_{\alpha})g_{A,\alpha}\otimes g_{B,\alpha}.
\]

Applying one projector in each party gives \(A_iB_jR_0=\delta_{ij}
u_i\otimes v_j\). A word with projectors from just one party, called
a pure word, leaves a private direction in the other factor. This
direction gives zero pairing with \(S_0\). We supply the missing
pairings in one auxiliary block for each party, where the other
party's outcome-zero projectors vanish and every mixed word is zero.
In Alice's block \(A_i\) projects onto \(u_i\) and \(B_i=0\). Bob's
block interchanges the roles. In each of these local spaces let
\(g_{\alpha}\) denote its common directions. Take \(q_{\alpha}>0\) on the modes
with \(\nu_{\alpha}\ne0\), \(q_{\alpha}=0\) otherwise, and \(\sum_{\alpha} q_{\alpha}^2=1\).
The reference amplitude at the endpoint of a pure word is
\(q_{\alpha}\sigma_{i,\alpha}\). The replacement adds an endpoint term
\((2\nu_{\alpha}/q_{\alpha})\sigma_{i,\alpha}\), so the two amplitudes supply
\(2\nu_{\alpha}\) when the endpoint modes agree. Use
\[
R_{\mathrm{aux}}=\sum_{\alpha:\nu_{\alpha}\ne0}\frac{q_{\alpha}}{\sqrt{\lambda_{\alpha}}}g_{\alpha},\qquad S_{\mathrm{aux}}=\sum_{\alpha:\nu_{\alpha}\ne0}\frac{q_{\alpha}+2\nu_{\alpha}/q_{\alpha}}{\sqrt{\lambda_{\alpha}}}g_{\alpha}.
\]

For each sign assignment use the direct sums
\(R_0\oplus R_{\mathrm{aux}}\oplus R_{\mathrm{aux}}\) and
\(S_0\oplus S_{\mathrm{aux}}\oplus S_{\mathrm{aux}}\).
Take a further direct sum over all assignments, weighting each
vector in an assignment by the square root of its probability.
The resulting reference and replacement vectors are denoted by
\(R_N\) and \(S_L\). The projectors act separately on each
assignment and commute across parties. An inner product of the
complete vectors is the average of the sum of the three block
pairings. The squared norm of \(S_L\) is independent of \(N\) and determined by
\(\lambda_{\alpha},\nu_{\alpha},q_{\alpha}\).

For a pure Alice word \(w\), idempotence removes consecutive equal
settings. Read the remaining labels from left to right as the walk
\(W=(i_0,\ldots,i_e)\). The rightmost projector acts first, selecting
the private direction with label \(i_e\). For a fixed assignment,
\(wR_0=\sqrt2 M_Wu_{i_0}\otimes f_{i_e}\). Pairing with \(R_0\) and
averaging gives \(2\mathbf{1}_{i_0=i_e}\mathbb E M_W\). Pairing with
\(S_0\) gives zero because of the private Bob factor \(f_{i_e}\).
The difference \(\langle S_{\mathrm{aux}}-R_{\mathrm{aux}},wR_{\mathrm{aux}}\rangle\)
of the replacement and reference pairings in Alice's auxiliary block
must equal the missing reference pairing. For that difference,
each terminal projector supplies \(\sqrt{\lambda_\alpha}\sigma_{i,\alpha}\),
cancelling the denominator in the auxiliary vector. Insert the added
amplitude \((2\nu_\alpha/q_\alpha)\sigma_{i_0,\alpha}\) at the first
endpoint and the reference amplitude \(q_\beta\sigma_{i_e,\beta}\)
at the last endpoint to obtain
\[
2\sum_{\alpha,\beta}\frac{\nu_{\alpha}q_{\beta}}{q_{\alpha}}\mathbb E[\sigma_{i_0,\alpha}\sigma_{i_e,\beta}M_W],
\]
where both sums are over modes with nonzero coefficient. When \(\alpha\ne\beta\),
flipping the signs at every setting in mode \(\alpha\) preserves \(M_W\) and
changes the sign of the integrand. Only \(\alpha=\beta\) survives the average.
Equation~\eqref{eq:phase-walk-pairing} gives exactly the tensor-product
reference pairing. The pure Bob words obey the same identity after
interchanging the parties.

For a mixed word, cross-party commutation collects Alice's and Bob's
projectors into two ordered products. Idempotence removes consecutive
equal labels in each product, giving
\(w=A_{i_1}\cdots A_{i_p}B_{j_1}\cdots B_{j_q}\) with \(p,q\ge1\)
and \(p+q\le L\). Let \(M_A,M_B\) be the overlap products along the
two walks. The auxiliary blocks vanish. In the tensor-product block
the two rightmost local projectors must act on the same private label
of \(R_0\), so within an assignment
\[
wR_0=\mathbf{1}_{i_p=j_q} M_A M_B u_{i_1}\otimes v_{j_1}.
\]
If the terminal labels agree, append Bob's reversed walk to Alice's
walk. Because \(X_{ij}=X_{ji}\), the joined walk has overlap product
\(M_A M_B\). It runs from \(i_1\) to \(j_1\) in
\((p-1)+(q-1)\le L-2\) steps. In its pairing with \(R_0\), the
coefficient two cancels the two private endpoint amplitudes \(1/\sqrt2\).
Averaging gives \(\mathbf{1}_{i_1=j_1}\mathbb E(M_A M_B)\).
In the pairing with \(S_0\), the coefficient \(\nu_\alpha/\lambda_\alpha\)
cancels the two shared endpoint amplitudes \(\sqrt{\lambda_\alpha}\),
leaving the two endpoint signs with weight \(\nu_\alpha\).
Equation~\eqref{eq:phase-walk-pairing} gives
\(\langle S_L,wR_N\rangle=\langle R_N,wR_N\rangle\) for every nonempty
word of length at most \(L\).

\subsection{Normalization and positivity of the moment matrix}

To preserve the projector identities while changing the empty-word
diagonal, let \(\varepsilon\) be evaluation at the deterministic
assignment of outcome \(1\) to every setting. It is an algebra character, meaning
that it takes products to products. It respects Eq.~\eqref{supp:relations}
and maps the identity to one and every nonempty word in outcome-zero
projectors to zero. Write \(C_L=\|S_L\|^2\). The character lets us
change the empty-word entry from \(\|R_N\|^2\) to \(C_L\) while keeping
every nonempty moment at its reference value before normalization,
\begin{equation}
 C_L\ell(p)=\langle R_N,pR_N\rangle+
             (C_L-\|R_N\|^2)\varepsilon(p).
 \label{supp:adjusted-functional}
\end{equation}
For \(N\ge1\), the vector \(A_iB_iR_N\) has norm one and inner product
one with \(S_L\). Thus Cauchy--Schwarz gives \(C_L\ge1\). Evaluating
Eq.~\eqref{supp:adjusted-functional} at the identity gives \(\ell(1)=1\).
Both terms are functionals on the quotient algebra, so \(\ell\) respects
adjoints and every projector identity in every context.

The character's coefficient in Eq.~\eqref{supp:adjusted-functional}
may be negative. Positivity through level \(L\) comes from the
matched pairings. Take a complex polynomial \(f\) of degree at most
\(L\). Write \(f=c1+g\), where \(c=\varepsilon(f)\) and \(\varepsilon(g)=0\),
so \(g\) is a combination of nonempty words of degree at most \(L\).
Equation~\eqref{supp:identity-match} matches the terms linear in \(g\),
and \(\varepsilon(g^*g)=0\) leaves the quadratic term
\(\langle R_N,g^*gR_N\rangle\) from the reference vector. The constant
term has coefficient \(C_L\), so all three terms come from the
squared norm of the same linear combination of Gram vectors,
\begin{equation}
 C_L\ell(f^*f)=\|c S_L+gR_N\|^2\ge0.
 \label{supp:squared-norm}
\end{equation}
Equation~\eqref{supp:squared-norm} proves positivity of the moment matrix
indexed by words of length at most \(L\), whose pairwise products can have
length up to \(2L\).

For outcome \(a\), write \(A_i^0=A_i\), \(A_i^1=1-A_i\), and similarly
for Bob. A product of these projectors expands to words of at most its
original degree. Its moment matrix is therefore positive by the same
square identity. The full outcome projectors obey orthogonality,
completeness, self-adjointness, idempotence and cross-party commutation
in the algebra, so their moments satisfy the corresponding NPA equalities.
Each joint outcome \(E=A_i^aB_j^b\) is a self-adjoint projector of degree
at most two. Since \(L\ge2\), \(p(a,b\mid i,j)=\ell(E)=\ell(E^*E)\ge0\).
Completeness makes the joint projectors sum to one. Summing over
one party's outcomes gives the other party's marginal projector
regardless of the first party's setting, proving nonsignalling.

The tensor-product block gives unnormalized mixed moments \(\delta_{ij}\)
and singleton moments two. The auxiliary block for that singleton adds
\(\mathbb E(\sum_{\alpha} q_{\alpha}\sigma_{i,\alpha})^2=\sum_{\alpha} q_{\alpha}^2=1\).
The functional therefore gives
\begin{equation}
 \ell(A_iB_j)=\frac{\delta_{ij}}{C_L},\qquad
 \ell(A_i)=\ell(B_i)=\frac3{C_L}.
 \label{supp:probabilities}
\end{equation}
With \(s_L=1/C_L\), the marginals and normalization give
\begin{equation}
\begin{array}{c|cc}
 &b=0&b=1\\ \hline
 a=0&s_L\delta_{ij}&3s_L-s_L\delta_{ij}\\
 a=1&3s_L-s_L\delta_{ij}&1-6s_L+s_L\delta_{ij}
\end{array} .
\label{supp:full-box}
\end{equation}
We denote this correlation by \(p_L\). An upper bound on \(C_L\)
gives a lower bound on its diagonal rate \(s_L\).

\subsection{A cubic bound on the normalization}
\label{supp:rate-bound}

For $1\le j\le L$, we use interpolation nodes proportional to the
squares of the half-integers $t_j=j-\tfrac12$,
\[
 T=4\sum_{j=1}^L t_j^2=\frac{4L^3-L}{3},\qquad
 \lambda_j=\frac{3t_j^2}{2T}.
\]
The Lagrange coefficient for evaluation at zero is
\begin{equation}
 \mu_j=\prod_{k\ne j}\frac{\lambda_k}{\lambda_k-\lambda_j}
 =\frac{2(-1)^{j-1}P^2}{t_j(L-j)!(L+j-1)!},
 \qquad P=\frac{(2L)!}{4^L L!}.
 \label{supp:phase-weights}
\end{equation}
The factorial formula uses $t_k^2-t_j^2=(k-j)(k+j-1)$.
For every polynomial $F$ of degree at most $L-1$,
$\sum_j\mu_j F(\lambda_j)=F(0)$.
Since $(L-j)!(L+j-1)!$ increases with $j$,
$P^2/[(L-1)!L!]=L[4^{-L}\binom{2L}{L}]^2$ bounds the ratio
of $P^2$ to the factorial product for every node. To bound this last
expression, write $a=4^{-L}\binom{2L}{L}$. The integrals of
$\sin^{2L}\theta$ and $\sin^{2L-1}\theta$ over $[0,\pi/2]$ equal
$\pi a/2$ and $1/(2La)$ respectively. The first is at most the second,
so $La^2\leq1/\pi$. Using $\sum_{j\geq1}t_j^{-2}=\pi^2/2$ gives
\begin{equation}
 |\mu_j|\leq\frac{2}{\pi t_j},\qquad
 \sum_{j=1}^{L}\mu_j^2\leq2 .
 \label{supp:interpolation-coefficient-bound}
\end{equation}

Use $n_j$ independent modes of weight $\lambda_j$, each with signed
coefficient $\nu_\alpha=\mu_j/n_j$, where
\begin{equation}
 n_j=\left\lceil\frac{T}{60t_j^4}\right\rceil .
 \label{supp:mode-multiplicity}
\end{equation}
Their total weight satisfies
\[
 \sum_j n_j\lambda_j
 \leq \sum_j\lambda_j+\frac1{40}\sum_jt_j^{-2}
 \leq\frac38+\frac{\pi^2}{80}<\frac12 .
\]
Add one mode of the remaining weight with signed coefficient zero.
Each replicated node keeps the same aggregate signed coefficient, so the
conditions in Eq.~\eqref{eq:phase-power-moments} hold. Its squared
norm contribution is divided by the multiplicity, giving
\[
 \|S_0\|^2=\sum_j\frac{\mu_j^2}{n_j\lambda_j^2}
 \leq \frac{80}{3}T\sum_j\mu_j^2\leq\frac{160}{3}T.
\]

To give the auxiliary block unnormalized marginal one, set
$q_\alpha^2=|\nu_\alpha|/(H_*\sqrt{\lambda_\alpha})$ on the active
modes and zero on the added mode, with
\[
 H_*=\sum_\alpha\frac{|\nu_\alpha|}{\sqrt{\lambda_\alpha}}
     =\sum_j\frac{|\mu_j|}{\sqrt{\lambda_j}}
     \leq \pi\sqrt{\frac{2T}{3}} .
\]
These squared amplitudes sum to one. The exact auxiliary norm is
\[
 \|S_{\mathrm{aux}}\|^2
 =\sum_{\alpha:\nu_\alpha\ne0}\frac{q_\alpha^2}{\lambda_\alpha}
   +4\sum_\alpha\frac{\nu_\alpha}{\lambda_\alpha}+4H_*^2 .
\]
Here the first sum is at most $1/\lambda_1=8T/3$. To evaluate the
signed sum, interpolate the degree-$L-1$ polynomial
\[
 F(z)=\frac{1-\prod_{j=1}^L(1-z/\lambda_j)}{z}.
\]
It takes the value $1/\lambda_j$ at each node and has constant term
$\sum_j1/\lambda_j$. Hence
\[
 \sum_\alpha\frac{\nu_\alpha}{\lambda_\alpha}
 =\sum_j\frac{\mu_j}{\lambda_j}=\sum_j\frac1{\lambda_j}
 \leq\frac{\pi^2T}{3}.
\]
It follows that
$\|S_{\mathrm{aux}}\|^2\leq(8/3+4\pi^2)T$. Adding the tensor-product
block and the two auxiliary blocks gives
\begin{equation}
 C_L\leq\left(\frac{176}{3}+8\pi^2\right)T
       <\frac{1664}{9}L^3<185L^3,\qquad s_L>\frac1{185L^3}.
 \label{supp:cubic-rate}
\end{equation}
All the mode weights and signed coefficients are rational. For each
active mode of weight $\lambda_j$, the factor
$H_*\sqrt{\lambda_j}=t_j\sum_k|\mu_k|/t_k$ is rational, so
$q_\alpha^2$ is rational. The expanded norm is a rational expression
in these quantities, giving rational $C_L$, $s_L$ and probabilities
in Eq.~\eqref{supp:full-box}.

\subsection{A larger guaranteed rate with level-dependent marginals}
\label{supp:variable-marginals}

Allowing the unnormalized marginals to depend on $L$ gives a guaranteed
diagonal rate $\widetilde s_L>1/(165L^3)$. For each $L\geq2$ there are
rational $\widetilde s_L$ and $\beta_L=O(\log L)$ such that the following
correlation belongs to $Q_L(N,N,2,2)$ for every $N$,
\begin{align*}
 p(0,0\mid i,j)&=\widetilde s_L\delta_{ij},\\
 p(0,1\mid i,j)&=p(1,0\mid i,j)=\widetilde s_L(\beta_L-\delta_{ij}),\\
 p(1,1\mid i,j)&=1-2\beta_L\widetilde s_L+\widetilde s_L\delta_{ij}.
\end{align*}
The one-bit protocol using joint zero outcomes has the same score
bound with $\widetilde s_L$ in place of $s_L$.

Keep the expressions for the reference and replacement vectors and set
$\widetilde q_\alpha^2=2|\widetilde\nu_\alpha|$ on the active modes.
The difference between the auxiliary vectors has coefficient
$2\widetilde\nu_\alpha/(\widetilde q_\alpha\sqrt{\widetilde\lambda_\alpha})$
in mode $\alpha$. Its product with the corresponding reference
coefficient still gives the signed pairing after the endpoint
overlaps contribute $\widetilde\lambda_\alpha$. Therefore the
empty-word pairings agree by the same interpolation conditions.
The two unnormalized marginals are now
$\beta_L=2+\sum_\alpha\widetilde q_\alpha^2
=2+2\sum_\alpha|\widetilde\nu_\alpha|$.
The squared numerator of each auxiliary replacement coefficient is
$8\widetilde\nu_\alpha$ if $\widetilde\nu_\alpha>0$ and zero if
$\widetilde\nu_\alpha<0$. Thus
\[
 \widetilde C_L
 =\sum_\alpha\frac{\widetilde\nu_\alpha^2}{\widetilde\lambda_\alpha^2}
  +16\sum_{\alpha:\widetilde\nu_\alpha>0}
       \frac{\widetilde\nu_\alpha}{\widetilde\lambda_\alpha}.
\]

Use nodes $\widetilde\lambda_j=t_j^2/T$. A common rescaling of the nodes
leaves the Lagrange coefficients $\mu_j$ unchanged. Split $\mu_j$
equally among $\widetilde n_j=\lceil T/(2\pi^2t_j^4)\rceil$ independent
modes at $\widetilde\lambda_j$. Their total weight is at most
$1/4+\sum_{j=1}^L t_j^{-2}/(2\pi^2)<1/2$. Add one mode of the
remaining weight with signed coefficient zero. The first sum in
$\widetilde C_L$ is bounded by
\[
 \sum_j\frac{\mu_j^2}{\widetilde n_j\widetilde\lambda_j^2}
 \leq2\pi^2T\sum_j\mu_j^2\leq4\pi^2T .
\]
The positive coefficients occur at odd $j$. Equation~\eqref{supp:interpolation-coefficient-bound}
gives
\[
 16\sum_{j\ \mathrm{odd}}\frac{\mu_j}{\widetilde\lambda_j}
 \leq\frac{32T}{\pi}\sum_{j\ \mathrm{odd}}t_j^{-3}
 \leq\frac{32T}{\pi}\left(8+\frac29\right),
\]
where $t_1^{-3}=8$ and the contribution of $j\geq3$ is bounded by
$\int_2^\infty (z-\tfrac12)^{-3}\,\mathrm dz=2/9$.
Adding both terms gives
\[
 \widetilde C_L\leq\left(4\pi^2+\frac{2368}{9\pi}\right)T<165 L^3,
 \qquad\widetilde s_L=1/\widetilde C_L .
\]
The strict inequality uses $3.14<\pi<22/7$ and $T<4L^3/3$.
The weights and signed coefficients are rational, so
$\widetilde C_L$, $\widetilde s_L$ and $\beta_L$ are rational.
The bound $\sum_\alpha|\widetilde\nu_\alpha|=\sum_j|\mu_j|
\leq(2/\pi)\sum_j 1/t_j$ gives $\beta_L=O(\log L)$.

\subsection{Normalization required by the sign construction}

The signed-moment conditions already force a cubic cost in the
tensor-product term $\|S_0\|^2=\sum_\alpha\nu_\alpha^2/\lambda_\alpha^2$.
List the positive mode weights in decreasing order
$\lambda_{(1)}\geq\lambda_{(2)}\geq\cdots$, including repetitions.
There are at least $L$ distinct weights with nonzero coefficient.
Otherwise a polynomial of degree at most $L-1$ that vanishes at all such
weights and equals one at zero would violate
Eq.~\eqref{eq:phase-power-moments}.

Fix $1\leq u\leq L$, put $r=\lambda_{(u)}$ and $d=L-u$.
The total weight $1/2$ gives $r\leq1/(2u)$.
For $x=\sin^2\theta\in[0,1]$, define
\[
 q_d(x)=\frac{\sin((2d+1)\theta)}{(2d+1)\sin\theta}.
\]
The expansion $\sin((2d+1)\theta)/\sin\theta
=1+2\sum_{j=1}^d\cos(2j\theta)$ gives a polynomial of degree $d$
with $q_d(0)=1$. It satisfies $\sqrt{x}|q_d(x)|\leq(2d+1)^{-1}$.
Rescale it to $[0,r]$ and include factors that vanish on the first
$u-1$ weights,
\[
 p(x)=q_d(x/r)\prod_{j=1}^{u-1}(1-x/\lambda_{(j)}).
\]
This polynomial has degree at most $L-1$ and $p(0)=1$.
At every remaining weight the product lies in $[0,1]$. In particular
$\sqrt{\lambda_\alpha}|p(\lambda_\alpha)|\leq\sqrt r/(2d+1)$
at every mode weight. Signed evaluation and Cauchy--Schwarz give
\[
 1=\sum_\alpha\nu_\alpha p(\lambda_\alpha)
 \leq\frac{\sqrt r}{2d+1}\sum_\alpha
           \frac{|\nu_\alpha|}{\sqrt{\lambda_\alpha}}
 \leq\frac{\sqrt r}{(2d+1)\sqrt2}\|S_0\| .
\]
Using $r\leq1/(2u)$ and maximizing over $u$ gives
\[
 C_L\geq\|S_0\|^2
 \geq\max_{1\leq u\leq L}4u(2L+1-2u)^2 .
\]
Choosing $u=1$ gives $C_L\geq36$ for all $L\geq2$. An admissible
integer nearest to $(2L+1)/6$ gives
$\liminf_{L\to\infty}\|S_0\|^2/L^3\geq64/27$.

\subsection{A cubic cost for rank-one projectors with independent settings}
\label{supp:rank-one-converse}

Keeping the mixed-word Gram block and its original pairings with the
empty word imposes a cubic normalization cost for the rank-one projectors
below. The bound allows arbitrary auxiliary blocks annihilated by mixed
words.

Consider unit vectors $u_i=h_i+\sqrt{\alpha_i}e_i$, with orthonormal
private directions $e_i$ orthogonal to a common subspace containing all $h_i$,
and $\alpha_i=1-\|h_i\|^2>0$. For each $N$ these subspaces are fixed before averaging. The $h_i$ may be random, with inner products
averaged over their joint law. The two parties project onto $u_i$ and
$\overline{u_i}$ in conjugate local spaces, and the reference vector in
this tensor-product block is
$R_0=\sum_i\alpha_i^{-1}e_i\otimes\overline{e_i}$. We assume the reference
vector is square integrable. The signs of the preceding construction
are one choice of the $h_i$. Put
\begin{equation}
 K_N=\mathbb E\left[\frac1N\sum_i\lvert h_i\rangle\langle h_i\rvert\right].
\end{equation}
If the $h_i$ are independent across settings and $\operatorname{Tr}K_N\le\tau$
for a fixed $0<\tau\le1$, then every positive semidefinite completion of
the indicated Gram block with its original empty-word pairings obeys
\begin{equation}
 \liminf_{N\to\infty} C_{L,N}
 \ \ge\ \max_{1\le u\le L-1}\frac{u(2L-2u-1)^2}{\tau^2},
 \label{supp:rank-one-converse-bound}
\end{equation}
where $C_{L,N}$ is the unnormalized empty-word diagonal and $L\ge2$ is
fixed. The number of common directions and the distributions may vary
with $N$. In particular, when $\tau=1/2$, the bound on $C_{L,N}$ grows
asymptotically at least as $(64/27+o(1))L^3$ after taking $N$ large at
each fixed $L$.

To prove this bound, let $P_i=\lvert u_i\rangle\langle u_i\rvert$ be Alice's local
projector. For $0\le e\le L-2$, consider the row of the mixed word
$A_{i_0}\cdots A_{i_e}B_{i_e}$. On the reference vector it equals
$\operatorname{vec}(P_{i_0}\cdots P_{i_e})$, where
$\operatorname{vec}(\lvert v\rangle\langle w\rvert)=v\otimes\overline w$.
Indeed, the two projectors with label $i_e$ select only its private
reference term, and their two factors of $\sqrt{\alpha_{i_e}}$ cancel
its coefficient. Average these rows over all ordered, distinct labels
$i_0,\ldots,i_e$, and call the vector $Z_e$. The original empty-word
pairing of $Z_e$ is one for $e=0$ and zero for $e>0$. For $e>0$ the two
endpoint directions selected by the word are attached to distinct
private labels, which have zero pairing with the matched reference.

We bound the averaged row using the empirical projectors
$\overline P=N^{-1}\sum_iP_i$ and the empirical common block
$\overline K=N^{-1}\sum_i\lvert h_i\rangle\langle h_i\rvert$. Throughout, matrix norms
in the following estimates are Hilbert--Schmidt norms, with the
expectation of their squares taken over the shared randomness.
The average over tuples sampled with replacement is
$\operatorname{vec}(\overline P^{e+1})$. Every product of the rank-one
projectors has Hilbert--Schmidt norm at most one. Hence conditioning on
no repeated labels changes the average by at most twice the collision
probability, which is at most $e(e+1)/N$. Furthermore, the orthogonal
private directions give in each realization
\begin{equation}
 \|\overline P-\overline K\|_{\mathrm{HS}}^2
 =\frac{1}{N^2}\sum_i\bigl[\alpha_i^2+2\alpha_i(1-\alpha_i)\bigr]
 \le\frac1N.
\end{equation}
The matrices $\overline P$, $\overline K$ and $K_N$ are contractions.
For powers of order $e+1$, the difference of the powers telescopes,
giving a Hilbert--Schmidt norm at most $e+1$ times the norm of the
difference of the matrices. Thus, for $N\ge L-1$,
\begin{equation}
 \|Z_e-\operatorname{vec}(K_N^{e+1})\|
 \le \frac{e(e+1)}{N}+(e+1)\left(N^{-1/2}+\Delta_N\right),
 \quad
 \Delta_N^2=\mathbb E\|\overline K-K_N\|_{\mathrm{HS}}^2 .
 \label{supp:rank-one-row-approximation}
\end{equation}
The norm on the left is the averaged Hilbert-space norm.
Independence across settings gives $\Delta_N^2\le1/N$, by adding the
variances of the $N$ common rank-one terms. The approximation also
applies to correlated choices with $\Delta_N\to0$.

It remains to combine the averaged rows. Fix $1\le u\le L-1$, put
$d=L-u-1$ and $r=\tau/u$, and write $\lambda_a$ for the positive
eigenvalues of $K_N$ counted with multiplicity. There are at most $u-1$
eigenvalues strictly above $r$. For $x=\sin^2\theta\in[0,1]$, define
the degree-$d$ polynomial
\begin{equation}
 q_d(x)=\frac{\sin((2d+1)\theta)}{(2d+1)\sin\theta}.
\end{equation}
It has $q_d(0)=1$ and $\sqrt{x}|q_d(x)|\le1/(2d+1)$. The polynomial
\begin{equation}
 F(z)=q_d(z/r)\prod_{a:\lambda_a>r}(1-z/\lambda_a)
     =\sum_{e=0}^{L-2} f_e z^e
\end{equation}
has $F(0)=1$. At every eigenvalue at most $r$ the linear factors lie in
$[0,1]$, while the other eigenvalues are zeros of $F$. Therefore
\begin{equation}
 \|K_N F(K_N)\|_{\mathrm{HS}}^2
 =\sum_a\lambda_a^2|F(\lambda_a)|^2
 \le \frac{r\operatorname{Tr}K_N}{(2d+1)^2}
 \le \frac{\tau^2}{u(2d+1)^2}.
 \label{supp:rank-one-filter-bound}
\end{equation}
The sum of the absolute coefficients of $F$ is bounded by
$D=(1+1/r)^{u-1}\|q_d(\cdot/r)\|_{\mathrm{coef},1}$, independently of
$N$ and the eigenvalues. By Eq.~\eqref{supp:rank-one-row-approximation},
the norm of $V=\sum_e f_e Z_e$ therefore satisfies
\begin{equation}
 \|V\|\le \frac{\tau}{\sqrt u(2d+1)}
 +D\left[\frac{(L-1)(L-2)}{N}+(L-1)(N^{-1/2}+\Delta_N)\right].
 \label{supp:rank-one-finite-n}
\end{equation}
The original empty-word pairing of $V$ is its constant coefficient
$f_0=1$. The principal $2\times2$ matrix on the empty word and this
linear combination must be positive semidefinite. It gives
$1\le C_{L,N}\|V\|^2$. Taking $N\to\infty$ at fixed $L$ gives
Eq.~\eqref{supp:rank-one-converse-bound}. To get the cubic coefficient,
take an admissible integer $u$ nearest to $(2L-1)/6$.
\section{Information causality with one copy}\label{supp:ic}

Alice holds a database \(R=(R_1,\ldots,R_{n_R})\) of \(n_R\) independent
bits. Bob alone receives an index \(j\in\{1,\ldots,n_R\}\), chosen independently of the database. Alice sends her message without knowing this index. Write \(G_j\) for Bob's output when he is asked for entry \(j\).
The information-causality (IC) score
\(I_{\mathrm{IC}}=\sum_j I(R_j:G_j)\), in bits, sums the Shannon mutual
information \(I(R_j:G_j)\) for each possible request.

Let \(F_L(N)\) be the supremum of this score for level-\(L\) correlations
with \(N\) settings per party and two outcomes, over finite databases
whose independent bits may have arbitrary biases. The database and the
request are independent of the supplied correlation. Alice and Bob may
use shared randomness independent of these inputs and of the correlation,
as well as private randomness. They have at most one use of the correlation
and at most one bit of communication from Alice to Bob. The message has two
possible values, including on runs where the correlation is unused. Bob can
choose his setting after receiving the message and keep his full transcript.

\subsection{Information causality from information axioms}\label{supp:ic-axioms}

Consider a nonsignalling theory with a finite, symmetric and nonnegative
mutual information on the relevant systems and their composites. It
agrees with Shannon mutual information when both systems are classical,
and local transformations cannot increase it. The chain rule requires
that \(I(A:BC)-I(A:C)=I(B:AC)-I(B:C)\), with the common difference
denoted by \(I(A:B\mid C)\) for systems \(A,B,C\)~\cite{Pawlowski2009,AlSafiShort2011,WakakuwaMurao2012}.
Juxtaposition denotes the composite of the indicated systems. Classical
registers can be copied and discarded.

Let \(B\) contain Bob's resources before he receives the classical message
\(M\). Bob may defer earlier operations to his decoder, with their
choices made independently of \(M\). After Alice forms \(M\), the state of
\(B\) averaged over the message is independent of \(R\) by nonsignalling.
Consequently \(I(R:B)=0\). The two independent systems can be prepared
locally from constant classical registers, whose mutual information is
zero, and this preparation cannot increase mutual information.

The information added by the message is \(I(R:M\mid B)\). By the
symmetric chain rule this is \(I(RB:M)-I(B:M)\). Take \(M'\) to be an
exact classical copy of \(M\). Adding it to \(RB\) can only increase
the mutual information with \(M\). The chain rule gives
\begin{align}
 I(R:BM)&=I(R:M\mid B)\le I(RB:M)\le I(RBM':M),\nonumber\\
 I(RBM':M)&=I(M':M)+I(RB:M\mid M')=H(M).
 \label{supp:ic-axiom-budget}
\end{align}
Data processing for copying \(M'\) and for discarding the copy gives
\(I(RB:M'M)=I(RB:M')\) in both directions, so the conditional term
vanishes. Classical consistency gives \(I(M':M)=H(M)\).

For each request \(j\), Bob obtains \(G_j\) by processing his resources
\(W=BM\). All requests thus admit the same bound \(I(R_j:G_j)\le I(R_j:W)\).
To sum these bounds, put \(R_{<j}=(R_1,\ldots,R_{j-1})\). Independence
gives \(I(R_j:R_{<j})=0\), so
\[
 I(R:W)=\sum_j I(R_j:W\mid R_{<j})=\sum_j I(R_j:W R_{<j}).
\]
Discarding \(R_{<j}\) bounds each \(I(R_j:W)\) by the corresponding
term on the right. Hence \(\sum_j I(R_j:G_j)\le I(R:W)\le H(M)\),
and a two-valued message has \(H(M)\le1\). Classical and quantum
mutual information obey the required rules in their respective
theories, so the upper bound applies to both.

\subsection{Lower bound from equality certification}\label{supp:ic-lower}

The equality certificate can tell Bob that his requested bit has the
rare value one. Let Alice's \(N\) independent bits each be one with
probability \(r\in(0,1)\). When the database contains a one, Alice
uses the correlation of Eq.~\eqref{supp:full-box} with setting \(x\)
equal to the index of the first one. She sends her outcome \(a\).
If the database contains only zeros, she sends one. Bob uses his
requested index \(i\) as the setting. He raises the flag \(G_i=1\)
when the message and his outcome are both zero, and sets \(G_i=0\)
on every other run. A flag requires \(x=i\), so it certifies
\(R_i=1\). Its probability is \(t_i=s_Lr(1-r)^{i-1}\), since the
entries before \(i\) must be zero and entry \(i\) must be one.

To count the information in a flag, average the relative entropy
\(D(P_{R_i\mid G_i=g}\Vert P_{R_i})\) of each posterior from the prior
over Bob's outputs \(g\in\{0,1\}\). This gives
\(I(R_i:G_i)=\sum_g\Pr(G_i=g)D(P_{R_i\mid G_i=g}\Vert P_{R_i})\).
The flagged posterior is
concentrated on one, an event of prior probability \(r\), and hence
has relative entropy \(\log_2(1/r)\) bits. Its contribution gives a
lower bound because the unflagged posterior also has nonnegative
relative entropy,
\begin{equation}
 I(R_i:G_i)\ge t_i\log_2(1/r).
 \label{supp:mutual-information}
\end{equation}
The events that select the first one are disjoint. The sum of their
certificate probabilities is \(s_L\) times the probability of any
one in the database, so the total score satisfies
\begin{equation}
 I_{\mathrm{IC}}\ge s_L\bigl[1-(1-r)^N\bigr]\log_2(1/r).
 \label{supp:ic-sparse}
\end{equation}
Decreasing \(r\) increases the information per flag. Increasing \(N\)
lets the chance of finding a one remain appreciable at the smaller
value of \(r\).

For \(N\ge2\), choose \(r=1/N\). On average there is one entry equal
to one. The probability of at least one such entry is at least
\(1-e^{-1}\), so
\begin{equation}
 F_L(N)\ge (1-e^{-1})s_L\log_2 N.
 \label{supp:ic-log-lower}
\end{equation}
Taking \(r=(\ln N)/N\), where \(\ln\) denotes the natural logarithm,
makes the probability of finding a one at least \(1-N^{-1}\).
Each flag contributes
\(\log_2 N-\log_2\ln N\) bits when it occurs. Equation~\eqref{supp:ic-sparse}
gives
\begin{equation}
 F_L(N)\ge s_L(1-N^{-1})\bigl(\log_2 N-\log_2\ln N\bigr)
          =s_L\log_2 N-O_L(\log\log N).
 \label{supp:ic-leading}
\end{equation}
At each fixed \(L\) the score thus grows without bound as \(N\)
increases. The constants in \(O_L\) can depend on \(L\). With
tensor-product quantum correlations, the bound of one bit follows
from Eq.~\eqref{supp:ic-axiom-budget} and is attained by sending a
fair database bit~\cite{Pawlowski2009}.

\paragraph{Fair independent bits.}
With fair bits the rare event can be an entire block of ones, as in
the protocol of Das Bhowmik and Banik~\cite{DasBhowmikBanik2026}.
Divide the database into \(N\) blocks of \(k\) independent fair bits.
The all-ones string occurs in a block with probability \(r=2^{-k}\).
Alice uses the index of the first all-ones block and sends her outcome,
or sends one if there is no such block. Bob uses the index of the
block containing his requested bit. He returns one precisely when
his outcome and the message are both zero. This flag certifies
that every bit of the requested block is one. The flag for block
\(i\) occurs with probability \(t_i=s_Lr(1-r)^{i-1}\).
For each of its \(k\) possible requests the flag
contributes \(t_i\) bits of mutual information in the score, for a
total of \(t_i\log_2(1/r)=kt_i\) bits from that block. Summing gives
\begin{equation}
 I_{\mathrm{IC}}\ge k s_L\bigl[1-(1-2^{-k})^N\bigr].
 \label{supp:ic-geometric}
\end{equation}

Taking \(k=\lfloor\log_2N\rfloor\) keeps the expected number of all-ones
blocks at least one and bounds the score from below by
\((1-e^{-1})s_L\lfloor\log_2N\rfloor\). For \(x=\log_2N\) sufficiently
large, \(k=\lfloor x-\log_2 x\rfloor\) gives the lower bound
\(s_L(x-\log_2x-1)(1-e^{-x})\), with the same leading term as
Eq.~\eqref{supp:ic-leading}. When no block is all ones, Alice can omit
the query or ignore its outcome.

\subsection{Nonsignalling upper bound}\label{supp:ic-upper}

To bound the growth of the score, ask how much classical communication
would reproduce a protocol using one nonsignalling correlation
\(p(a,b\mid x,y)\). There are \(N\) Alice settings \(x\), and Alice's
message lies in the finite alphabet \(\mathcal M\). Sending the setting
along with the original message lets Bob pick out one row of a table
sampled in advance. Before either party gets an input,
sample the shared seed and private controls. For each setting \(x\),
use additional shared randomness to draw an outcome \(a_x\) from
\(p_A(\cdot\mid x)\). Nonsignalling makes this distribution independent
of Bob's setting. For each drawn \(a_x\), sample a response row \(g_x\)
with an independent entry for each Bob setting \(y\). Each entry has
Bob's conditional outcome law given the sampled Alice outcome,
\[
 \Pr\bigl[g_x(y)=b\mid a_x\bigr]
 =p(a_x,b\mid x,y)/p_A(a_x\mid x).
\]
The denominator is positive on every sampled outcome. If the shared seed
selects from a family of nonsignalling correlations, the table is sampled
from the selected correlation.

Alice chooses \(X\) from her database and the controls, computes her
original message \(M\) using \(a_X\), and sends \((X,M)\) in the
simulation. Bob uses his original rule to choose \(y\) from his request
and, when available at that step of the original protocol, \(M\). He
then uses \(g_X(y)\) as his outcome. Conditional on \(X,a_X,M\) and the
controls, the setting choice uses no entry of the row. Its selected
entry therefore has the required conditional distribution, so the
simulation reproduces the joint law of the database and Bob's entire
transcript for each request. If Alice skips her query, she may send a
dummy setting and ignore \(a_X\). If Bob skips it, he omits the lookup.

Let \(\Lambda\) contain the random controls and the table, which are
independent of the database and request. For each request Bob's
output is a function of \((X,M,\Lambda)\). Data processing, independence
of the database bits, and the alphabet size of the simulated message
give
\begin{equation}
 \sum_j I(R_j:G_j)
 \le \sum_j I(R_j:X,M,\Lambda)
 \le I(R:X,M\mid\Lambda)
 \le \log_2N+\log_2|\mathcal M|.
 \label{supp:ic-upper-bound}
\end{equation}
The entries are independent conditional on \(\Lambda\), giving
\(H(R\mid\Lambda)=\sum_j H(R_j)\). Conditional entropy is subadditive,
\(H(R\mid X,M,\Lambda)\le\sum_j H(R_j\mid X,M,\Lambda)\), which gives
the middle inequality. The final bound counts the possible pairs
\((X,M)\). For a two-symbol message, Eqs.~\eqref{supp:ic-log-lower} and
\eqref{supp:ic-upper-bound} give \(F_L(N)=\Theta_L(\log N)\), with
multiplicative constants independent of \(N\) at fixed \(L\).

The bound is attained under nonsignalling at \(N=2^k\) for \(k\ge1\).
Alice has \(k+1\) independent fair bits \(R_0,\ldots,R_k\). Her setting
encodes each bit's parity with a reference bit \(R_0\). With \(\oplus\)
denoting addition modulo two, she inputs
\(x=(R_1\oplus R_0,\ldots,R_k\oplus R_0)\) and sets \(x_0=0\). For
Bob's setting \(y\in\{0,\ldots,k\}\), the correlation supplies a uniform
\(a\) and \(b=a\oplus x_y\), so both marginals are uniform. There are
enough settings because \(k+1\le2^k\), and unused Bob settings duplicate
setting zero. Alice sends \(M=a\oplus R_0\) and Bob recovers
\(M\oplus b=R_y\) on every request. The score is \(k+1\).

\subsection{Optimal one-copy information growth for equality correlations}
\label{supp:optimal-equality-score}

Let \(P_=\) and \(P_{\ne}\) be laws on finite outcome alphabets
\(\mathcal A\times\mathcal B\), with the same Alice marginals and the
same Bob marginals. Alice and Bob each have \(N\geq2\) settings. Their
joint outcomes follow \(P_=\) for equal settings and \(P_{\ne}\) for
unequal settings. The database \(R=(R_i)_i\) consists of a finite
number of independent bits with arbitrary biases. Bob's request is
independent of the database and the random controls, and Alice does
not know it. The parties may use the correlation once and send one
message from Alice to Bob in an alphabet \(\mathcal M\). The message
alphabet includes any distinguishable silence and accounts for all
communication. Shared and private random controls are independent
of \(R\), the request and the correlation. Bob may choose his setting
after receiving the message and take his full local transcript
as \(G_i\). Write \(J_N\) for the supremum of \(\sum_i I(R_i:G_i)\)
over these protocols.

An outcome pair outside the support of \(P_{\ne}\) certifies that
the settings agree. Its total probability on equal settings is
\begin{equation}
 \rho=\sum_{a,b:\,P_{\ne}(a,b)=0}P_=(a,b).
 \label{eq:support-mass}
\end{equation}
For each Alice outcome \(a\) of positive common marginal probability
\(\alpha_a\), let \(V_a=P_{\ne}(B=\cdot\mid A=a)\) and
\(W_a=P_=(B=\cdot\mid A=a)\) be Bob's conditional outcome laws.
Write \(\eta_a=W_a(\mathcal B\setminus\operatorname{supp}V_a)\)
for the probability of certification conditional on \(a\). If
\(\eta_a<1\), let \(U_a\) be \(W_a\) conditioned on \(\operatorname{supp}V_a\).
The information from these supported outcomes is bounded by the
relative entropy \(D(U_a\Vert V_a)\), in bits. The resulting upper
bound is
\begin{equation}
 J_N\leq \rho\log_2N+\log_2|\mathcal M|
 +\sum_{a:\,\alpha_a>0,\ \eta_a<1}
     \alpha_a(1-\eta_a)D(U_a\Vert V_a).
 \label{eq:equality-info-upper}
\end{equation}
If \(|\mathcal M|\geq|\mathcal A|\), Alice can send her outcome and achieve
\begin{equation}
 J_N\geq\frac{\rho}{N}\log_2(N!)
 \geq \rho\log_2N-\rho\log_2 e.
 \label{eq:equality-info-lower}
\end{equation}
Thus \(J_N=\rho\log_2N+O(1)\) when the outcome laws and message
alphabet are fixed, and the alphabet can carry Alice's outcome.

\paragraph{Upper bound.}
To bound the sum over alternative requests, we reproduce each
request's transcript from a single collection of classical
variables. Sample the original random controls \(\Lambda\)
independently of the database and request. Conditioned on their
values, Alice chooses her setting \(X\) from \(R\). She generates
her message \(M\) from \(R\) and her outcome \(A\). A skipped query
can be represented by a fixed dummy setting whose outcome Alice
ignores. The marginal of \(A\) is \((\alpha_a)_a\) at every setting,
so we can draw \(A\) independently of \((R,\Lambda)\) and then
compute \(M\) by Alice's rule.

Given \(A=a\) and \(X=x\), draw a table \(T=(T_1,\ldots,T_N)\) of
independent entries. Draw \(T_x\) from \(W_a\) and every other entry
from \(V_a\). At request \(i\), Bob's original setting rule
selects an entry using his random controls and \(i\), and also
\(M\) if he queries after receiving it. Each choice precedes
reading the entry. If Bob queries, the selected entry has the
required law conditional on \(A\) and the settings. If he skips
the query, his transcript needs no entry. Thus each request
has the original joint law of \(R\) and \(G_i\). In this coupling
\(G_i\) is a function of \((\Lambda,A,M,T)\) for fixed \(i\), even
when it contains Bob's entire transcript.

Data processing and independence of the \(R_i\) bound the score by
the information that this collection of variables carries about \(R\).
Since \((\Lambda,A)\) is independent of \(R\), the chain rule gives
\begin{equation}
 \sum_i I(R_i:G_i)
 \leq I(R:\Lambda,A,M,T)
 \leq \log_2|\mathcal M|+I(R:T\mid\Lambda,A,M).
 \label{eq:equality-info-sim}
\end{equation}
Indeed, for any common variable \(Z\),
\(\sum_i I(R_i:Z)\leq I(R:Z)\) by
\(H(R)=\sum_iH(R_i)\) and \(H(R\mid Z)\leq\sum_iH(R_i\mid Z)\).
The message term in Eq.~\eqref{eq:equality-info-sim} is bounded by
\(H(M\mid\Lambda,A)\leq\log_2|\mathcal M|\).

Let \(E\) indicate whether \(T_X\) lies outside \(\operatorname{supp}V_A\),
and put \(C=(\Lambda,A,M)\). Given \(C\) with \(A=a\), the probability
of \(E=1\) is \(\eta_a\) for every database, so \(I(R:E\mid C)=0\).
We may bound \(I(R:T\mid C)\) by \(I(R:T,E\mid C)=I(R:T\mid C,E)\).
On \(E=1\), the conditional table law depends on \(R\) only through \(X\),
so the conditional information is at most \(\log_2N\).

On \(E=0\), the table differs from the reference law \(V_a^{\otimes N}\)
in one entry. At each fixed \(R\) and \(C\), its relative entropy from
this reference is \(D(U_a\Vert V_a)\). To bound the information in
the table, fix a value \(z\) of \(C\) with \(A=a\) and average this
relative entropy over \(R\) conditional on \(C=z,E=0\).
The relative-entropy chain rule gives
\begin{align*}
 &\mathbb E\!\left[
 D(P_{T\mid R,C=z,E=0}\Vert V_a^{\otimes N})
 \,\middle|\, C=z,E=0\right]\\
 &\qquad=I(R:T\mid C=z,E=0)
       +D(P_{T\mid C=z,E=0}\Vert V_a^{\otimes N}).
\end{align*}
The last term is nonnegative, so \(D(U_a\Vert V_a)\) bounds
the conditional information for every posterior distribution of
\(X\) after the message. Averaging over \(C\) and \(E\), and using
\(\sum_a\alpha_a\eta_a=\rho\), gives Eq.~\eqref{eq:equality-info-upper}.

\paragraph{Achievability.}
Use \(N-1\) independent bits with \(r_i=\Pr(R_i=1)=1/(N-i+1)\)
for \(1\leq i\leq N-1\). Alice chooses \(X\) as the first index
with \(R_i=1\), with setting \(N\) assigned to the all-zero database.
For \(i\leq N-1\), the first \(i-1\) bits are zero with
probability \((N-i+1)/N\), and the all-zero probability is \(1/N\).
Thus \(X\) is uniform on \([N]\). Alice sends \(A\). For request
\(i\), Bob queries setting \(i\) and flags an outcome pair outside
\(\operatorname{supp}P_{\ne}\). The flag occurs with probability
\(\rho/N\) and certifies \(X=i\) and \(R_i=1\). For \(\rho>0\), its
contribution to \(I(R_i:G_i)\) is \((\rho/N)\log_2(1/r_i)\).
The other output contributes a nonnegative term. Summing over
the \(N-1\) requests and using \(\ln(N!)\geq\int_1^N\ln t\,\mathrm dt\)
gives Eq.~\eqref{eq:equality-info-lower}, which also holds at \(\rho=0\).

\paragraph{Application to the binary equality correlations.}
Suppose \(p(0,0\mid x,y)=s\mathbf 1_{x=y}\) and the two zero-outcome
marginals are \(ms\), where \(s>0\), \(m\geq1\) and \(2ms\leq1\).
When \(2ms<1\), the outcome \((0,0)\) is the only certifying pair
of positive probability, so \(\rho=s\). Conditional on \(A=0\),
the unequal-setting law of \(B\) is a point mass at one, and the
supported part of the equal-setting law is the same point mass
whenever it has positive weight. For \(A=1\), the equal- and
unequal-setting probabilities of \(B=0\) are
\((m-1)s/(1-ms)\) and \(ms/(1-ms)\). Substituting into
Eqs.~\eqref{eq:equality-info-upper} and \eqref{eq:equality-info-lower}
for a one-bit message gives
\begin{equation}
 \frac{s}{N}\log_2(N!)
 \leq J_N
 \leq s\log_2N+1
 +(1-ms)D_{\mathrm{Ber}}\!\left(
    \frac{(m-1)s}{1-ms}\,\middle\Vert\,\frac{ms}{1-ms}\right),
 \label{eq:optimal-binary-equality-info}
\end{equation}
where \(D_{\mathrm{Ber}}(p\Vert q)\) is the relative entropy in bits
of the Bernoulli laws of parameters \(p\) and \(q\).
For Eq.~\eqref{supp:full-box}, \(m=3\) and the normalization satisfies
\(C_L\geq\|S_0\|^2\geq4(2L-1)^2\), so \(s_L\leq1/36\).
The inequality \(\ln t\leq t-1\) gives
\(D_{\mathrm{Ber}}(p\Vert q)\leq(p-q)^2/[q(1-q)\ln2]\) for
\(0<q<1\). The last term in Eq.~\eqref{eq:optimal-binary-equality-info}
is therefore at most
\(s_L(1-3s_L)/[3(1-6s_L)\ln2]\leq s_L\log_2 e\).

At \(2ms=1\), the pair \((1,1)\) also certifies equality, with
probability \(s\). Conditional on either value of \(A\), Bob's
unequal-setting law is a point mass, so the relative-entropy
term vanishes. With \(\rho=2s\), the one-bit score satisfies
\begin{equation}
 \frac{2s}{N}\log_2(N!)\leq J_N\leq2s\log_2N+1.
 \label{eq:optimal-binary-equality-boundary}
\end{equation}

\subsection{The optimal information score under local product noise}
\label{supp:noise-exponent}

Let \(p_s(0,0\mid x,y)=s\mathbf 1_{x=y}\), with zero-outcome
marginals \(3s\) on both sides and \(0<s<1/6\). Mix \(p_s\)
with a setting-independent product correlation \(U_N\) whose
zero-outcome marginal on each side is \(0<u_N\leq1/2\).
For a fixed mixture weight \(0<\epsilon<1\), the parties use
\(q_N=(1-\epsilon)p_s+\epsilon U_N\) with the mixture label hidden.
Write \(c=(1-\epsilon)s\) for the diagonal excess and
\(t_N=\epsilon u_N^2\) for the unequal-setting joint-zero
probability. The equal-setting joint-zero probability is \(c+t_N\).
Over the one-copy, one-bit protocols defined above, the optimal
score for \(N\geq2\) is
\begin{equation}
 J_N(q_N)=c\log_2\!\left(\frac{N}{1+Nt_N/c}\right)
              +O_{s,\epsilon}(1),
 \label{eq:supp-noise-crossover}
\end{equation}
with an additive bound uniform in \(N\) and \(u_N\).

\paragraph{Upper bound.}
The mixture label is independent of the database. Append it to
Bob's output in each request. The score of these enlarged outputs
is the weighted sum of the two scores conditional on the label
and upper bounds the original score by data processing. The
product branch has one-bit score at most one. The certificate
branch obeys Eq.~\eqref{eq:optimal-binary-equality-info}, so
\(J_N(q_N)\leq c\log_2 N+O_s(1)\).

A second bound follows from the equal- and unequal-setting
outcome laws \(P_=\) and \(P_{\ne}\) of \(q_N\). Both laws have
full support, so \(\rho=0\) in Eq.~\eqref{eq:equality-info-upper}.
Their Alice marginal is common. The relative-entropy chain
rule thus gives \(J_N(q_N)\leq 1+D(P_=\Vert P_{\ne})\).
In this divergence the \(01\) and \(10\) terms are nonpositive
since each equal-setting probability is lower by \(c\).
The \(00\) term is at most
\(c\log_2(1+c/t_N)+c\log_2 e\), by applying
\(\ln(1+c/t_N)\leq c/t_N\) to the part weighted by \(t_N\).
The \(11\) term is
\(P_=(11)\log_2[1+c/P_{\ne}(11)]\leq
c\log_2 e/[(1-\epsilon)(1-6s)]\).
The two bounds together give
\(J_N(q_N)\leq c\log_2\min\{N,1+c/t_N\}+O_{s,\epsilon}(1)\).
For \(B=c/t_N>0\),
\[
 \min\{N,1+B\}
 \leq (1+1/B)\min\{N,B\}
 \leq 2(1+t_N/c)\frac{N}{1+Nt_N/c}.
\]
The logarithm of the prefactor contributes at most
\(c+t_N\log_2e\) to the bound on \(J_N(q_N)\).

For the constructed family, \(s\leq1/36\) and
\(P_{\ne}(11)\geq1/4\), since both \(1-6s\) and \((1-u_N)^2\)
are at least \(1/4\). The mixture-label bound has additive
term at most \(1+c\log_2e\) by
Eq.~\eqref{eq:optimal-binary-equality-info}, and the
relative-entropy bound has additive term at most
\(1+5c\log_2e\) since \(P_=(11)\leq1\).
Including the prefactor gives a total additive upper bound of
\(1+c+(5c+t_N)\log_2e\), uniform in the hierarchy level.

\paragraph{Lower bound.}
Use the \(N-1\) independent bits of
Sec.~\ref{supp:optimal-equality-score}, with
\(r_i=\Pr(R_i=1)=1/(N-i+1)\).
The first-one setting \(X\) is uniform on \([N]\), including
setting \(N\) for an all-zero database. Alice sends her outcome
for every setting. For request \(i\), Bob queries \(i\) and
flags the joint-zero outcome. The flag probability is
\(t_N+c\mathbf 1_{X=i}\) conditional on \(X\), and \(X=i\)
implies \(R_i=1\). Consequently the marginal flag probability
is \(v=\Pr(G_i=1)=t_N+c/N\) and its joint probability with
\(R_i=1\) is \(w_i=\Pr(G_i=1,R_i=1)=t_Nr_i+c/N\).

The flag contributes \(vD_{\mathrm{Ber}}(w_i/v\Vert r_i)\) to
the mutual information. The term weighted by \(v-w_i\) in
this divergence is bounded below by
\(-(w_i-vr_i)\log_2e\) using \(a\ln(a/b)\geq a-b\).
The unflagged output contributes a nonnegative amount, so
\begin{equation}
 I(R_i:G_i)\geq w_i\log_2\frac{w_i}{vr_i}
                    -(w_i-vr_i)\log_2e .
\end{equation}
Here \(w_i-vr_i=c(1-r_i)/N\). With \(k=N-i+1\), the first
logarithm has argument \((t_N+ck/N)/(t_N+c/N)\geq1\).
We may therefore replace its coefficient \(w_i\) by \(c/N\)
in the lower bound. Its argument is at least
\(k/(1+Nt_N/c)\). Summing over \(k=2,\ldots,N\) and using
\(\ln(N!)\geq N\ln N-N\) yields
\begin{equation}
 \begin{aligned}
 J_N(q_N)
 &\geq \frac{c}{N}\log_2(N!)
       -c\left(1-\frac1N\right)\log_2(1+Nt_N/c)-c\log_2e\\
 &\geq c\log_2\!\left(\frac{N}{1+Nt_N/c}\right)-2c\log_2e .
 \end{aligned}
\end{equation}

\paragraph{Noise that decreases as a power of \(N\).}
For \(u_N=\tfrac12N^{-\alpha/2}\) with \(\alpha\geq0\), the
denominator in Eq.~\eqref{eq:supp-noise-crossover} is
\(1+(\epsilon/4c)N^{1-\alpha}\). The optimal score is then
\begin{equation}
 J_N(q_N)=c\min\{\alpha,1\}\log_2N+O_{s,\epsilon}(1).
 \label{eq:supp-noise-power-law}
\end{equation}
All outcome probabilities are at least
\(t_N=(\epsilon/4)N^{-\alpha}\), the smallest contribution
from the product branch. At \(\alpha=0\), the uniform
noise bounds the score independently of \(N\). At every
\(\alpha>0\) the score diverges, with coefficient \(c\)
when \(\alpha\geq1\). If \(p_s\in Q_L\), then \(q_N\in Q_L\) by convexity. For \(N>K\) and
\((K+5)s\leq1\), every restriction to at most \(K\) settings
per party stays local by Eq.~\eqref{eq:supp-rectangular-locality}.

\subsection{Strict violations with positive outcome probabilities}
\label{supp:noise}

Fix a finite protocol of Sec.~\ref{supp:ic-lower}. Mix \(p_L\) with
the local correlation \(U\) with \(U(a,b\mid x,y)=1/4\) at every setting pair,
using a mixing weight \(0\le\delta\le1\). By convexity,
\(p_\delta=(1-\delta)p_L+\delta U\) is at level \(L\), with every
outcome probability positive for \(\delta>0\). The mixing adds false flags and changes Bob's output probabilities. For each database string,
and hence conditional on any bit value, the guess law changes by at
most \(\delta\) in total variation, the half-sum of the absolute
differences of probabilities. For
\(0\le\delta\le 1/2\), couple the old and new binary guesses so their
probability of disagreement is at most \(\delta\). Either guess plus
the disagreement indicator determines the other, so the entropy changes
by at most the binary entropy
\(h_2(\delta)=-\delta\log_2\delta-(1-\delta)\log_2(1-\delta)\),
continued to zero. Apply this bound to the unconditional and
conditional guess entropies for each of the \(n_R\) database bits,
\begin{equation}
 I_{\mathrm{IC}}(p_\delta)\ \ge\
 I_{\mathrm{IC}}(p_L)-2n_R h_2(\delta).
 \label{supp:noise-bound}
\end{equation}

Every strict finite violation persists for positive \(\delta\) sufficiently
close to zero. Here \(n_R=N\) for the sparse bits and \(n_R=kN\) for the
fair bits in blocks. Choosing \(k=\lceil4/s_L\rceil\) and \(N=k2^k\)
gives a score above \(2\) by Eq.~\eqref{supp:ic-geometric}. Hence
\(0<\delta\le1/2\) with \(2kN h_2(\delta)<1\) preserves a violation.

\subsection{A noise tolerance independent of the setting count}
\label{supp:uniform-noise}

Conditioning on the requested bit gives a noise tolerance independent
of the number of settings. Use the protocol of Sec.~\ref{supp:ic-lower}
with $r=1/N$, and write $s=s_L$ for the exact certificate's diagonal
probability. Let $\widetilde p$ be any nonsignalling correlation within
total variation distance $\delta$ of $p_L$ at every setting pair. The
score on $\widetilde p$ is at least $2$ whenever
\begin{equation}
 N\ge\left\lceil4\mathrm e\,2^{4/s}\right\rceil,
 \qquad
 \delta\le \frac{s}{4\mathrm e}\,2^{-4/s} .
 \label{supp:uniform-noise-radius}
\end{equation}
At fixed $L$ this choice allows the same positive noise tolerance for
every sufficiently large $N$. For $\widetilde p=(1-\eta)p_L+\eta U$,
where $U$ is the uniform local
correlation and $0<\eta\le\delta$, convexity gives
$\widetilde p\in Q_L$, and every outcome probability is positive.

We compare the flag probability given a one with the unconditional
flag probability. Under $\widetilde p$, write $a_i=\Pr(G_i=1\mid R_i=1)$
and $b_i=\Pr(G_i=1)$ for these probabilities. Conditional on any nonzero database
string, the flag probability is $\widetilde p(0,0\mid x,i)$, where $x$
is the position of the first one. On the all-zero string Alice sends
one without needing the box, so the flag probability is zero for both
correlations. The per-setting total variation bound thus controls the
change in each flag probability by $\delta$, even after conditioning on
$R_i=1$. In particular, when $R_i=1$ and the preceding entries are zero,
the diagonal flag probability is at least $s-\delta$. Hence
\begin{align}
 a_i &\ge (s-\delta)(1-r)^{i-1}, &
 b_i &\le sr(1-r)^{i-1}+\delta\le s/N+\delta .
 \label{supp:uniform-noise-probabilities}
\end{align}
The mean $\bar a=N^{-1}\sum_i a_i$ is at least
$(s-\delta)[1-(1-1/N)^N]$. Under Eq.~\eqref{supp:uniform-noise-radius},
$\delta\le s/8$, so $\bar a\ge (7s/8)(1-\mathrm e^{-1})>s/2$.

Write $b_*=s/N+\delta$ and let ${\rm Ber}(a)$ denote the Bernoulli law
with parameter $a$. The contribution from $R_i=1$ gives
$I(R_i:G_i)\ge rD({\rm Ber}(a_i)\Vert{\rm Ber}(b_i))$.
In bits, $(1-a)\ln(1-a)\ge -a$ gives
$D({\rm Ber}(a)\Vert{\rm Ber}(b))\ge a\log_2[a/(\mathrm e b)]$.
To sum the contributions, replace each $b_i$ by its upper bound
$b_*$ and use convexity of $a\log_2 a$ to lower bound the sum by its
value at the mean $\bar a$,
\begin{equation}
 I_{\rm IC}(\widetilde p)\ge
 \bar a\log_2\frac{\bar a}{\mathrm e b_*}
 \ge \frac{s}{2}\log_2\frac{s}{2\mathrm e b_*}\ge 2 .
\end{equation}
Here $b_*\le (s/(2\mathrm e))2^{-4/s}$, and
$a\log_2[a/(\mathrm e b_*)]$ is increasing for $a\ge b_*$.

At a fixed positive uniform mixing weight $\eta$, the score of this
protocol stays bounded as $N$ grows, even if Bob keeps the record
$T_i=(M,b)$ of the message and his outcome for request $i$. The
uniform part puts a floor under every transcript probability.
With $c=1-\mathrm e^{-1}$, every value
of $T_i$ has unconditional probability at least $c\eta/4$, from the
uniform outcomes on nonzero database strings. The divergence of the
$R_i=1$ conditional law from the unconditional law is therefore at
most $\log_2(4/(c\eta))$. The unconditional law contains the $R_i=0$
conditional law with weight $1-r$, so the divergence for that branch
is at most $-\log_2(1-r)$. Summing over requests bounds the score for
the full records and, by data processing, for the flags,
\begin{equation}
 I_{\rm IC}(\widetilde p)\le
 \log_2\frac{4}{c\eta} + \log_2\mathrm e ,
\end{equation}
where $-(1-r)\ln(1-r)\le r$ and $Nr=1$.

\section{Zero-error communication with independent copies}\label{supp:copies}

Alice and Bob receive labels \(i,j\in[N]\). Bob may abstain, and must
give the correct answer whenever he declares the labels equal or unequal.
The success criterion is the minimum over input pairs of the
probability that he gives either conclusive answer.
For one copy Alice uses setting \(i\) and sends her outcome, while Bob
uses setting \(j\). The joint zero outcome certifies equality, and by
Eq.~\eqref{supp:probabilities} it occurs with probability \(s_L\) on equal
inputs. To certify inequality they share a random map from \([N]\) to
\(\{0,1\}\) with independent fair values. Alice sends the hash at \(i\).
The hashes must agree on equal inputs and differ with probability
\(1/2\) on unequal inputs.

Alice sends the outcomes of \(q\) independent copies and her hash bit in
one message. On equal labels Bob has a joint zero outcome in at least one
copy with probability \(1-(1-s_L)^q\). On unequal labels the differing
hashes are conclusive with probability \(1/2\). Bob abstains if neither event occurs. On equal labels the conclusive
probability is at least \(1/2\) when \((1-s_L)^q\leq1/2\). The least
such integer is
\[
 q=\left\lceil\frac{\ln 2}{-\ln(1-s_L)}\right\rceil .
\]
The number of copies is \(q\) and the message has \(q+1\) bits, both
independent of \(N\). The rational level-two choice \(s_2=7/424\) of
Sec.~\ref{supp:optimized-level-two} gives \(q=42\) and \(43\) bits.

Independent copies give another correlation at the same NPA level. The parallel
resource has a local outcome for each tuple of local outcomes in the
constituent copies. Represent its projector by the tensor product of
the component projectors. Completeness and same-setting orthogonality
follow from the corresponding relations in each factor, and tensor
products also preserve idempotence and cross-party commutation. Each
retained word of length at most \(L\) maps to a tensor of component words
of length at most \(L\). The \(u,v\) entry of the product moment matrix
is the product of the entries for the corresponding component words.
Taking tensor products of the component Gram vectors realizes this
matrix and proves its positivity. The parallel resource is therefore
at level \(L\). The hash is sampled independently of these copies.

For \(N=2^n\), the labels are \(n\)-bit strings. Corollary~4 of
Gazi \emph{et al.}~\cite{Gazi2026} gives the exact one-way cost \(n\) bits
at the same worst-case conclusive probability throughout the
commuting-operator model, including tensor-product quantum theory.
Their Proposition~4 gives a lower bound of \(n-1\) bits for interactive
classical communication.
\section{An optimized certificate at level two}\label{supp:optimized-level-two}

At level two the retained words contain at most two projectors.
It suffices to match singletons, pure words with distinct labels,
and the mixed words \(A_iB_j\). Use reference and replacement
vectors \(R,S\) as in Sec.~\ref{supp:certificate}, requiring
\(\langle S,wR\rangle=\langle R,wR\rangle\) on these words.
Each local space has one common direction in each summand, with
a distribution of within-party overlaps. The tensor-product block
supplies the mixed-word pairings and two auxiliary blocks supply
the missing pure-word pairings. The rate is the reciprocal of
\(C=\|S\|^2\). The construction below gives \(C=35+18\sqrt{2}\)
with the original equality and marginal relations.

In the tensor-product block choose the within-party overlap \(d\)
for distinct settings from \(\{0,1/2\}\)
with \(\Pr(d=1/2)=m_0-1\), where \(1<m_0\leq 2\). Inner products in this
block are averages over \(d\).
Let \(g_A,g_B\) be unit common directions and \(e_i,f_i\) private
orthonormal directions in their respective spaces, orthogonal to the common
directions. In this block \(A_i,B_i\) project onto
\[
 u_i=\sqrt d\,g_A+\sqrt{1-d}\,e_i,\qquad
 v_i=\sqrt d\,g_B+\sqrt{1-d}\,f_i
\]
in the respective tensor factors. The reference vector pairs the private
directions at each setting. Its replacement lies along the product of
the common directions, with amplitude chosen to match the average
inner product with the mixed row,
\[
 R_0=\frac{1}{1-d}\sum_{i=1}^N e_i\otimes f_i,
 \qquad S_0=\frac{d}{\mathbb E d^2}\,g_A\otimes g_B.
\]

The mixed row \(A_iB_jR_0=\delta_{ij}u_i\otimes v_j\) has reference pairing
\(\delta_{ij}\) for each \(d\). Pairing this vector with \(S_0\) and
averaging gives \(\delta_{ij}\mathbb E[d^2]/\mathbb E[d^2]=\delta_{ij}\).
With Alice's projector alone,
\(A_iR_0=u_i\otimes f_i/\sqrt{1-d}\) has reference pairing
\(1/(1-d)\), whose average is \(m_0\).
For distinct \(i,j\),
\(A_iA_jR_0=d\,u_i\otimes f_j/\sqrt{1-d}\), whose reference pairing
vanishes by orthogonality of the private directions. The private Bob factor
in both Alice rows is orthogonal to the common factor of \(S_0\). The same
holds for pure Bob rows with the factors interchanged. A repeated label
reduces to a word of length one by idempotence. The tensor-product
replacement costs its mean squared norm,
\[
 C_0=\frac{1}{\mathbb E d^2}=\frac{4}{m_0-1}.
\]

The auxiliary blocks must supply the rest of the pure-word pairings.
The desired singleton moment is three before normalization. Since
the tensor-product reference vector contributes \(m_0\), each auxiliary
reference vector must supply \(\beta=3-m_0\) for its party. In Alice's
block all \(B_i\) are zero and \(A_i\) projects onto
\(\sqrt{t}\,g+\sqrt{1-t}\,e_i\), where \(g\) is a unit common direction,
the \(e_i\) are private orthonormal directions orthogonal to \(g\), and
\(0<t\leq 1\). Here \(t\) is the inner product between vectors
for distinct settings. Let \(R_A=\sqrt{\beta/t}\,g\) and
\(S_A=f(t)R_A\) for real \(f\), with inner products averaged over \(t\).
The reference pairings for \(A_i\) and \(A_iA_j\) with \(i\ne j\) are
\(\beta\) and \(\beta t\), respectively. The corresponding replacement
pairings are \(\beta f(t)\) and \(\beta t f(t)\). Bob's block
interchanges the parties. The total replacement pairing for a
singleton comes entirely from its auxiliary block, so
\(\beta\mathbb E f=3\). At distinct labels the tensor-product pairings
are both zero, so \(\mathbb E[t f]=\mathbb E t\).

The following two-point distribution satisfies both matching conditions
and minimizes the auxiliary squared norm \(\beta\mathbb E[f^2/t]\),
\begin{align*}
 r&=\frac{9}{36-8\beta}, & t&\in\{r,1\}, &
 \Pr(t=r)&=\frac{3-4r}{3-3r},\\
 f(r)&=\frac{4r}{4r-1}, & f(1)&=\frac{4(2r-1)}{4r-1}.
\end{align*}
For \(1\leq\beta<2\) this defines probabilities and gives
\(\mathbb E t=4r/3\), \(\mathbb E f=8r/[3(4r-1)]=3/\beta\), and
\(\mathbb E[t f]=4r/3\). The squared norm in each auxiliary block is
\(C_A=C_B=\beta\mathbb E[f^2/t]=81/\beta-24\).

The direct sums \(R=R_0\oplus R_A\oplus R_B\) and
\(S=S_0\oplus S_A\oplus S_B\) match every required pairing. Dividing
the singleton moments \(3\) and the mixed moments \(\delta_{ij}\) by
the total squared norm gives the zero-outcome marginals \(3/C\)
and \(p(0,0\mid i,j)=\delta_{ij}/C\). The three blocks' costs add,
\[
 C=\frac{4}{m_0-1}+\frac{162}{3-m_0}-48.
\]
Raising \(m_0\) lowers the tensor-product cost and increases the
auxiliary costs. Minimizing their sum over \(1<m_0\leq2\)
gives \(35+18\sqrt{2}\) at
\(m_0=1+4/(2+9\sqrt{2})\). The rational choice \(m_0=5/4\) gives
\(r=9/22\), \(\Pr(t=r)=10/13\), \(f(r)=18/7\), \(f(1)=-8/7\),
and \(s_2=7/424\).

\paragraph{Optimality within the three-block rank-one construction.}
For \(N\geq 2\), allow any distribution on \(0\leq d<1\) in the
tensor-product block and a variable amplitude in
\(R_0=q(d)(1-d)^{-1}\sum_i e_i\otimes f_i\).
The reference pairing for \(A_iB_i\) requires \(\mathbb E q^2=1\).
The singleton reference pairing gives \(m_0=\mathbb E[q^2/(1-d)]\).
A replacement along \(g_A\otimes g_B\) must have averaged mixed pairing
one. Cauchy--Schwarz then gives \(C_0\geq 1/\mathbb E[q^2d^2]\).
Using \(d^2\leq d/[4(1-d)]\) gives \(C_0\geq 4/(m_0-1)\).

In the auxiliary blocks allow any distribution on \(0<t\leq1\) and
any reference amplitude along the common direction. The measure
weighted by \(t\|R_A\|^2\) has total mass \(\beta=3-m_0\).
Normalize it to a probability measure and write \(S_A=f R_A\) on
the support of the reference amplitude. This gives the same matching
conditions and the cost \(\beta\mathbb E[f^2/t]\) on that support.
Where the reference amplitude is zero, the replacement can only add to the squared norm.

To prove a lower bound on the quadratic cost, subtract a square built
from \(f(t)/t\) and a linear function \(\alpha+\gamma t\). The cross
term depends only on the known averages of \(f\) and \(tf\).
For equality, choose the line interpolating \(f(t)/t\) at the two
proposed support points. Put \(n=4r-1\),
\(\alpha=4(1+2r)/n\), and \(\gamma=-8/n\) for the same \(r\) as above.
Completing the square with \(\alpha+\gamma t\) and using the two
matching conditions gives
\begin{align*}
 \beta\mathbb E[f^2/t]
 &\geq \beta\mathbb E\big[2(\alpha+\gamma t)f-t(\alpha+\gamma t)^2\big]\\
 &=6\alpha+\beta\mathbb E\big[2\gamma t-t(\alpha+\gamma t)^2\big].
\end{align*}
The final integrand is \(64[(1-t)(t-r)^2-r^2]/n^2\), bounded below by
\(-64r^2/n^2\) because \(t\leq1\). Substituting \(r\) gives
\(C_A\geq81/\beta-24\) and the same bound for \(C_B\). At the two
support points \(t=r,1\) the integrand attains its lower bound and
\(f(t)=t(\alpha+\gamma t)\) saturates the square inequality.
The sum of the three bounds on \(C\) is minimized over \(1<m_0<3\) at the
same \(m_0\) as above, which lies in the range of the construction.
\section{Dependence on the hierarchy level}\label{supp:level-dependence}

\subsection{An upper bound on exact equality certification}
\label{supp:rate-upper}

At higher levels, positivity constrains sequences of projectors that
involve more settings. Suppose a level-\(L\) moment matrix has
\(p(0,0\mid i,j)=0\) for \(i\ne j\). We use this condition to compare
every diagonal certificate to a different branch of a single sequence
of tests. Choose \(m\) distinct settings numbered \(1,\ldots,m\), with
\(2\le m\le\min(N,L)\). Write \(Q_i=1-A_i\) for the projector onto
outcome one. Alice's \(j\)th test is affirmative if its outcome is
zero, in the range of \(A_j\). The first affirmative result among
the first \(m-1\) tests gives branch \(j<m\), and all negative
results give branch \(m\). The polynomial for a branch places the
earliest test at the right,
\[
 F_j=A_j Q_{j-1}\cdots Q_1\quad (j<m),\qquad
 F_m=Q_{m-1}\cdots Q_1 .
\]
The branch probabilities have a common budget of one. With
\(U_j=Q_j\cdots Q_1\) and \(U_0=1\), the relation
\(A_j^2+Q_j^2=1\) gives
\(F_j^*F_j+U_j^*U_j=U_{j-1}^*U_{j-1}\) for \(j<m\). Since
\(F_m=U_{m-1}\), summing gives \(\sum_{j=1}^{m} F_j^*F_j=1\).

To compare an individual diagonal certificate with its branch, work
with the seminorm \(\|w\|_\ell^2=\ell(w^*w)\) of polynomials of
degree at most \(L\), where \(\ell\) is the moment functional.
Positive semidefiniteness makes every vector
of seminorm zero orthogonal to all retained vectors. If \(P\) is an
outcome projector and \(w,Pw\) both have degree at most \(L\), then
\(\|Pw\|_\ell^2+\|(1-P)w\|_\ell^2=\|w\|_\ell^2\). Thus multiplying
a zero-seminorm vector on the left by \(P\) preserves its zero seminorm
within the degree bound.

For \(i\ne j\), the joint projector \(A_iB_j\) has squared seminorm
\(p(0,0\mid i,j)=0\). On Bob's outcome-zero branch at setting \(j\),
every earlier affirmative outcome in Alice's sequence has such a
zero vector. Expand the difference between the sequence of earlier
negative tests and the identity to isolate those zero vectors,
\begin{equation}
 Q_{j-1}\cdots Q_1-1
 =-\sum_{i<j} Q_{j-1}\cdots Q_{i+1}A_i .
 \label{supp:telescoping-rows}
\end{equation}
For \(j<m\), each term of \(F_jB_j-A_jB_j\) is obtained from
\(A_iB_j\) by left multiplication by the indicated \(Q\)'s and \(A_j\).
The degree is at most \(j+1\le m\), so the difference has zero seminorm.
At \(j=m\), apply the same expansion with \(F_m=U_{m-1}\). It gives
\(F_mB_m=B_m\) up to a zero-seminorm vector, with degrees at most \(m\).

Since \(B_j\) and \(1-B_j\) commute with \(F_j\), they split
\(\|F_j\|_\ell^2\) into two nonnegative squared seminorms, one of which
is \(\|F_jB_j\|_\ell^2\). The latter seminorm is the diagonal
probability for \(j<m\) and the marginal for \(j=m\). All the
polynomials being squared have degree at most \(m\). The branch
budget \(\sum_j F_j^*F_j=1\) therefore gives
\begin{equation}
 \sum_{j<m}p(0,0\mid j,j)+p_B(0\mid m)\le1 .
 \label{supp:packing-bound}
\end{equation}
If every diagonal probability is \(s\), then \(p_B(0\mid m)\ge s\),
so \(ms\le1\). A family with a uniform rate for every \(N\) satisfies
\(s_L\le 1/L\). The marginal in Eq.~\eqref{supp:full-box} is \(3s_L\),
which gives \(C_L\ge L+2\) for \(N\ge L\). For \(L\ge N\ge2\), take
\(m=N\) in Eq.~\eqref{supp:packing-bound}. Since
\(p_B(0\mid m)\ge p(0,0\mid m,m)\), the sum of all \(N\) diagonal
probabilities is at most one even when they are unequal. The upper
bound \(1/N\) on the average is attained by drawing a shared uniform
label and outputting \(0\) precisely when the local input equals
that label.

\subsection{A quadratic bound on the probability of exact equality certification}
\label{supp:quadratic-packing}

The sequential argument uses one degree for each tested setting.
A polynomial in the sum of Alice's projectors gives a bound using
all settings at once. Take a level-$L$ correlation, with $L\ge2$ and
$N$ settings per party, whose joint zero outcome vanishes whenever
$i\ne j$. Write $s_i=p(0,0\mid i,i)$ for the diagonal rate at setting
$i$. The positive odd integer $K$ controls the polynomial degree.
For every $K\le2L-3$ with $K^2<6N+1$, the sum of the diagonal
rates obeys
\begin{equation}
 \sum_{i=1}^N s_i\le
 \frac{N}{K^2}\left(1-\frac{K^2-1}{6N}\right)^{-2} .
 \label{eq:quadratic-packing}
\end{equation}
With $K=2L-3$, this gives $s_L\le (2L-3)^{-2}$ for any common rate
feasible for arbitrarily many settings at level $L$. For finite $N$,
choose the largest odd $K\le\sqrt{2N}$ and take $L\ge(K+3)/2$.
Equation~\eqref{eq:quadratic-packing} bounds the average diagonal rate
by $(9/8+O(N^{-1}))/N$.
Indeed, $K^2/N=2+O(N^{-1/2})$, and $[y(1-y/6)^2]^{-1}$ has its
stationary point at $y=2$, where the value is $9/8$. For quantum
correlations the sequential argument can use the entire list of
settings, giving the exact average rate $1/N$, attained by the
shared-label strategy above.

Put $J=\sum_{j=1}^{N}A_j$. Its first two moments on Bob's outcome-zero
branch at setting $i$ are fixed by $s_i$. To derive them, let $\ell$ be
the level-$L$ moment functional. Use the inner product on moment
vectors, $\langle f,g\rangle_\ell=\ell(f^\dagger g)$ for polynomials of
degree at most $L$, with $\dagger$ denoting the adjoint.
Write $\|f\|_\ell^2=\ell(f^\dagger f)$ for the squared seminorm.
Every term $A_jB_i$ of $JB_i$ with $j\ne i$ is a zero moment vector
because it is a projector with zero probability. Thus $JB_i$ and
$A_iB_i$ have the same pairings. Pairing $JB_i$ with $B_i$ and with
itself gives
\begin{equation}
 \ell(B_i J B_i)=\ell(B_i J^2 B_i)=s_i .
 \label{eq:quadratic-packing-moments}
\end{equation}

We seek a polynomial $q$ for which $F_i=A_iq(J)$ has a small total
squared seminorm over all $i$ and still pairs strongly with
$A_iB_i$ on Bob's branch. The total is $\ell(Jq(J)^2)$. We can
bound it by choosing $q$ so that $xq(x)^2$ is uniformly small
on $[0,N]$ and proving the bound by squares of low-degree
polynomials after substituting $x=J$. Set $d=(K-1)/2\le L-2$
and write $x=N\sin^2\theta$ with $0\le\theta\le\pi/2$. A sine
quotient makes $xq(x)^2$ proportional to a bounded sine squared,
and its cosine counterpart supplies the difference from the bound,
\begin{align}
 q(x)&=\frac{\sin(K\theta)}{K\sin\theta}
     =\frac{1+2\sum_{k=1}^{d}\cos(2k\theta)}{K},
 \qquad
 t(x)=\frac{\cos(K\theta)}{\cos\theta} .
\end{align}
Both quotients extend continuously to the endpoints and are polynomials
in $x$ of degree at most $d$. The sine and cosine squares yield a
polynomial identity for the difference between $xq(x)^2$ and its
upper bound,
\begin{equation}
 \frac N{K^2}-xq(x)^2 =\frac{N-x}{K^2} t(x)^2 .
 \label{eq:quadratic-packing-filter-upper}
\end{equation}
The identity is useful because $N-J=\sum_j(1-A_j)^2$. At $x=J$,
the right side is the sum of squares of
$K^{-1}(1-A_j)t(J)$, whose degrees are at most $d+1$. Hence
$\ell(Jq(J)^2)\le N/K^2$. This is the required upper bound on
$\sum_i\|F_i\|_\ell^2$ since $A_i^2=A_i$.

For the diagonal pairing, keep the linear and quadratic terms of
$xq(x)$, since their moments at $B_i$ are already fixed by
Eq.~\eqref{eq:quadratic-packing-moments}. Put $a=(K^2-1)/6$,
$\Delta=1-a/N>0$ and let $h(x)=x[q(x)-1+ax/N]$ be the
remainder after subtracting $x-(a/N)x^2$. A sum of squares for
$B_i h(J)B_i$ within the allowed degree would give the required
lower bound. The cosine expansion of $q$ and
$a=(4/K)\sum_{k=1}^{d}k^2$ give
$h(x)=\frac{4x^2}{KN}\sum_{k=1}^d[k^2-\sin^2(k\theta)/\sin^2\theta]$.
The squared sine quotient is the squared modulus of a geometric
sum. Group the terms by the difference of their indices to write
$h$ as a nonnegative sum,
\begin{align}
 k^2-\left|\sum_{j=0}^{k-1} e^{2ij\theta}\right|^2
   &=4\sum_{r=1}^{k-1}(k-r)\sin^2(r\theta),\\
 h(x)&=\frac{16x^2}{KN}\sum_{k=1}^{d}\sum_{r=1}^{k-1}
                   (k-r)\sin^2(r\theta) .
 \label{eq:quadratic-packing-filter-lower}
\end{align}

Each sine squared in Eq.~\eqref{eq:quadratic-packing-filter-lower}
must be converted to a polynomial square of known degree.
For odd $r$, $P_r(x)=\sin(r\theta)/\sin\theta$ is a polynomial of
degree at most $(r-1)/2$. For even $r$,
$P_r(x)=\sin(r\theta)/(\sin\theta\cos\theta)$ has degree at most
$(r-2)/2$. Expand the imaginary part of
$(\cos\theta+\mathrm{i}\sin\theta)^r$ and substitute
$\sin^2\theta=x/N$ and $\cos^2\theta=1-x/N$ to obtain these degrees.
Accordingly, the summands $x^2\sin^2(r\theta)$ in
Eq.~\eqref{eq:quadratic-packing-filter-lower} are
$x^3P_r(x)^2/N$ for odd $r$ and $x^3(N-x)P_r(x)^2/N^2$ for even $r$.

The remaining factors $x$ and $N-x$ are converted to sums of squares
by the projector relations. For an odd $r$, substitute $J$ and use
$J=\sum_j A_j^2$.
Then $B_i J^3 P_r(J)^2 B_i$ is the sum of the squares of
$A_j J P_r(J)B_i$. For an even $r$, use
\begin{equation}
 J(N-J)=\sum_{j<k}(A_j-A_k)^2
\end{equation}
to express $B_i J^3(N-J)P_r(J)^2 B_i$ as the sum of the squares of
$(A_j-A_k)J P_r(J)B_i$. In both cases the polynomials being squared
have degree at most $\deg P_r+3\le d+2\le L$, since $r\le d-1$.
Thus $\ell(B_i h(J)B_i)\ge0$. The contribution from the two
known moments in Eq.~\eqref{eq:quadratic-packing-moments}
is $(1-a/N)s_i$, which gives the diagonal lower bound
\begin{equation}
 \ell(B_i Jq(J) B_i)\ge\ell(B_i[J-(a/N)J^2]B_i)=\Delta s_i .
\end{equation}

The operator $A_i$ is a projector and $A_iB_i$ has the same
pairings as $JB_i$, so the overlap is
\begin{align}
 \langle A_iB_i,F_i B_i\rangle_\ell
 &=\langle A_iB_i,q(J)B_i\rangle_\ell
  =\langle JB_i,q(J)B_i\rangle_\ell
  =\ell(B_i Jq(J)B_i) .
\end{align}
Since $\|A_iB_i\|_\ell^2=s_i$, Cauchy--Schwarz gives
$\|F_i B_i\|_\ell^2\ge\Delta^2 s_i$ when $s_i>0$. The bound also
holds at $s_i=0$ by positivity. Cross-party commutation gives
$\|F_i\|_\ell^2=\|F_i B_i\|_\ell^2+\|F_i(1-B_i)\|_\ell^2$,
with all degrees at most $d+2$. Summing and using
Eq.~\eqref{eq:quadratic-packing-filter-upper} gives
\begin{equation}
 \Delta^2\sum_i s_i\le\sum_i\|F_i\|_\ell^2
    =\ell(Jq(J)^2)\le N/K^2 .
\end{equation}
Dividing by $\Delta^2$ gives Eq.~\eqref{eq:quadratic-packing}.
\subsection{Loss of positivity at higher degree}\label{supp:higher-degree}

For large enough \(N\), the functional in Eq.~\eqref{supp:adjusted-functional}
takes a negative value on a square beyond the tested degree. The
term from the character can be negative because the norm of the
reference vector grows with \(N\). We choose a polynomial that
keeps this term fixed while reducing the contribution from the
reference vector. Let \(J=\sum_i A_i\) in the reference representation
and put \(f_m=(1-J/N)^m\) for a positive integer \(m\). The character
sends \(J\) to zero, so \(\varepsilon(f_m)=1\). On positive
eigenvalues of \(J\), increasing \(m\) attenuates the corresponding
component of \(R_N\).

It suffices to quantify the attenuation in the tensor-product block.
There \(J\) acts on Alice's factor as \(\sum_i |u_i\rangle\langle u_i|\).
If \(U\) has columns \(u_i\), then
\(UU^*\) and \(U^*U\) have the same nonzero eigenvalues. The latter is
the Gram matrix \((X_{ij})_{i,j=1}^N+I_N/2\), where \(I_N\) is the
\(N\)-dimensional identity matrix. The common directions give
the positive semidefinite first term and the private directions give
\(I_N/2\), so the nonzero eigenvalues are at least \(1/2\). They are at
most \(N\) because \(\|A_i\|\le1\). To bound the reference weight
supported on those eigenvalues, let \(\Pi\) project onto the span of
the \(u_i\) on Alice's factor. Each unit vector \(u_i\) lies in the span,
so \(\|\Pi e_i\|\ge|\langle u_i,e_i\rangle|=1/\sqrt 2\). The \(f_i\)
on Bob's factor are orthonormal, giving
\[
 \|(\Pi\otimes I)R_0\|^2
   =4\sum_i\|\Pi e_i\|^2\ge 2N.
\]

On this subspace, the squared operator norm of \(f_m\) is at most
\((1-1/(2N))^{2m}\). On the full reference space \(f_m\) is a contraction
because \(0\le J\le N I\). Thus \(\|f_mR_N\|^2\) is smaller than
\(\|R_N\|^2\) by at least \(2N[1-(1-1/(2N))^{2m}]\). Substituting
this estimate and \(\varepsilon(f_m^*f_m)=1\) in
Eq.~\eqref{supp:adjusted-functional} gives
\begin{equation}
 \ell(f_m^*f_m)
   \le 1-\frac{2N}{C_L}\bigl[1-(1-1/(2N))^{2m}\bigr].
 \label{supp:negative-square}
\end{equation}
For \(N>C_L/2\), choose \(m>L\) sufficiently large to make this bound
negative. This square lies beyond those tested by level \(L\).
At fixed \(N\), Eq.~\eqref{supp:packing-bound} bounds the common
certificate rate by \(1/N\) at level \(N\).

\subsection{Information causality beyond every prescribed restriction of settings}
\label{supp:local-restrictions}

Fix an integer NPA level \(L\geq1\) and an integer \(K\geq1\). We
construct a binary-outcome correlation at level \(L\) with \(N\) settings
per party for which any choice of at most \(K\) settings on each side
admits a local hidden-variable model. For \(N\geq\max\{4,K+1\}\), every
joint outcome can have probability at least \(1/(2N)\), and the
information score with one use of the full correlation and one classical
bit grows without bound as \(N\) increases at fixed \(L\) and \(K\).

Consider a binary-outcome correlation \(p_s\) on settings
\(x,y\in\{1,\ldots,N\}\) with joint zero probability \(s\delta_{xy}\)
and both zero-outcome marginals \(ms\). Here \(\delta_{xy}\) is the
Kronecker delta, and \(s\geq0\) and \(m\geq1\) are such that \(p_s\) is
a valid correlation. The certificate in Eq.~\eqref{supp:full-box} has
\(m=3\). Let \(X\) and \(Y\) be the nonempty sets of labels retained by
Alice and Bob, respectively, and write \(c=|X\cap Y|\) for the number
of common labels. In a deterministic local strategy, let
\(U\subseteq X\) and \(W\subseteq Y\) be the sets on which Alice and
Bob output zero. A strategy of positive weight with both sets nonempty
must have \(U=W=\{j\}\) for some \(j\in X\cap Y\). Otherwise it would
assign a joint zero to a pair of unequal settings. This classification
of strategies follows the argument of Gazi \emph{et al.} for their
almost-quantum example~\cite{Gazi2026}. At each common label \(j\) the
joint zero probability fixes the weight of the corresponding singleton
strategy to \(s\).

The singleton strategies use weight \(cs\) and supply a zero-outcome
marginal \(s\) at each common label. Alice's remaining marginal is
\((m-1)s\) at common labels and \(ms\) at labels in \(X\setminus Y\).
It must be supplied by strategies with \(U\ne\varnothing\) and
\(W=\varnothing\), in which Bob always outputs one.
They need weight at least Alice's largest remaining
marginal, \(s[m-1+\mathbf{1}_{X\setminus Y\ne\varnothing}]\), where
\(\mathbf{1}_E\) equals one when condition \(E\) holds and zero
otherwise. Bob's one-sided strategies, with \(U=\varnothing\) and
\(W\ne\varnothing\), need at least the weight obtained by exchanging
\(X\) and \(Y\). The three classes are disjoint,
so locality requires
\begin{equation}
 s\big[c+2(m-1)+\mathbf 1_{X\setminus Y\ne\varnothing}
                     +\mathbf 1_{Y\setminus X\ne\varnothing}\big]\leq 1.
 \label{eq:supp-rectangular-locality}
\end{equation}
When this inequality holds, assign weight \(s\) to each singleton and
give each one-sided class exactly the weight equal to its largest
remaining marginal. For a class of positive weight, the active party
outputs zero at a given setting when a uniform random variable on
\([0,1]\) lies below the remaining marginal for that setting divided by
the class weight. A single draw thus specifies a deterministic response
at every setting in the class. Omit any class of zero weight and assign
the unused probability to the strategy in which both parties always
output one. The constructed model reproduces the marginals and joint
zero probabilities, which determine all four binary outcome
probabilities. Equation~\eqref{eq:supp-rectangular-locality} is therefore
necessary and sufficient.

For \(m=3\), the rate that makes every restriction local is determined
by the largest bracket in Eq.~\eqref{eq:supp-rectangular-locality} over
\(|X|,|Y|\leq K\). If the sets are identical, the bracket is at most
\(K+4\). Distinct sets have at most \(K-1\) common labels and can
contribute at most two further units from the indicators. For
\(1\leq K<N\), two \(K\)-element sets differing in one label attain the
resulting maximum \(K+5\). Thus all these restrictions are local
exactly when \(s\leq 1/(K+5)\).

To keep the diluted correlation at level \(L\) and meet the locality
condition, set \(L'=\max\{2,L\}\),
\(D=\max\{185(L')^3,K+5\}\), and \(s=1/D\). For \(N\geq K+1\),
mix the level-\(L'\) certificate of rate
\(s_{L'}>1/[185(L')^3]\) with the deterministic correlation returning
\((1,1)\), giving the certificate weight \(s/s_{L'}\). The joint zero
probability and the zero-outcome marginals are reduced in the same
proportion, so the resulting correlation is \(p_s\) with \(m=3\).
Convexity and nesting give \(p_s\in Q_L(N,N,2,2)\), and the bound on
\(s\) gives locality for all prescribed restrictions. The first-one
protocol in Sec.~\ref{supp:ic-lower}, with independent database bits
equal to one with probability \(1/N\), gives a score of at least
\((1-\mathrm e^{-1})s\log_2N\) by Eq.~\eqref{supp:ic-sparse}.

To give every joint outcome positive probability while keeping joint
zeros rare on unequal settings, let \(U_N\) be the setting-independent
local product correlation in which Alice and Bob each output zero with
probability \(N^{-1/2}\). Supply the parties with \(q=(p_s+U_N)/2\)
without disclosing which component was drawn. For
\(N\geq\max\{4,K+1\}\), zero is each party's less probable outcome,
possibly tied with one. Hence the smallest product probability is the
joint zero probability \(1/N\), and \(q(a,b\mid x,y)\geq1/(2N)\) for
each outcome pair \(a,b\in\{0,1\}\) and setting pair \(x,y\). The
mixture remains at level \(L\), and every prescribed restriction
remains local because \(U_N\) itself is local.

Use the first-one protocol on \(q\) with the same database bias.
Alice sends her outcome at the index of the first one, or sends one
if her database consists of zeros. For a request \(i\), Bob uses
setting \(i\) and raises the flag \(G_i=1\) exactly when both his
outcome and Alice's message are zero. Let \(R_i\) be the requested bit
and \(A_N=1-(1-1/N)^N\) the probability that the database contains a one.
Write \(v_i=\Pr(G_i=1)\) for the flag probability and
\(w_i=\Pr(G_i=1,R_i=1)\) for the probability of a flag on a bit that
is one. In the \(p_s\) component, a flag occurs with probability
\(s\) when the requested index is the first one, and every such flag
certifies \(R_i=1\). In the \(U_N\) component, each request has flag
probability \(1/N\) conditional on the database containing a one.
The first-one index, when it exists, is unique, so summing these
contributions over requests gives
\[
 \sum_{i=1}^{N} w_i\geq \frac{sA_N}{2},\qquad
 \sum_{i=1}^{N} v_i=\frac{(s+1)A_N}{2}\leq 1.
\]

To bound the information in a flag, compare the conditional
one-probability \(w_i/v_i\) with the prior \(1/N\). Write
\(\operatorname{Ber}(t)\) for the Bernoulli law with one-probability
\(t\in[0,1]\), \(h_2(t)\) for its entropy in bits, and \(D_{\mathrm{KL}}\)
for the relative entropy in bits. The relative entropy of the posterior
from the prior can be bounded using \(h_2(t)\leq1\),
\[
 D_{\mathrm{KL}}\!\left(\operatorname{Ber}(t)\middle\Vert
                    \operatorname{Ber}(1/N)\right)
 = t\log_2N+(1-t)\log_2\!\frac{1}{1-1/N}-h_2(t)
 \geq t\log_2N-1.
\]
The flag contributes this relative entropy at \(t=w_i/v_i\), weighted
by \(v_i\). The outcome \(G_i=0\) has a nonnegative contribution,
so \(I(R_i:G_i)\geq w_i\log_2N-v_i\). Summing over requests and
using \(A_N\geq1-\mathrm e^{-1}\) gives
\[
 \sum_{i=1}^{N} I(R_i:G_i)\geq
 \frac{(1-\mathrm e^{-1})\log_2N}{2\max\{185(L')^3,K+5\}}-1.
\]
For \(N\geq\max\{4,K+1\}\) this bound exceeds any target \(V\geq0\)
once \(\log_2N>2D(V+1)/(1-\mathrm e^{-1})\). It also diverges when
\(L=o((\log N)^{1/3})\) and \(K=o(\log N)\) simultaneously. A score
above one certifies that the full correlation is Bell nonlocal, since
a local hidden-variable model gives the classical one-bit bound of
Eq.~\eqref{supp:ic-axiom-budget}.
\end{document}